\documentclass [article]{nature}
\usepackage{float}
\usepackage{units}
\usepackage{graphicx}
\usepackage{color}
\usepackage{epstopdf}
\usepackage{amsmath}
\usepackage{amsfonts}
\usepackage{amssymb}
\usepackage{dcolumn}
\usepackage{soul}
\usepackage{multirow}
\usepackage{bm}
\def\onlinecite{\citeonline}

\title{Negative magnetoresistance without well-defined chirality in the Weyl semimetal TaP} 

\author{Frank Arnold$^{1~\ast}$, Chandra Shekhar$^{1~\ast}$, Shu-Chun Wu$^1$, Yan Sun$^1$, Ricardo Donizeth dos Reis$^1$, Nitesh Kumar$^1$, Marcel Naumann$^1$, Mukkattu O. Ajeesh$^1$, Marcus Schmidt$^1$, Adolfo G. Grushin$^2$, Jens H. Bardarson$^2$, Michael Baenitz$^1$, Dmitry Sokolov$^1$, Horst Borrmann$^1$, Michael Nicklas$^1$,  Claudia Felser$^1$, Elena Hassinger$^{1~\dag}$, \& Binghai Yan$^{1,2~\dag}$}

\date{\today}

\begin{document}

\maketitle

\begin{affiliations}
\item Max Planck Institute for Chemical Physics of Solids, 01187 Dresden, Germany
\item Max Planck Institute for the Physics of Complex Systems, 01187 Dresden, Germany
\\
$^{\ast}$ These authors contributed equally to this work.

\end{affiliations}

\begin{abstract}
Weyl semimetals (WSMs) are topological quantum states~\cite{Wan2011} wherein the electronic bands linearly disperse around pairs of nodes, the Weyl points, of fixed (left or right) chirality.
The recent discovery of WSM materials~\cite{Liu:2014bf,Weng:2014ue,Huang:2015uu,Lv2015TaAs,Xu2015TaAs,Yang2015TaAs} triggered an experimental search for the exotic quantum phenomenon known as the chiral anomaly~\cite{Adler1969,Bell1969}.
Via the chiral anomaly nonorthogonal electric and magnetic fields induce a chiral density imbalance that results in an unconventional negative longitudinal magnetoresistance~\cite{Nielsen1983}, the chiral magnetic effect. Recent theoretical work suggests that this effect does not require well-defined Weyl nodes~\cite{Chang2015,Ma2015,Zhong2015}. Experimentally however, it remains an open question to what extent it survives when chirality is not well-defined, for example when the Fermi energy is far away from the Weyl points.
Here, we establish the detailed Fermi surface topology of the recently identified WSM TaP~\cite{Liu2015NbP} via a combination of angle-resolved quantum oscillation spectra and band structure calculations. The Fermi surface forms spin-polarized banana-shaped electron and hole pockets attached to pairs of Weyl points. Although the chiral anomaly is therefore ill-defined, we observe a large negative magnetoresistance (NMR) appearing for collinear magnetic and electric fields as observed in other WSMs~\cite{Huang2015anomaly,Zhang2015ABJ}. 
In addition, we show experimental signatures indicating that such longitudinal magnetoresistance measurements can be affected by an inhomogeneous current distribution inside the sample in a magnetic field~\cite{Yoshida1976}. Our results provide a clear framework how to detect the chiral magnetic effect.
\end{abstract}

In a semimetal the conduction and valence bands touch at isolated points in the three-dimensional (3D) momentum ($k$) space at which the bands disperse linearly. 
Depending on whether the bands are nondegenerate or doubly degenerate, 
such a 3D semimetal is called a Weyl semimetal (WSM)~\cite{volovik2007quantum ,Wan2011,Burkov:2011de} or a Dirac semimetal (DSM)~\cite{murakami2007 ,Young:2012kz, Fang:2012ga}, respectively.
Correspondingly, the band touching point is referred to as a Weyl point or a Dirac point. 
The Dirac point can split into one or two pairs of Weyl points by breaking either time-reversal symmetry or crystal inversion symmetry. 
At energies close to the Weyl points, electrons behave effectively as Weyl fermions, a fundamental kind of massless fermions that has never been observed as an elementary particle~\cite{bertlmann2000anomalies}.
In condensed-matter physics, each Weyl point acts like a singularity of the Berry curvature in the Brillouin zone (BZ), equivalent to magnetic monopoles in $k$ space, 
and thus  always occur in pairs with opposite chirality or handedness~\cite{Nielsen1981}. 
In the presence of nonorthogonal magnetic (\textbf{B}) and electric (\textbf{E}) fields (i.e., $\textbf{E}\cdot \textbf{B}$ is nonzero), 
the particle number for a given chirality is not conserved quantum mechanically, 
a phenomenon known as the Adler-Bell-Jackiw anomaly or chiral anomaly in high-energy physics~\cite{Adler1969,Bell1969, bertlmann2000anomalies}. 
In Weyl semimetals, the chiral anomaly is predicted to lead to a negative magnetoresistance (MR) due to the suppressed backscattering of electrons of opposite chirality~\cite{Nielsen1983,Son:2013jz}.

Theoretically, the chiral anomaly strictly appears only, when the chirality is well-defined, i.e., the Fermi energy is close enough to the Weyl nodes such that the two Fermi surface pockets around Weyl nodes of a pair of opposite chirality are independent~\cite{Nielsen1983}.
Observing the chiral-anomaly-induced MR requires that the applied magnetic field and current be as parallel as possible, as otherwise the negative MR will easily be overwhelmed by the positive contribution of the Lorentz force.
In addition to a negative MR, the chiral anomaly is predicted to induce an anomalous Hall effect~\cite{Xu2011,Yang2011QHE,Burkov:2011de,Grushin2012}, nonlocal transport properties~\cite{Parameswaran2014,Zhang:2015ub} and sharp discontinuities in ARPES signals~\cite{Behrends:2015ux} in WSMs.

%%%%%%% Experiment
The discovery of different WSM materials has stimulated experimental efforts to confirm the chiral anomaly in condensed-matter physics. Recently, negative MR has been reported in two types of WSMs:  WSMs induced by time reversal symmetry breaking, i.e., Dirac semi metals in an applied magnetic field, for example Bi$_{1-x}$Sb$_x$ ($x\approx3 \%$)~\cite{Kim2013}, ZrTe$_5$,~\cite{Li2014ZrTe5} and Na$_3$Bi~\cite{Xiong2015}, and the inversion-asymmetric WSMs TaAs~\cite{Huang2015anomaly,Zhang2015ABJ}, NbP,~\cite{Wang:2015wm} and NbAs~\cite{Yang:2015vz}. 
However, a clear verification of whether the Fermi surface topology supports the chiral anomaly or not, is still lacking in most of the above systems.

In the non-centrosymmetric WSM of the TaAs family two types of Weyl nodes exist at different positions in reciprocal space \cite{Huang:2015uu,Weng:2014ue} and at different energies. Therefore, the Weyl electrons  
will generally coexist with normal electrons. Small changes of the Fermi energy by doping or vacancies can in principle change the topology of the Fermi surface significantly due to the smallness of the carrier density. Therefore, the Fermi energy and resulting Fermi surface topology in a crystal have to be known precisely when linking the negative MR to the chiral magnetic effect.
Extensive angle-resolved photoemission spectroscopy (ARPES) measurements in the TaAs family have shown the existance of Fermi arc surface states and linear band crossings in the bulk band structure in all four materials \cite{Xu2015TaAs,Lv2015TaAs,Yang2015TaAs,Liu2015NbP,Xu2015NbAs}. However, because of the insufficient energy resolution (> 15 meV \cite{Liu2015NbP}), these measurements are not able to make any claims about the presence or absence of a well-defined chirality at the Fermi level.
Quantum oscillation measurements have the advantage of an meV resolution of the Fermi energy level.

%% This work
In this work, we reconstruct the 3D Fermi surface of TaP by combining sensitive Shubnikov--de Haas (SdH) and de Haas--van Alphen (dHvA) oscillations with \textit{ab~initio} band structure calculations reaching a good agreement between theory and experiment. We reveal that the Fermi energy is such that electron and hole Fermi surfaces contain pairs of Weyl nodes. 
Although the chiral anomaly is not well-defined, a large negative MR is measured. We discuss possible explanations for this result considering also the current distribution in our samples \cite{Yoshida1975}. 

\begin{figure}[tb!]
\includegraphics[angle=0,width=16cm,clip]{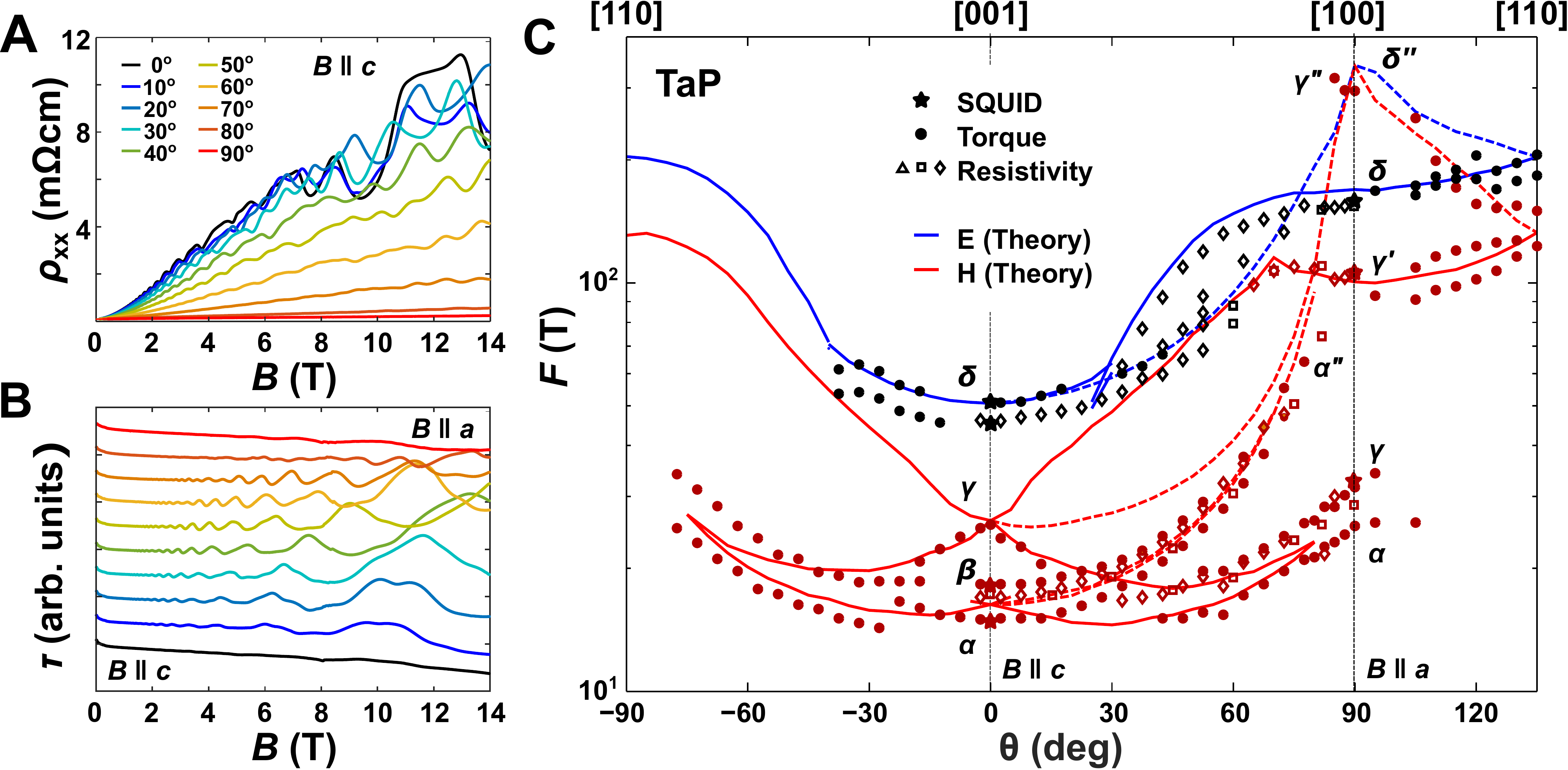}
\caption{{\bf  Quantum oscillations and angular dependence of oscillation frequencies in TaP.} 
{\bf A}. Shubnikov-de Haas oscillations in resistivity for different angles in steps of 10$^{\circ}$. 
{\bf B}. Quantum oscillations from magnetic torque measurements for the same angles.
{\bf C}. Full angular dependence of the measured and theoretical quantum oscillation frequencies. Open and closed symbols refer to SdH and dHvA data of five different samples from two different batches. Lines show the extremal orbits calculated from the banana-shaped 3D Fermi surface topology given in Fig. 3 (solid lines for the pockets lying in the tilting plane of the magnetic field, dashed lines for the pockets lying perpendicular to it). 
} 
\end{figure}

%%%%%%%%%  Methods
We synthesized high-quality single crystals of TaP by using chemical vapor transport reactions and verified 
TaP as a noncentrosymmetric compound in a tetragonal lattice (space group $I4_1 md$, No. 109).
The temperature-dependent resistivity exhibits typical semimetallic behavior. 
For more details see the Supplementary Information (SI). 

%%%%%%%% Fermi surface topology

The Fermi surface topology of TaP was investigated by means of quantum oscillations. Typically, for a semimetal with light carriers and high mobility, such as bismuth~\cite{Edel1976}, prominent oscillations appear in all measurable properties sensitive to the density of states at the Fermi energy. Here, we measured Shubnikov-de Haas (SdH) oscillations in transport (Fig. 1A) and de Haas-van Alphen (dHvA) oscillations in torque (Fig. 1B) and magnetization (Fig. 2A) for different field orientations. These oscillations are periodic in 1/B and their frequency ($F$) is proportional to the extremal Fermi surface cross section ($A_k$) that is perpendicular to \textbf{B} following the Onsager relation $ F = (\Phi_0 / 2\pi^2) A_k$, where $\Phi_0 = h/2e $ is the magnetic flux quantum and $h$ is the Planck constant.
Figure 1A shows the resistivity as a function of the magnetic field for different field orientations. When the electric current and magnetic field are perpendicular ($\textbf{B} || c$), the magnetoresistance is very high as typical for other WSM (for example ref. \onlinecite{Shekhar2015}) and normal semimetals \cite{Kapitza1928,Kopelevich:2003aa}. This implies a very high mobility of the charge carriers.

\begin{figure}
\includegraphics[angle=0,width=16cm,clip]{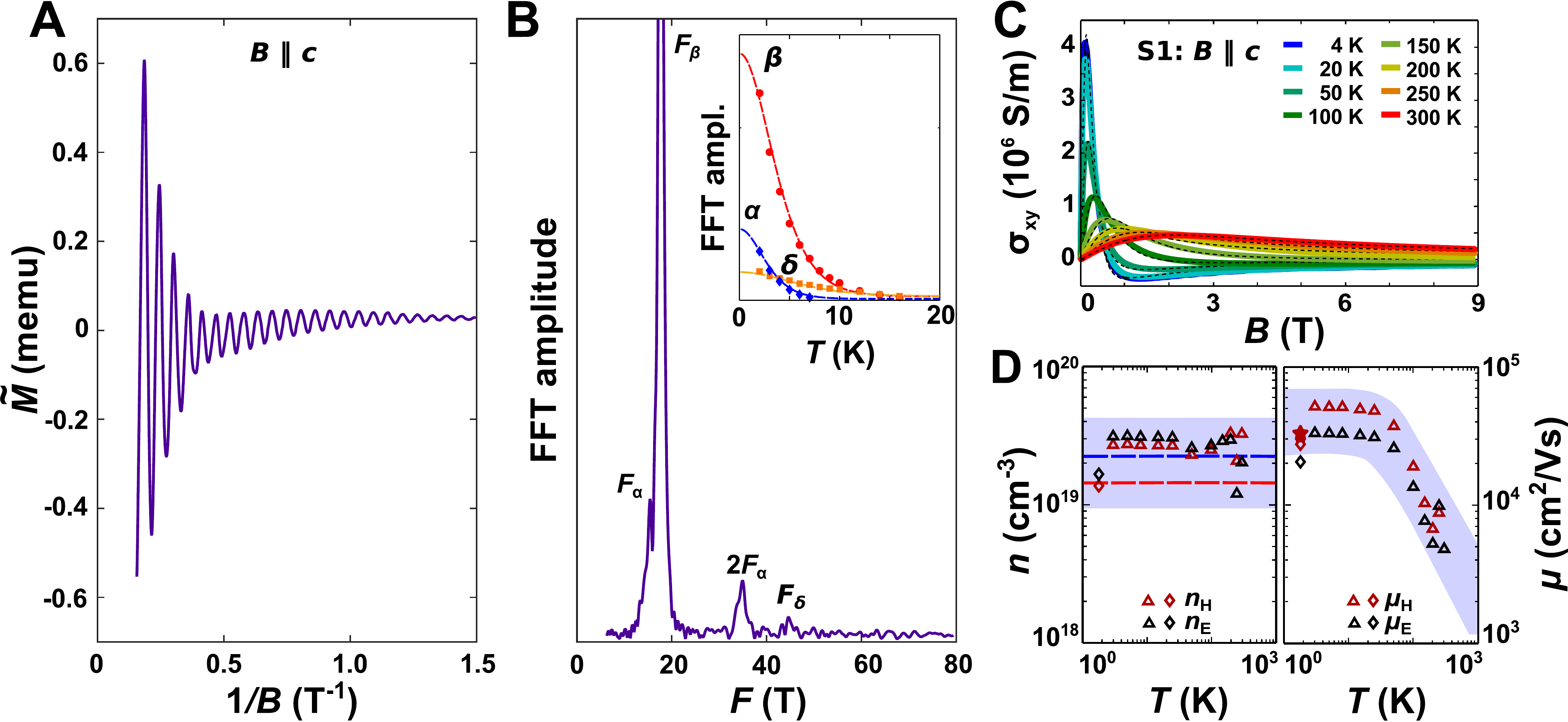}
\caption{{\bf Quantum oscillations, effective masses, carrier density and mobility.}
{\bf A}. Quantum oscillations in the magnetization as a function of the inverse field for $T = 1.85$\,K.
{\bf B}. FFT of the curve in A.
{\bf C}. Hall resistivity of sample S1 for different temperatures as indicated and the fits with a two band model (dashed lines).
{\bf D}. Carrier concentration and mobility of the hole (H) and electron (E) pockets as obtained by fitting the temperature-dependent Hall resistivity of samples S1 and S3, which are labeled by triangles and diamonds, respectively. The gray-shaded areas give the confidence intervals of the densities and mobilities. The blue and red dashed lines mark the theoretical electron and hole densities based on the fitted quantum oscillation and DFT Fermi surface. The hole mobility labeled by a star is determined from dHvA analysis.
}
\end{figure}

Figure 1B depicts the magnetic torque oscillations for the same field orientations.
Figure 2A represents the magnetic dHvA oscillations in the magnetization as a function of the inverse magnetic field and their corresponding Fourier transform (Fig. 2B) for $\textbf{B} || c$. The observable fundamental frequencies from these measurements are $F_{\alpha} = 15$\,T, $F_{\beta} = 18$\,T, $F_{\gamma} = 25$\,T, and $F_{\delta} = 45$\,T. 
The frequencies are consistent within the error bar (given in Table I in the SI) for all three measurement techniques and different sample batches. 
This indicates that all samples have a similar chemical potential to within 1 meV. Additionally, we can conclude that the resistivity measurements are sensitive to the bulk Fermi surface as well.
For $\textbf{B} || a$ the main frequencies are $F_{\alpha} = 26$\,T, $F_{\gamma} = 34$\,T, $F_{\gamma'} = 105$\,T $F_{\delta} = 147$\,T, and $F_{\gamma''} = 320$\,T  clearly indicating anisotropic 3D Fermi surface pockets. For most of the detected oscillation frequencies, we derive their cyclotron effective masses ($m^{\ast}$) of the carriers by fitting the temperature dependence of the oscillation amplitude (insert of Fig. 2B) with the Lifshitz-Kosevich formula (see Ref.~\onlinecite{shoenberg1984magnetic} and the SI). The values of the effective masses are $m^{\ast}_{\alpha} = (0.021\pm0.003)\,m_0$, $m^{\ast}_{\beta} = (0.05\pm0.01)\,m_0$, and $m^{\ast}_{\delta} = (0.11\pm0.01)\,m_0$ for $\textbf{B}||c,$ whereas they are a factor of 4--10 greater for $\textbf{B}||a$ with $m^{\star}_{\gamma} = (0.13\pm0.03)\,m_0$, $m^{\star}_{\gamma'} = (0.35\pm0.03)\,m_0$, and $m^{\ast}_{\delta} = (0.4\pm0.1)\,m_0$, where $m_0$ is the mass of a free electron (see Table I in the SI). These values are small and comparable to the effective masses in other slightly doped Dirac materials such as Cd$_3$As$_2$ or graphene~\cite{Rosenman1969,novoselov2005two} . These low masses, together with long scattering times are the reason for the high mobility and the huge transverse magnetoresistances seen in semimetals.

\begin{figure}[tb!]
\centering
\includegraphics[width=0.6\textwidth,angle =90]{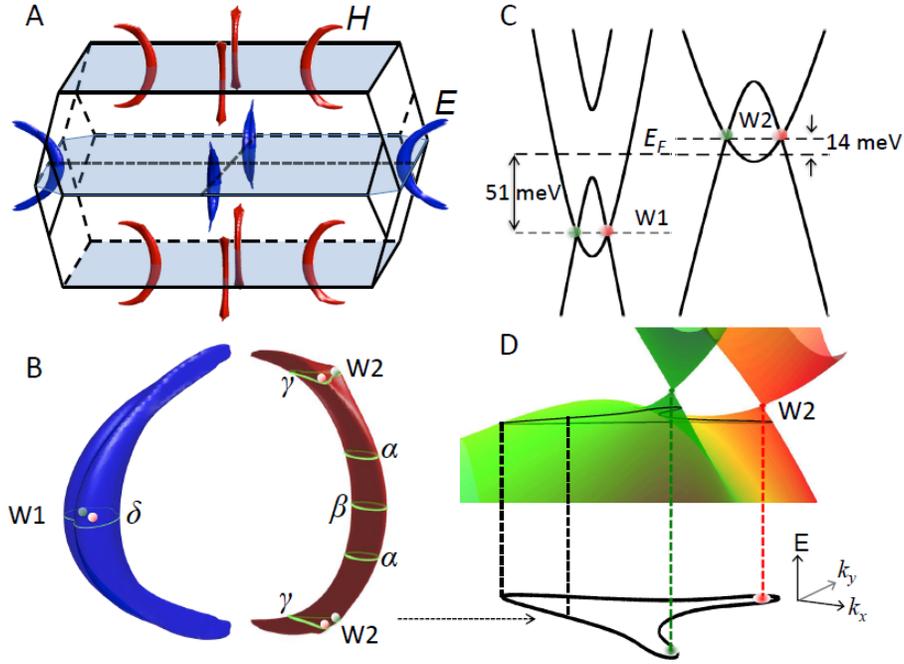}%{Figure3.eps}
\caption{{\bf 3D Fermi pockets and Weyl points.} 
{\bf A}. Fermi pockets in the first Brillouin zone at the Fermi energy ($E_F$) detected in the experiment. 
The electron ($E$) and hole ($H$) pockets are represented by blue and red colors, respectively. 
{\bf B}. Enlargement of the banana-shaped $E$ and $H$ pockets. The pink and green points indicate the Weyl points with opposite chirality. 
W1- and W2-type Weyl points can be found inside $E$ and close to $H$ pockets, respectively. Green loops represent some extremal  $E$ and $H$ cross sections, corresponding to the oscillation frequencies measured, $F_{\alpha,\beta,\gamma,\delta}$ for $H || c$ (see the text). 
 {\bf C}. Energy dispersion along the connecting line between a pair of Weyl points with opposite chirality for W1 (left) and W2 (right). 
The deduced experimental $E_F$ (thick dashed horizontal line) is 14 meV below the W2  Weyl points and 51 meV above the W1 Weyl points.
{\bf D}.  Strongly anisotropic Weyl cones originating from a pair of W2-type Weyl points on the plane of $F_\gamma$. Green and red Weyl cones represent opposite chirality.
}
\end{figure}

\clearpage

To reconstruct the shape of the Fermi surface, the full angular dependence of the quantum oscillation frequencies is measured and compared to band-structure calculations (Fig. 1C). The exact position of the Fermi energy ($E_F$) is determined by matching the calculated frequencies and their angular dependence to the measured ones (see the SI for more details). The best fit is obtained when $E_F$ lies 5 meV above the ideal electron-hole compensation point, in agreement with the resulting carrier concentration from the Hall measurements (Fig. 2C and D). 
Calculations reveal two banana-shaped Fermi surface pockets at this $E_F$, a hole pocket ($H$) and a slightly larger electron pocket ($E$).
These two pockets reproduce the angular dependence of the measured dHvA frequencies with great accuracy (see lines in Fig.~1C and the Fermi surface in Fig.~3). $E$ and $H$ are almost semicircular and are distributed along rings~\cite{Weng:2014ue,Huang:2015uu} on the $k_x = 0$ and $k_y = 0$ mirror-planes in the BZ. The rather isotropic frequencies $F_{\alpha}$ and $F_{\gamma}$ result from a neck and extra humps (``head with horns'') at the end of the hole pocket (see Fig.~3B).
The splitting of all frequencies with field angles departing from $\textbf{B}||c$ in the (100)-plane is explained by the existence of four banana-shaped pockets in the BZ, two for each mirror plane for both $E$ and $H$ pockets.
The splitting of the frequency $F_\delta$ seen in the experiment is not reproduced by the calculation. One possibility for this discrepancy is that a waist may appear in the $E$-pocket at $k_z = 0$.

The Berry phase for the largest $\beta$ oscillations from our magnetization in Fig. 2A is trivial (see SI).
Additionally, in the dHvA experiment the mobility of the hole orbits ($F_{\alpha}$ and $F_{\beta}$) was extracted via the width of the Fourier transform peaks, which is given by the exponential decrease of the oscillations with $1/B$ (the so-called Dingle term; see the SI). The deduced mobility is $\mu_h = 3.2 \times 10^4$ ($\pm 20 \%$) cm$^2$ V$^{-1}$s$^{-1}$ (star in Fig.~2D).

We extract information on the carrier density and mobility from field- and temperature-dependent Hall measurements in sample S1 (full temperature range, Fig.~2C) and only at low temperature in sample S3 from a different batch. We employ a two-carrier model (see Ref.~\onlinecite{Singleton2001}) to fit the Hall conductivity ($\sigma_{xy}$) by making use of the longitudinal conductivity at zero field ($\sigma_{xx}$) as an additional condition (see the SI). As shown in the left panel of Fig. 2D, the carrier concentration for both electrons and holes at low temperature is around $n = (2 \pm 1) \times 10^{19}$  cm$^{-3}$ with an absolute error bar given approximately by the scattering of the data for different samples and shown as a gray-shaded area.  Although TaP is almost a compensated metal, the electron density is slightly larger than the hole density for each sample at low temperature, in agreement with the fact that $E_F$ lies 5 meV above the charge-neutral point determined from the Fermi surface topology.
The theoretical values of the carrier densities,  given in Fig.~2D as dashed lines, are in good agreement with the experimental data. Note that above 150 K the hole density becomes larger than the electron density for sample S1. This inversion is reflected in the sign change of the high field Hall resistivity (see SI) and is similar to the Hall effect observed in TaAs~\cite{Huang2015anomaly,Zhang2015ABJ} and NbP~\cite{Shekhar2015,Wang:2015wm}.
The mobility of both carriers at low temperature lies in the range of $\mu = (2$--$5) \times 10^4$ cm$^2$ V$^{-1}$s$^{-1}$, with the hole mobility higher than the electron one. The high mobility indicates the very high quality of the single crystals with only a few defects and impurities. 
The Hall mobility agrees well with the mobility from dHvA measurements (the star in Fig.~2D).

The measured Fermi surface topology and carrier concentration described above converge to the same statement that the Fermi energy $E_F$ of our samples is 5~meV above the ideal charge-neutral point in the calculated band structure. This slight doping is not expected for a completely stoichiometric sample. However, a slight appearence of defects/vacancies in this type of samples is possible \cite{Besara2015} and can explain this small shift of the Fermi energy.
The consistency between experiment and theory strongly suggests that this is the true Fermi surface of our TaP samples.  
We plot the corresponding 3D Fermi surfaces from the \textit{ab initio} calculations in Fig.~3. 
One can see that $E$ and $H$ pockets are the only two pockets at the Fermi energy.

%%%%%%%%%  Band structure
We investigate the Weyl points in the band structure.  
In the 3D BZ, there are twelve pairs of Weyl points with opposite chirality:  four pairs lie in the $k_z = 0$ plane (labeled as W1) and eight pairs are located in planes close to $k_z = \pm \pi / c$ (where $c$ is the lattice constant) (labeled as W2). 
One can see that the W1 points are far below $E_F$ by 51 meV and are included in the $E$ pocket.
The W2-type Weyl points are 14 meV above $E_F$ and are included in the $H$ pocket. 
Such pairs of W2 points merge slightly into the head position of the $H$ pocket, leading to a two-horn-like cross section (see $F_{\gamma}$ in Fig. 3B). 
Energetically, they are separated by a 16 meV barrier along the line connecting 
the Weyl points of a pair. We plot the energy dispersion of Weyl bands over the $F_{\gamma}$ plane in Fig. 3D. There are no independent Fermi surface pockets around the W2 Weyl points and therefore the chiral anomaly is not well-defined.
The Weyl cone is strongly anisotropic in the lower cone region below the Weyl point.

\begin{figure}[htb!]
\centering
\includegraphics[angle=0,width=16cm,clip]{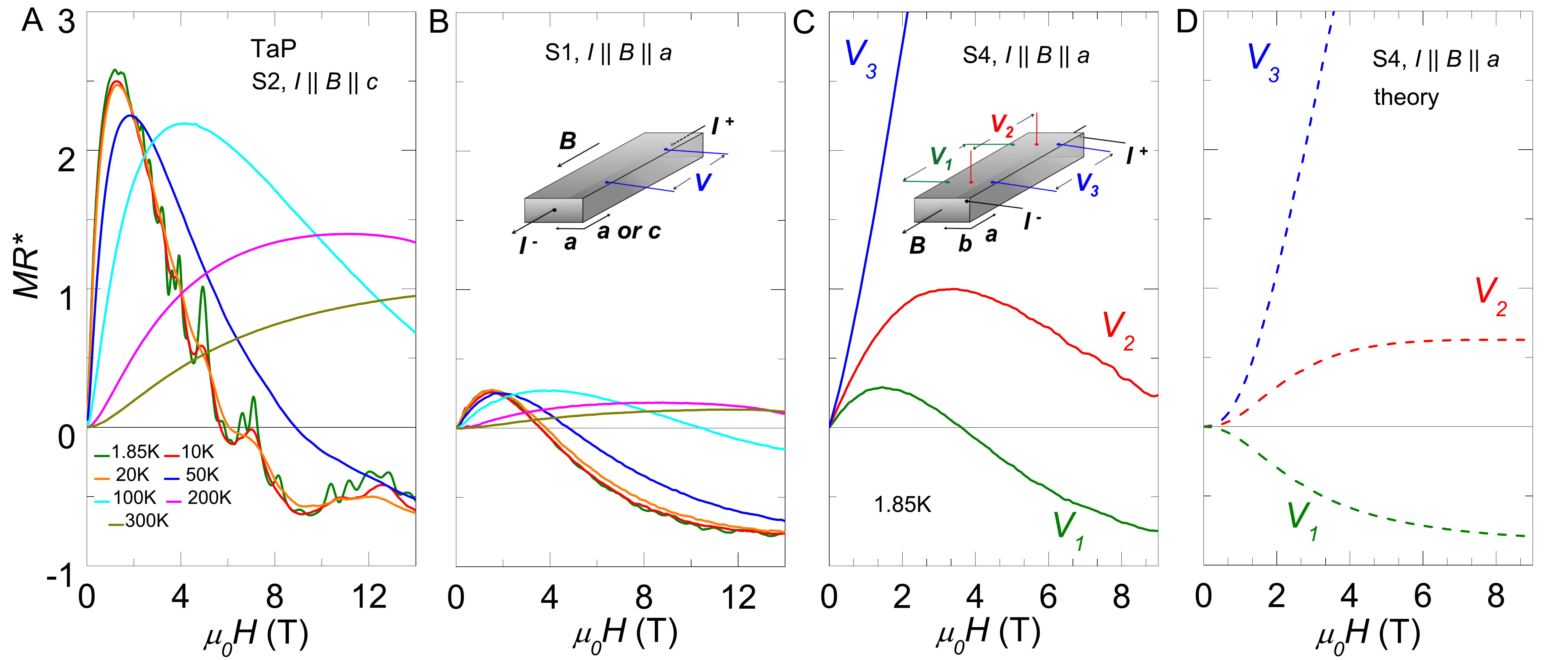}%{FIG4c.eps}
\caption{{\bf Negative longitudinal magnetoresistance in TaP.} 
{\bf A}. Longitudinal MR$^{\star} = [V - V(B = 0)]/V(B = 0)$ for $B || I || c$ for different temperatures.
{\bf B}. Same for $B || I || a$. The temperatures are the same as in {\bf A}.
{\bf C}. Same for $B || I || a$ and three pairs of contacts. The difference in the curves can be explained by an inhomogeneous current distribution induced by the magnetic field (see text).
The contact geometry is shown in the inserts: panel {\bf B} for S1 and S2, and panel {\bf C} for S4.
{\bf D}. Theoretical curves for S4 as in panel {\bf C}. For details about the calculation see the text and the SI \cite{Yoshida1980}.
} 
\end{figure}

%%%%%     Longitudinal MR
Finally, we discuss the longitudinal magnetotransport properties of TaP. 
We measure the longitudinal magnetoresistance of three samples from the same crystal batch where the current (\textbf{I}) is along the crystallographic $c$ (sample S2) and $a$ (samples S1 and S4) axes, respectively.
Figure 4A and B represent the longitudinal magnetoresistance (MR$^{\star} = [V - V(B = 0)]/V(B = 0)$ where $V$ is the measured voltage drop with a fixed electric current) measured for the two crystallographic directions. MR$^{\star}$ is equal to the usual MR = [$\rho - \rho(B = 0)]/\rho(B = 0)$ as long as the current flows homogeneously in the sample.
At 1.85 K, the MR$^{\star}$ first increases slightly and soon drops steeply from positive to negative values.
The negative MR$^{\star}$ is very robust with increasing temperature.
Note that the MR$^{\star}$ is negative only in a very narrow window of $\theta \le 2 ^\circ$ and becomes positive otherwise (see SI).
The small $\theta$ window of negative MR$^{\star}$ is similar to that observed in Na$_3$Bi~\cite{Xiong2015}.
In light of the given Fermi energy position in our crystals, it is not possible to link the negative MR$^{\star}$ to the chiral anomaly, 
simply because the former is not a well-defined concept when the Fermi surface connects both Weyl nodes. 
Although this does not rule out the presence of a non-trivial negative MR~\cite{Chang2015,Ma2015,Zhong2015}, we find that the  observed negative MR$^{\star}$ is strongly affected by the geometric configuration.

This becomes apparent when we measure the longitudinal MR$^{\star}$ for three different voltage contact configurations on sample S4 as illustrated in Fig. 4C. A clear voltage decrease in magnetic field is observed in pair V1, similar to the low temperature curves in Fig. 4A and B, whereas the two other pairs, denoted V2 and V3, show a higher MR$^{\star}$. This points to an underlying inhomogeneous current distribution in the sample becoming important in high magnetic fields. As typical for high mobility semimetals, TaP has a large transverse MR arising from the orbital effect, whereas the longitudinal MR most likely stays of the same order of magnitude. For current contacts smaller than the cross section of the sample, this leads to a field-induced stearing of the current to the direction of the magnetic field, which is along the line connecting the current contacts when current and magnetic field are parallel. This effect is known as ``current jetting''~\cite{Pippardbook, Reed1971, Yoshida1975,Yoshida1976, Yoshida1976b, Yoshida1980}. As a consequence, a voltage pair close to this line (V3) detects a higher MR$^{\star}$ than the intrinsic longitudinal MR whereas a voltage pair far away from it (V1) detects a smaller MR$^{\star}$ than the intrinsic one.
This effect is confirmed by calculations of the voltage distribution for sample S4, taking into account the current jetting by following ref.~\onlinecite{Yoshida1980} (see SI). Using the measured transverse MR and assuming a field-independent intrinsic longitudinal MR, the model qualitatively reproduces the three observed MR$^{\star}$ curves without any free parameters (see Fig. 4D).
Therefore, the field dependence of the longitudinal measured voltage is largely induced by the strong transverse MR, if the current is not homogeneously injected into the whole cross section of the sample.

%%%%%% conclusions
In summary, we determined the Fermi surface topology of the inversion-asymmetric WSM TaP.
The Fermi surface consists of banana-shaped spin-polarized electron and hole pockets with very light carrier effective masses. Despite the absence of independent Fermi surface pockets around the Weyl points, an apparent negative longitudinal MR is detected. We show that in such measurements, special care is needed to avoid a decoupling of the voltage contacts from the current jet in longitudinal magnetic fields. 

\noindent {\bf Methods} 

{\it Single crystal growth.} High-quality single crystals of TaP were grown via a chemical vapor transport reaction using iodine as a transport agent.  Initially, polycrystalline powder of TaP was synthesized by a direct reaction of tantalum (Chempur 99.9\%) and red phosphorus (Heraeus 99.999\%) kept in an evacuated fused silica tube for 48 hours at 800 $^{\circ}\mathrm{C}$. Starting from this microcrystalline powder, the single-crystals of  TaP were synthesized by chemical vapor transport in a temperature gradient starting from 850 $^{\circ}\mathrm{C}$ (source) to 950 $^{\circ}\mathrm{C}$ (sink) and a transport agent  with a concentration of 13.5 mg/cm$^3$ iodine (Alfa Aesar 99,998\%)~\cite{martin1988chemischen}. 
The orientation and crystal structure of the present single crystals were investigated using the diffraction data sets collected on a Rigaku AFC7 diffractometer equipped with a Saturn 724+ CCD detector (monochromatic Mo$K_{\alpha}$ radiation, $\lambda =$ 0.71073~\AA). Structure refinement was performed by full-matrix least-squares on $F$ within the program package WinCSD.

{\it Magnetization.} The magnetism and magnetic quantum oscillations of TaP along the main crystallographic axes were measured in a Quantum Design SQUID-VSM. The angular dependencies were measured using the Quantum Design piezo-resistive torque magnetometer (Tq-Mag \cite{Rossel1996}) in a PPMS with installed rotator option. 
Magnetization and torque measurements were performed on two large 4.4 and 21.7 mg TaP single crystals. The samples were mounted on the sample holder and torque lever by GE varnish or Apiezon N grease and aligned along their visible crystal facets, which was confirmed by X-ray diffraction measurements. The crystal alignment was verified by photometric methods and showed typical misalignments of less than 2$^\circ$. 
SQUID magnetization measurements were performed for static positive and negative magnetic fields up to 7 T in the temperature range of 2 to 50 K. 
Magnetic torque measurements were performed up to 14 T in the same temperature range. Here the sample is mounted on a flexible lever which bends when torque is applied. The sample magnetization M induces a magnetic torque $\tau = M \times B$, which bends the lever and can be sensed by piezo-resistive elements which are micro-fabricated onto the torque lever. These elements change their resistance under strain and are thus capable of sensing the bending of the torque lever. The unstrained resistance of these piezo-resistors is typically 500 $\Omega$ and can change by up to a few percent when large magnetic torques are applied. Measuring quantum oscillations, however, requires a higher torque resolution. Thus the standard torque option was altered and extended by an external balancing circuit and SR830 lock-in amplifiers as read-out electronic. This way a resolution of one in $10^7$ was achieved. The magnetic torque was measured for magnetic fields in the (100), (001) and (110)-plane in angular steps of 2.5$^\circ$ and 5$^\circ$. Measurements were taken during magnetic field down sweeps from 14 to 0 T. The magnetic field sweep rate was adjusted such that the sweep rate in $1/B$ was constant. The resultant magnetic torque signals are a superposition of the sample diamagnetism, de Haas-van Alphen oscillation and uncompensated magneto-resistance of the piezo-resistors. Due to the vector product of the magnetization and magnetic field, this method can only be applied to samples with strong Fermi surface anisotropy and is insensitive when the magnetic field and magnetization are aligned parallel, e.g. along the crystallographic $c$-direction in TaP. The quantum oscillation frequencies are extracted from the measured resistivities and magnetizations by subtracting all back ground contributions to those signals and performing a Fourier transformation of the residual signal over the inverse magnetic field. The resulting spectra show the de Haas-van Alphen frequencies and their amplitude. 

{\it Band structure calculations.} The \textit{ab initio} calculations were performed using density-functional theory with the Vienna \textit{ab-initio} simulation package~\cite{kresse1996}. Projector-augmented-wave potential represented core electrons.
The modified Becke-Johnson exchange potential~\cite{Becke2006,Tran2009} was employed for accurate band structure calculations. 
Fermi surfaces were interpolated using maximally localized Wannier functions (MLWFs)~\cite{Mostofi2008} in dense $k$-grids (equivalent to $300\times 300 \times 300$ in the whole BZ). Then angle-dependent extremal cross-sections of Fermi surfaces are calculated to compare with the oscillation frequencies according to the Onsager relation. 

{\it Magnetoresistance measurement.} 
Resistivity measurements were performed in a physical property measurement system (PPMS, Quantum Design) using the DC mode of the AC-Transport option. Samples with two different crystalline orientations, \textit{i.e.}, bars with their long direction parallel to the crystallographic  $a$ and $c$-axis, were cut from large TaP single crystals using a wire saw. The orientation of these crystals was verified by x-ray diffraction measurements. A detailed analysis of their crystalline orientations is given in the SI. The samples were named S1 ($\textbf{I}||a$), S2 ($\textbf{I} || c$), S3 ($\textbf{I}||a$) and S4($\textbf{I}||a$). The physical dimensions of S1, S2, S3 and S4 are (width $\times$ thickness $\times$ length) 0.42$\times$0.16$\times$1.1 mm$^3$, 0.48$\times$0.27$\times$0.8 mm$^3$, 0.5$\times$0.2$\times$3.0 mm$^3$ and 0.79$\times$0.57$\times$3.2 mm$^3$, respectively. Contacts to the crystals were made by spot welding 25 micrometer platinum wire (S1 and S2) or gluing 25 micrometer gold wire to the sample using silver loaded epoxy (Dupont 6838). %
The resistance and Hall effect were measured in 6-point geometry using a current of about 3 milliampere at temperatures of 1.85 to 300 K and magnetic fields up to 14 T. Crystals were mounted on a PPMS rotator option. Special attention was paid to the mounting of the samples on the rotator puck to ensure a good parallel alignment of the current and magnetic direction. The Hall contributions to the resistance and vice versa were accounted for by calculating the mean and differential resistance of positive and negative magnetic fields. Almost symmetrical resistivities were obtained for positive and negative magnetic fields when current and magnetic field were parallel showing the excellent crystal and contact alignment of our samples. Otherwise, the negative MR$^{\star}$ was overwhelmed by the transverse resistivity. %
In order to increase the sensitivity of the angular dependent Shubnikov-de Haas measurements at low magnetic fields, external Stanford Research SR830 lock-in amplifiers were used to measure the resistance and Hall resistance. Here typical excitation currents of a 2 to 5 milliampere were applied at frequencies of about 20 Hz.

\begin{addendum}
\item[Acknowledgements]  We are grateful for Dr. L.-K. Lim, Prof. E. G. Mele, Prof. Z.-K. Liu, and Prof. Y.-L. Chen, and Prof. S.-Q. Shen for helpful discussions. This work  was financially supported  by the  Deutsche Forschungsgemein- schaft DFG (Project No. EB 518/1-1 of DFG-SPP 1666 ``Topological Insulators'', and SFB 1143) and by the ERC (Advanced Grant No. 291472 �Idea Heusler�).
AGG acknowledges insightful discussions with D. Varjas and J. Moore. R. D. dos Reis acknowledges financial support from the Brazilian agency CNPq.
\item[Competing Interests] The authors declare that they have no
competing financial interests.
\item[Correspondence] Correspondence should be addressed to B. Yan~(email: yan@cpfs.mpg.de) or E. Hassinger~(email: elena.hassinger@cpfs.mpg.de).
\end{addendum}
%%%%%%%%%%%%%%
%\bibliographystyle{naturemag}
%\bibliography{TaP-references}
%

\clearpage

%%%%%%%%%%%%%%%%%%%%%%%%%%%%%%%%%%%%%%%%%%%%%%%%%%%%%%%%%%%%%%%%%%%%%%%%%%%%%%%%%%%%%%%%%%

%SUPPLEMENTAL ONLINE MATERIAL

%%%%%%%%%%%%%%%%%%%%%%%%%%%%%%%%%%%%%%%%%%%%%%%%%%%%%%%%%%%%%%%%%%%%%%%%%%%%%%%%%%%%%%%%%%
\setcounter{figure}{0}   
\renewcommand{\thefigure}{S\arabic{figure}}
\noindent\Large\textbf{\textit{Supplemental Information}: Negative magnetoresistance without well-defined chirality in the Weyl semimetal TaP}\normalsize

% introduction paragraph
\section{Crystal structure}
\begin{figure}[tb!]
\centering
\includegraphics[angle=0,width=16cm,clip]{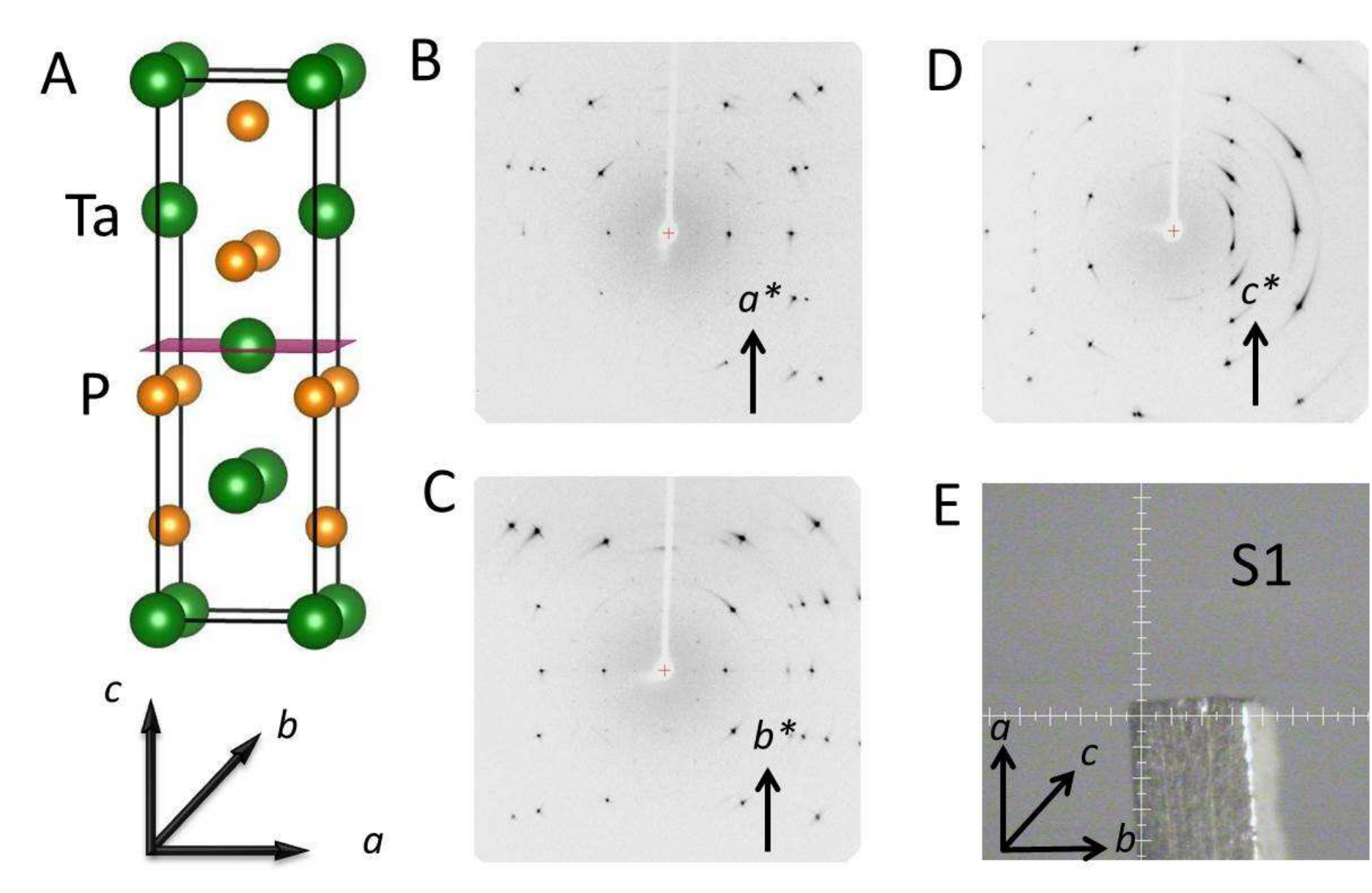}
\caption{{\bf Unit cell of TaP and x-ray diffraction patterns of crystal S1.} {\bf a}, non-centrosymmetric unit cell of TaP. {\bf b, c, d}, show the rotation diffraction patterns of the TaP crystal, S1, for the rotation about the crystallographic  \textit{a}, \textit{b}, and \textit{c}-axis respectively. The rotation axis in each case is vertical. {\bf e}, shows an image of S1 with its longest dimension parallel to the crystallographic \textit{a}-axis.}
\label{fig:LaueDiffractionOne}
\end{figure}

\begin{figure}[tb!]
\begin{center}
\includegraphics[angle=0,width=12cm,clip]{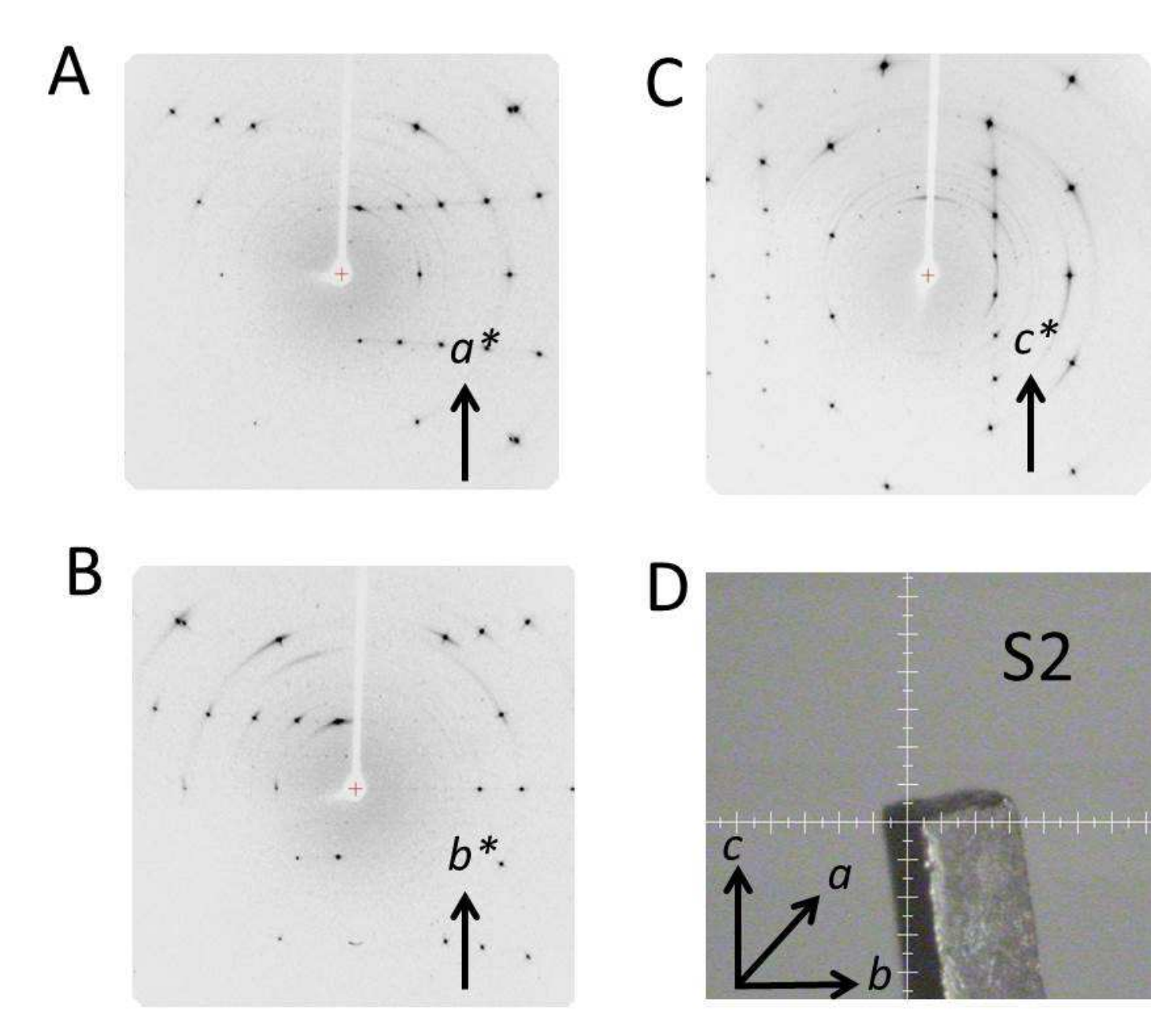}
\end{center}
\caption{{\bf X-ray diffraction patterns of crystal S2.} {\bf a, b, c}, show the rotation diffraction patterns of the TaP crystal, S2, for the rotation about the crystallographic  \textit{a}, \textit{b}, and \textit{c}-axis respectively. The rotation axis in each case is vertical. {\bf d}, shows an image of S2 with its longest dimension parallel to the crystallographic \textit{c}-axis.}
\label{fig:LaueDiffractionTwo}
\end{figure}

The crystal structure and orientation of TaP crystals were measured by x-ray diffraction at room temperature. For this, TaP single crystals were mounted on a four-circle Rigaku AFC7 x-ray diffractometer with a built-in Saturn 724+ CCD detector. A suitable sample edge was selected where the transmission of $Mo-K_\alpha$ ($\lambda = 0.71073$~ {\AA}) radiation seemed feasible. The intensities of the measured reflections were corrected for absorption by using a multi-scan technique. The unit cell was assigned by using a 30 images standard indexing procedure. Here oscillatory images about the crystallographic axes allowed the assignment of the crystal orientation, confirmed the appropriate choice of the unit cell and showed the excellent crystal quality. Figures \ref{fig:LaueDiffractionOne} and \ref{fig:LaueDiffractionTwo} show x-ray diffraction patterns of the S1 and S2 TaP crystals, which were used in our transport measurements. Their analysis revealed that TaP has a non-centrosymmetric crystal structure with space group $I4_1md$ and lattice parameters $a = b = 3.30$~ {\AA}, $c = 11.33$~ {\AA} at room temperature. 

\begin{figure}[tb!]
	\centering
		\includegraphics[width=0.75\textwidth]{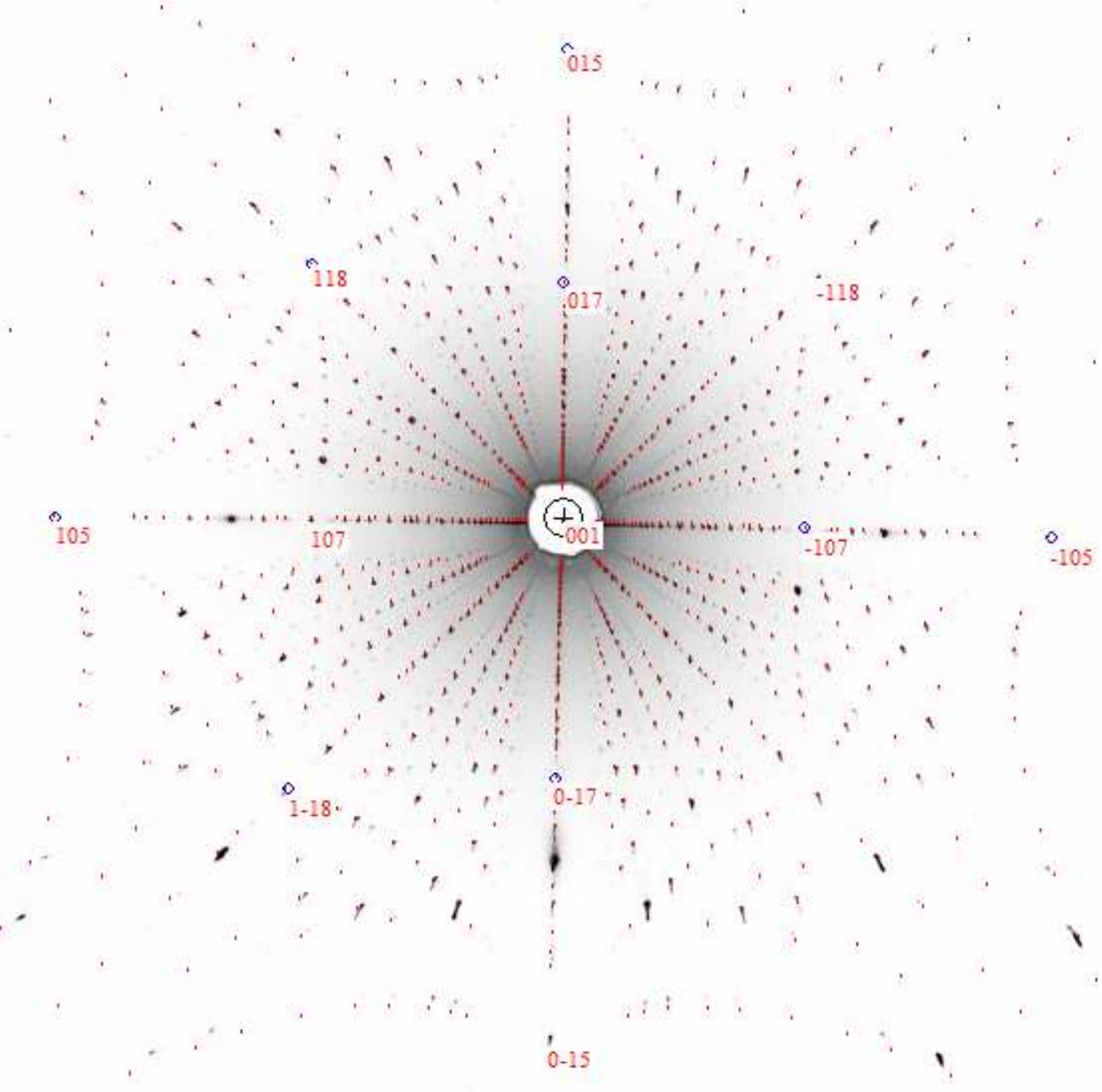}
	\caption{{\bf Room temperature Laue diffraction image of an unperturbed, as-grown (001)-facet of a TaP single crystal}. The red spots and assigned Miller indices show the calculated diffraction pattern of the $I4_1md$ tetragonal-space group.}
	\label{fig:LaueImage}
\end{figure}

In order to prove the quality of our TaP single crystals additional Laue images from an unperturbed as grown (001)-facet of a TaP crystal were taken. The single crystal was oriented using a white beam backscattering Laue x-ray diffraction method. Figure \ref{fig:LaueImage} shows the corresponding Laue diffraction image indexed with the $I4_1md$-structure and room temperature lattice parameters, a=3.30${\AA}$, c=11.33${\AA}$. The Laue diffraction images shows sharp reflections, which confirm the excellent quality of the sample. The presence of domains or twinning can be ruled out by indexing all reflections of the image by a single pattern. 

%%%%%%%%%%%%%%%%%%%%%%%%%%%%%%%%%%%%%%%%%%%%%%%%%%%%%%%%%%%%%%%%%%%%%%%%%%%%%%%%%%%%%%%%%%%%%%%%%%%%%%%%%%%%%%%%%%%%%%%%%%
\section{Resistivity}

The longitudinal, transverse and Hall resistivities of the TaP samples S1 to S4 were measured in four point configuration. Here, the current was passed along the length of the samples, i.e. the crystallographic \textit{a}-direction for samples S1, S3 and S4 and \textit{c}-direction for sample S2. Electrical contacts were made by spot welding platinum wire or gluing gold wire to the sample (see Methods section of the main text).

%%%%%%%%%%%%%%%%%%%%%%%%%%%%%%%%%%%%%%%%%%%%%%%%%%%%%%%%%%%%%%%%%%%%%%%%%%%%%%%%%%%%%%%%%%%%%%%%%%%%%%%%%%%%%%%%%%%%%%%%%%

\section{Transverse Resistivity}

\begin{figure}[tb!]
\begin{center}
\includegraphics[angle=0,width=16cm,clip]{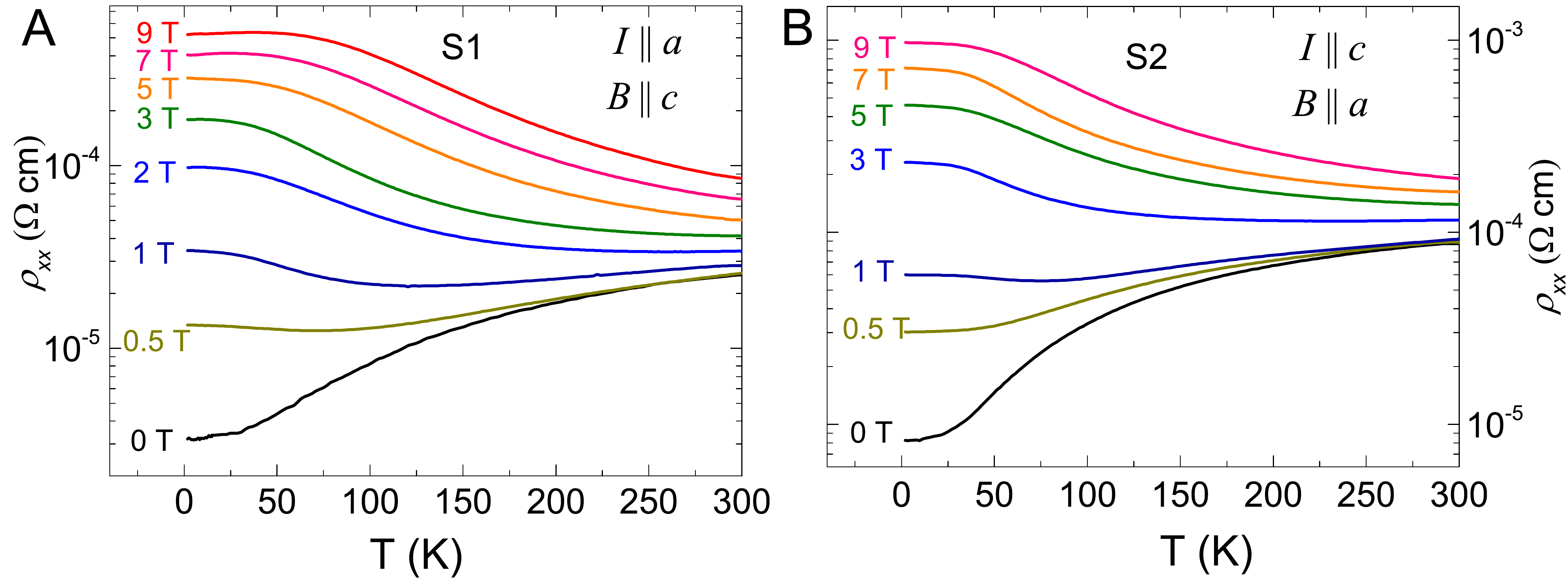}
\end{center}
\caption{{\bf Transverse resistivity.} Temperature dependence of the transverse resistivity, $\rho_\mathrm{xx}$ measured up to 9~T for {\bf a}, S1 along the \textit{a}-axis and {\bf b}, S2 along the \textit{c}-axis when $B \bot I$.}
\label{fig:TransverseResistivity}
\end{figure}

Figure \ref{fig:TransverseResistivity} (a, b) show the temperature dependence of the transverse resistivity, $\rho_\mathrm{xx}$, for different magnetic fields applied perpendicular to the current as indicated. At zero magnetic field, the resistivities decrease monotonically with decreasing temperature. The resistance anisotropy between the in and out-of plane transport is about four at high temperatures and three at low temperatures. At moderate magnetic fields of about 1~T and intermediate temperatures a crossover from a positive to a negative temperature coefficient takes place. This behavior is typical for high mobility semimetals and has been observed for example in graphite\cite{Uher83}.

\subsection{Positive Magnetoresistance}

Owing to the large charge carrier mobility of TaP (see also Fig. 2D of the main text), we observe a high magnetoresistance (MR) for $B\perp I$. As an example Fig. \ref{fig:PositiveMRandHall} shows the transverse resistivity of samples S1 for $B \bot I$ at selected temperatures between 2 and 300~K and magnetic fields up to 9~T. Up to 25~K, we observe an almost constant strong positive magnetoresistance of $\mathrm{MR}=(\rho_{xx}(B)-\rho_0)/\rho_0=1.8 \times 10^4 \% $ at 9~T. Above 25~K the MR is dropping to about 100 \% at 9~T and room temperature due to the decreasing charge carrier mobility (see Two Band Model section of the SI).

\subsection{Negative Magnetoresistance for $B || I$}

\begin{figure}[tb!]
	\centering
		\includegraphics[angle=0,width=16cm,clip]{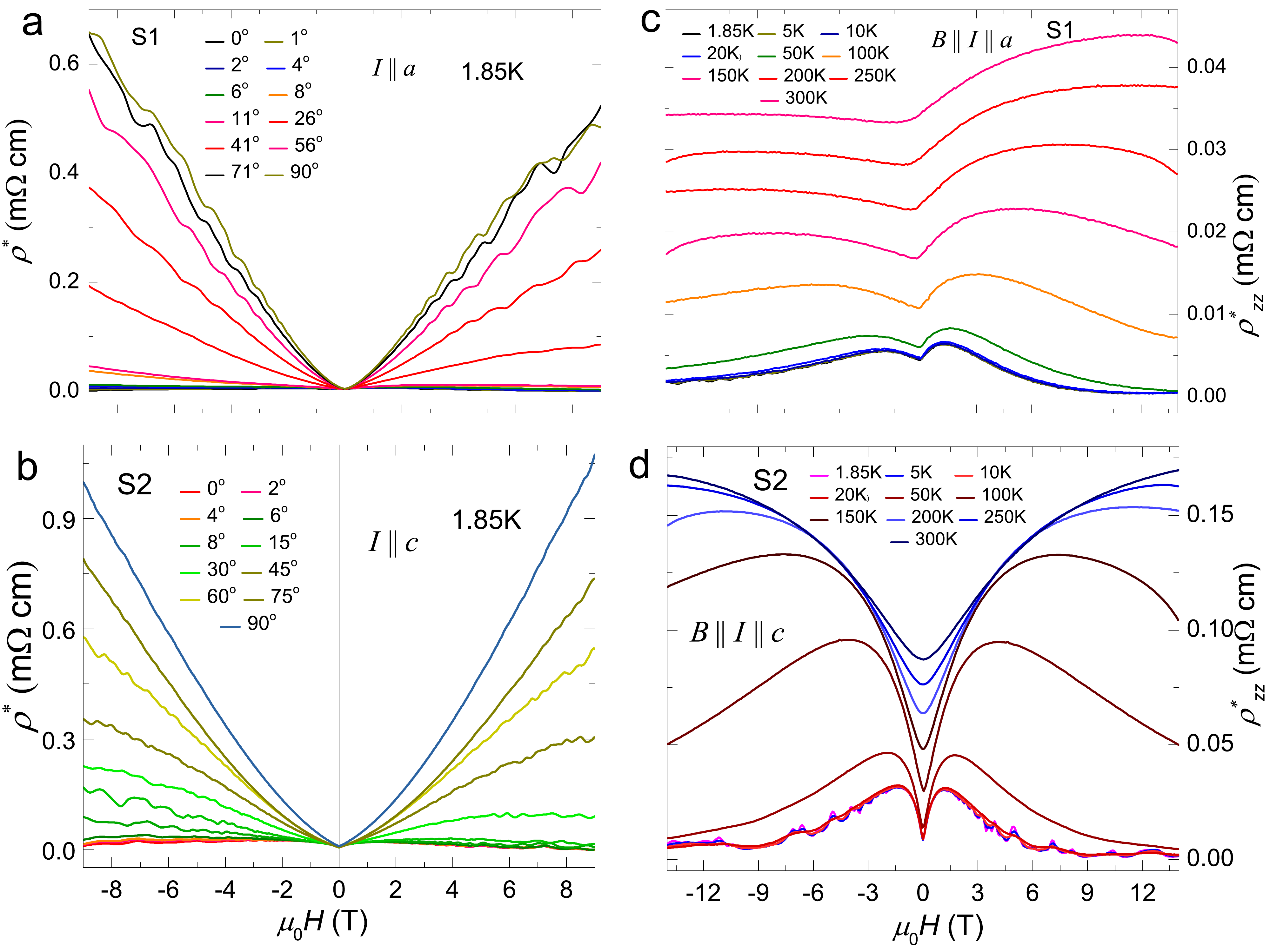}
	\caption{{\bf Unsymmetrized magnetic field dependences of the resistances.} a) and b) show the angular dependence of the magnetoresistance for sample S1 and S2. Here $0^o$ refers to $B||I$. c) and d) show the temperature dependences of the negative magnetoresistance of both samples for $B||I$.}
	\label{fig:FIG_NMR_unprocessed}
\end{figure}

The longitudinal resistivities, $\rho_{zz}$, were measured from room temperature down to $1.8\,\mathrm{K}$ for magnetic fields from -14 to 14 T. Figure \ref{fig:FIG_NMR_unprocessed} shows the temperature and angular dependence of the longitudinal resistivity of samples S1 and S2. Here $\rho^*_\mathrm{zz}=V_\mathrm{z}/I_\mathrm{z}\times A/l$ is the apparent longitudinal resistivity for an assumed homogenous current distribution in the sample.

For perfect alignment of magnetic field and current to better than $2^o$, a negative MR$^*$ can be observed. Its magnitude depends strongly on the precise location of the voltage contacts with respect to the current contacts. 

\begin{figure}[tb]
\begin{minipage}{0.54\textwidth}
	\centering
		\includegraphics[angle=0,width=9cm,clip]{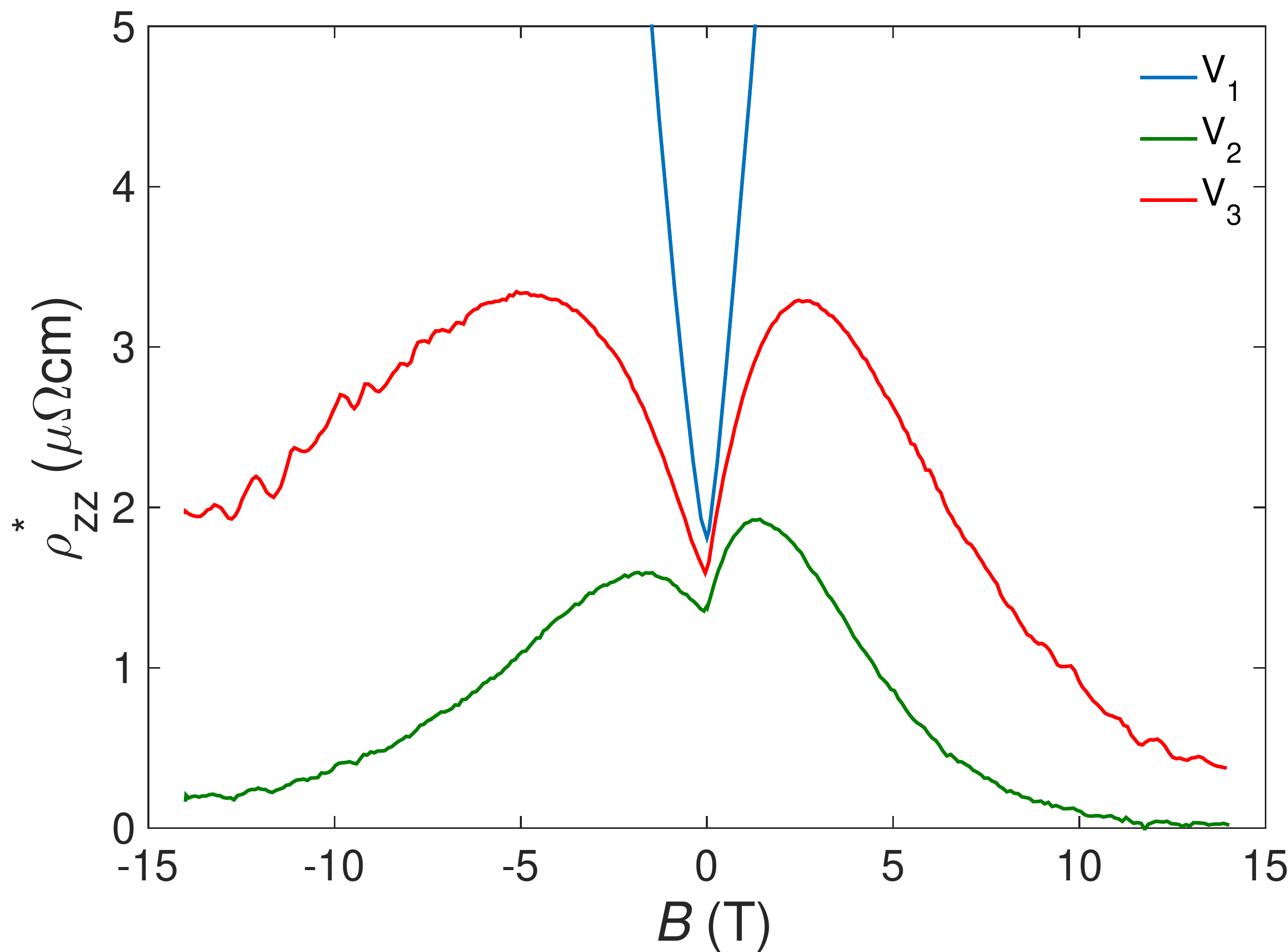}
\end{minipage}
\begin{minipage}{0.44\textwidth}
		\begin{center}
			\includegraphics[angle=0,width=9cm,clip]{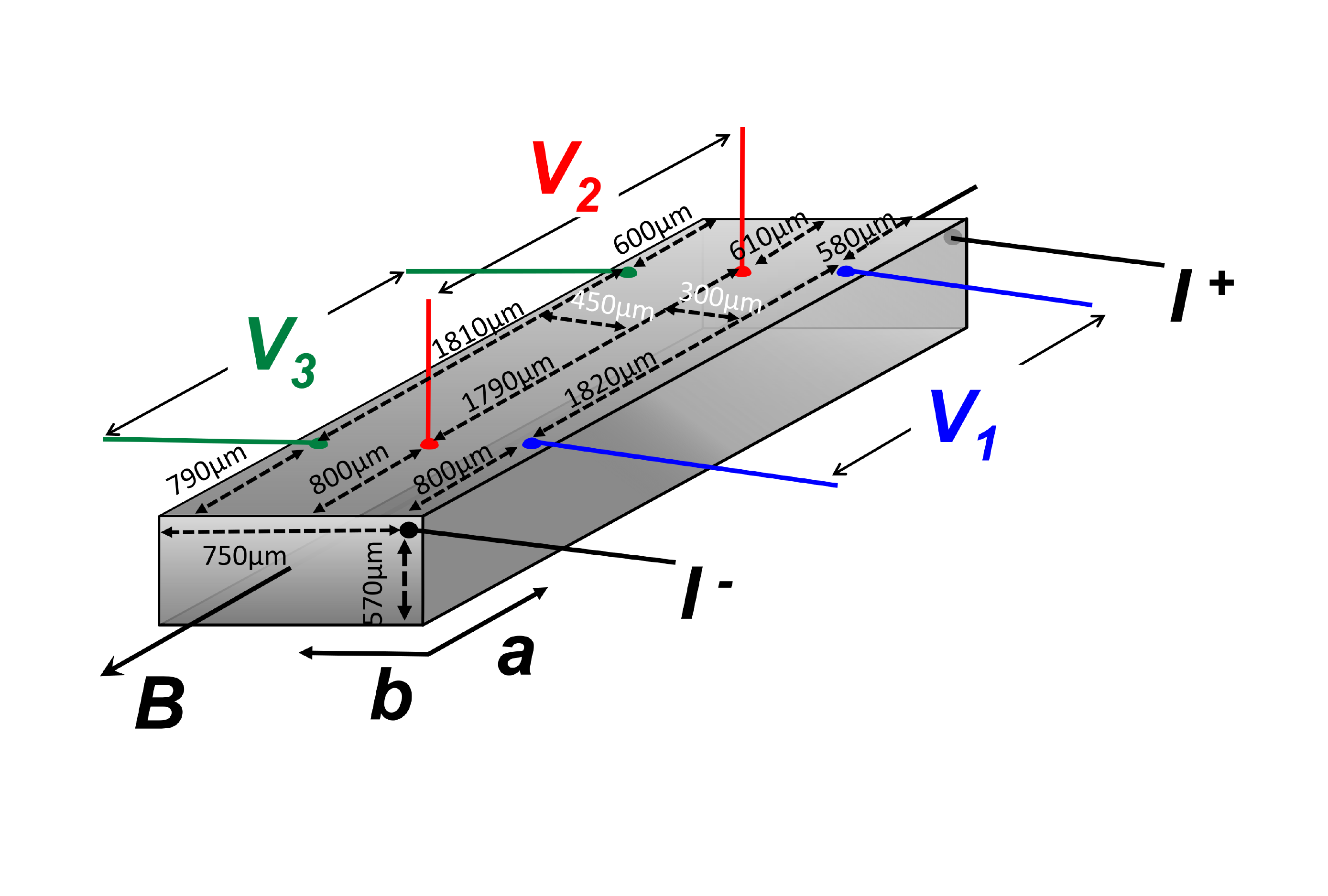}
		\end{center}
\end{minipage}
	\caption{{\bf Contact dependence of the apparent longitudinal resistivity}: The left graph shows the measured voltage drop of sample S4 for three voltage contact pairs. The geometry of the contacts is shown on the right. A simulation of the potential distribution can be found in Fig. \ref{fig:CurrentJetting}.}
	\label{fig:ContactDependence}
\end{figure}

Figure \ref{fig:ContactDependence} shows the longitudinal resistivity of sample S4 measured by three different voltage contact pairs. In zero field, all three contact pairs show approximately the same resistivity, indicating a homogenous current flow through the sample. On increasing the magnetic field, the measured resistivities diverge. It can be seen that the negative MR is largest for the contact pairs furthest away from the current contacts, whereas the contacts which are in-line with the current contacts show a strongly positive MR. Such a behavior can be explained by a magnetic field induced current redistribution in the sample, where the current flows homogeneously through the sample in zero field and concentrates on a narrow path directly between the current contacts in high fields (see also Fig. \ref{fig:CurrentJetting}). This phenomenon is called "`current jetting"' and was first observed in chromium and tungsten \cite{Reed71}. It arises due to a strong anisotropy of the longitudinal and transverse conductivity \cite{Reed71,Yoshida76,Ueda80}. Under a large magnetic field, the longitudinal conductivity $\sigma_{zz}\gg\sigma_{xx}$ of a high mobility metal becomes much larger than its transverse conductivity. A current injected through a single point into the sample will thus primarily follow the magnetic field and form a "`current jet"'. Its width depends to first approximation on the charge carrier mobility as: $\theta \approx 2\tan^{-1}\left(\sqrt{3}/\omega_c\tau\right)$ leading to an opening angle of the current jet of approximately $3^\circ$ at 14~T in TaP \cite{Pippard}. On applying a magnetic field the current is drawn from the outer parts of the sample and focused between the current contacts. Thus the voltage measured at the contact pair furthest away from the current contacts drops due to the vanishing current in this region of the sample, whilst the voltage increases between the current contacts. However neither of these contacts measure the correct intrinsic longitudinal resistivity.

\subsection{Theoretical Procedure of the Current Jetting Calculations}
\begin{figure}[tb]
	\centering
		\includegraphics[angle=0,width=16cm,clip]{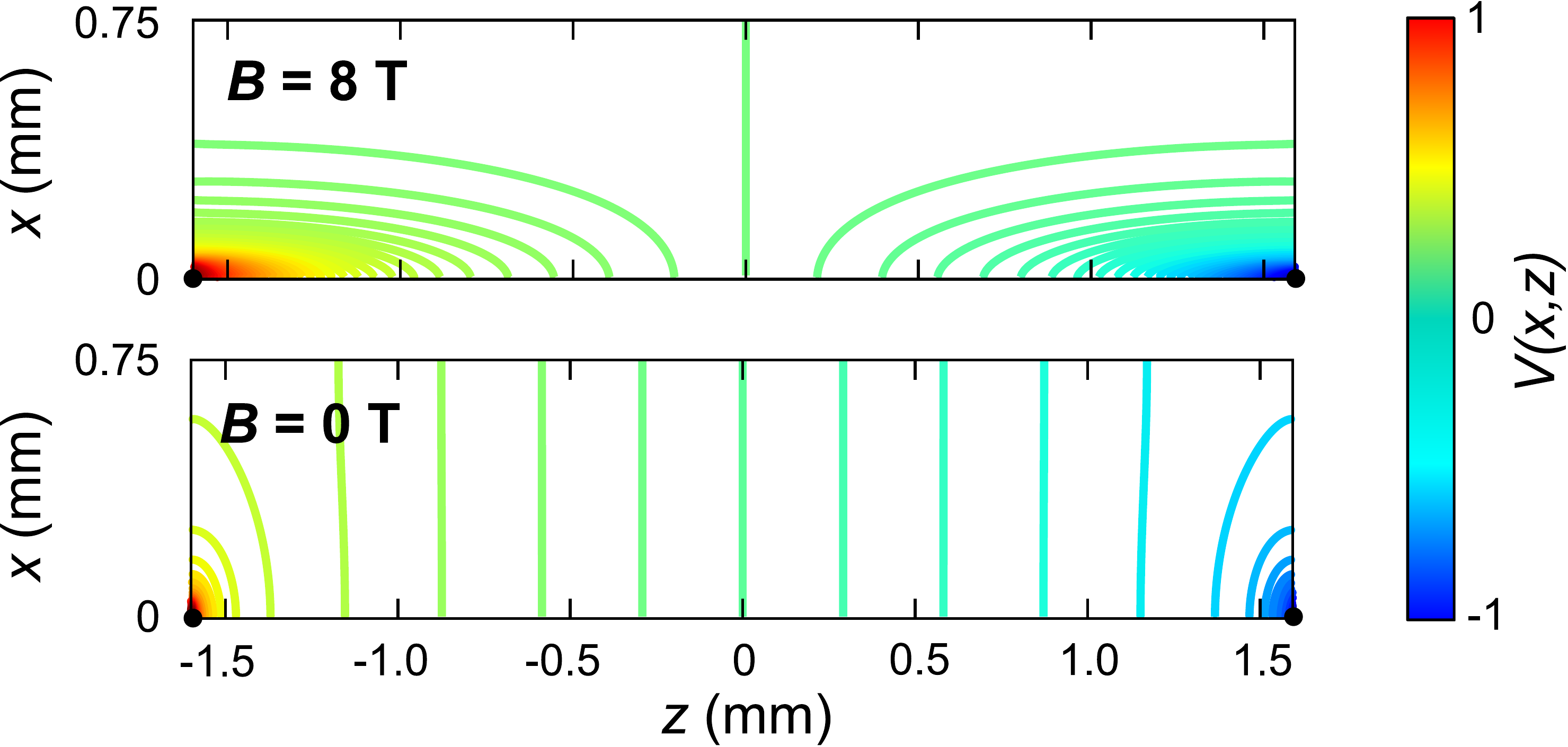}
	\caption{{\bf Current jetting in TaP.} The graph shows the simulated potential distributions of our S4 sample for magnetic fields applied parallel to the current contacts at zero field and $8\,\mathrm{T}$ respectively. The magnetic field drives a redistribution of the current density leading to a focused current jet between the current contacts parallel to the magnetic field and strong bending of the equipotential lines. The black dots show the current source and drain contacts.}
	\label{fig:CurrentJetting}
\end{figure}

We now outline the procedure to generate the theoretical curves shown in Fig. \ref{fig:CurrentJetting} and Fig. 4D of the main text. 
For the sample with the geometry shown in Fig. \ref{fig:ContactDependence} we first solve the Laplace equation following \cite{Yoshida80}, taking into account the position and size of the contacts. The resulting potential distribution depends on the transverse and longitudinal conductivity as well as the sample geometry. In this calculation we assume that the effect of the Hall effect contribution is small and can be neglected i.e. $\rho_{xy}=0$. The plotted curves for different contact configurations are shown in Fig. 4D of the main text. The corresponding potential distributions at the sample surface for magnetic fields of $0~\mathrm{T}$ and $8.02~\mathrm{T}$ are shown in Fig. \ref{fig:CurrentJetting}. In order to obtain these plots we have i) used the experimentally measured transverse conductivity as an input ii) assumed a constant longitudinal conductivity fixed to match the zero magnetic field value measured with the V$_2$ arrangement and iii) taken the experimentally measured geometry of the sample which is $3.2\times 0.75 \times 0.57~\mathrm{mm}^3$ and contact size of $25\times25~\mu\mathrm{m}^2$. We note that by applying i), ii) and iii) there are no free parameters. It is therefore remarkable how the qualitative features that are seen in experiments, i.e. positive and negative magnetoresistance depending on the contact position is captured by our modelling.

\section{Hall Resistivity and Two Band Model}

\begin{figure}[tb!]
\begin{center}
\includegraphics[angle=0,width=16cm,clip]{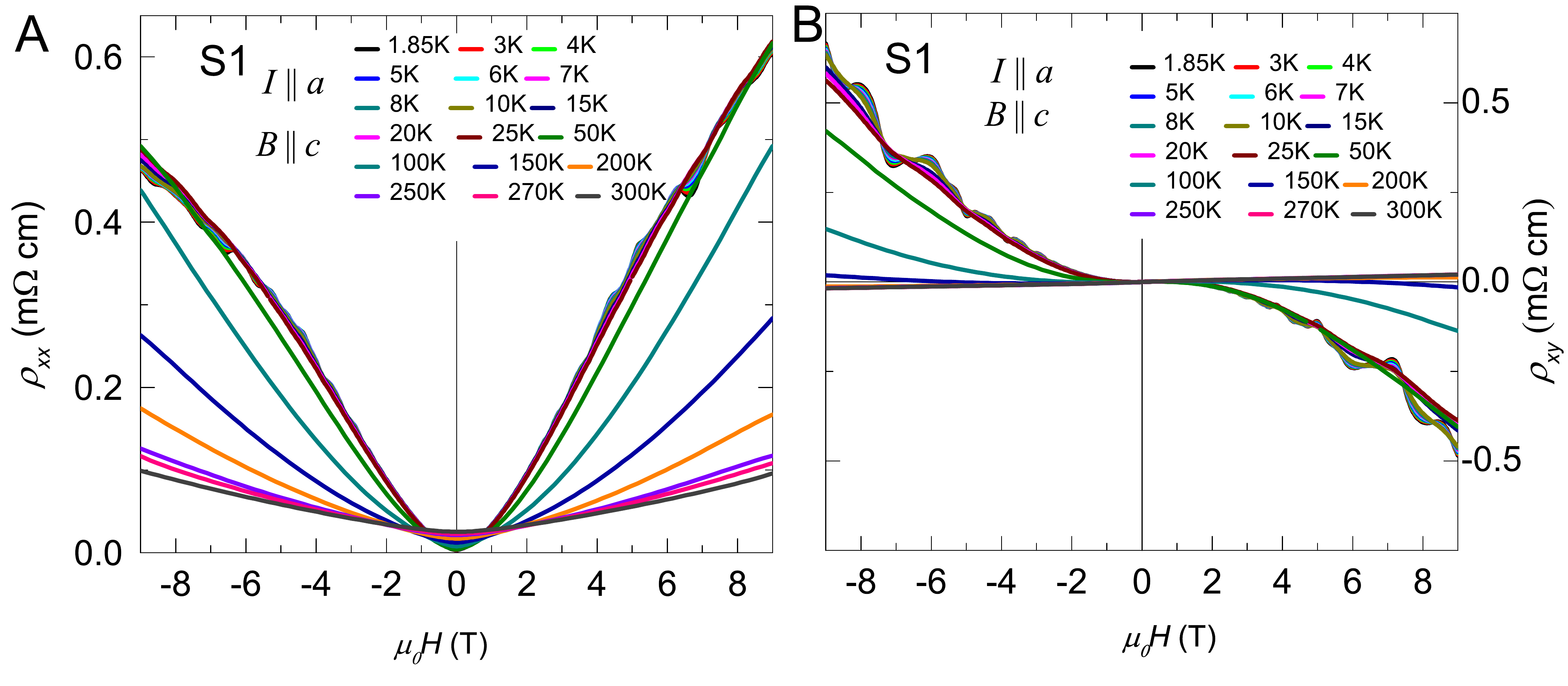}
\end{center}
\caption{{\bf Magneto- and  Hall resistance.} Field dependence of magnetoresistance (MR) for $B\perp I$ at different temperatures {\bf a}, along \textit{a}-axis and {\bf b}, along \textit{c}-axis. The insets of {\bf a} and {\bf b} show the respective high temperatures MR. {\bf c} and {\bf d} show the Hall resistivity $\rho_\mathrm{xy}$ of samples S1 and S2 at different temperatures.}
\label{fig:PositiveMRandHall}
\end{figure}

\begin{figure}[htb!]
\centerline{
\includegraphics[angle=-90,width=0.5\textwidth]{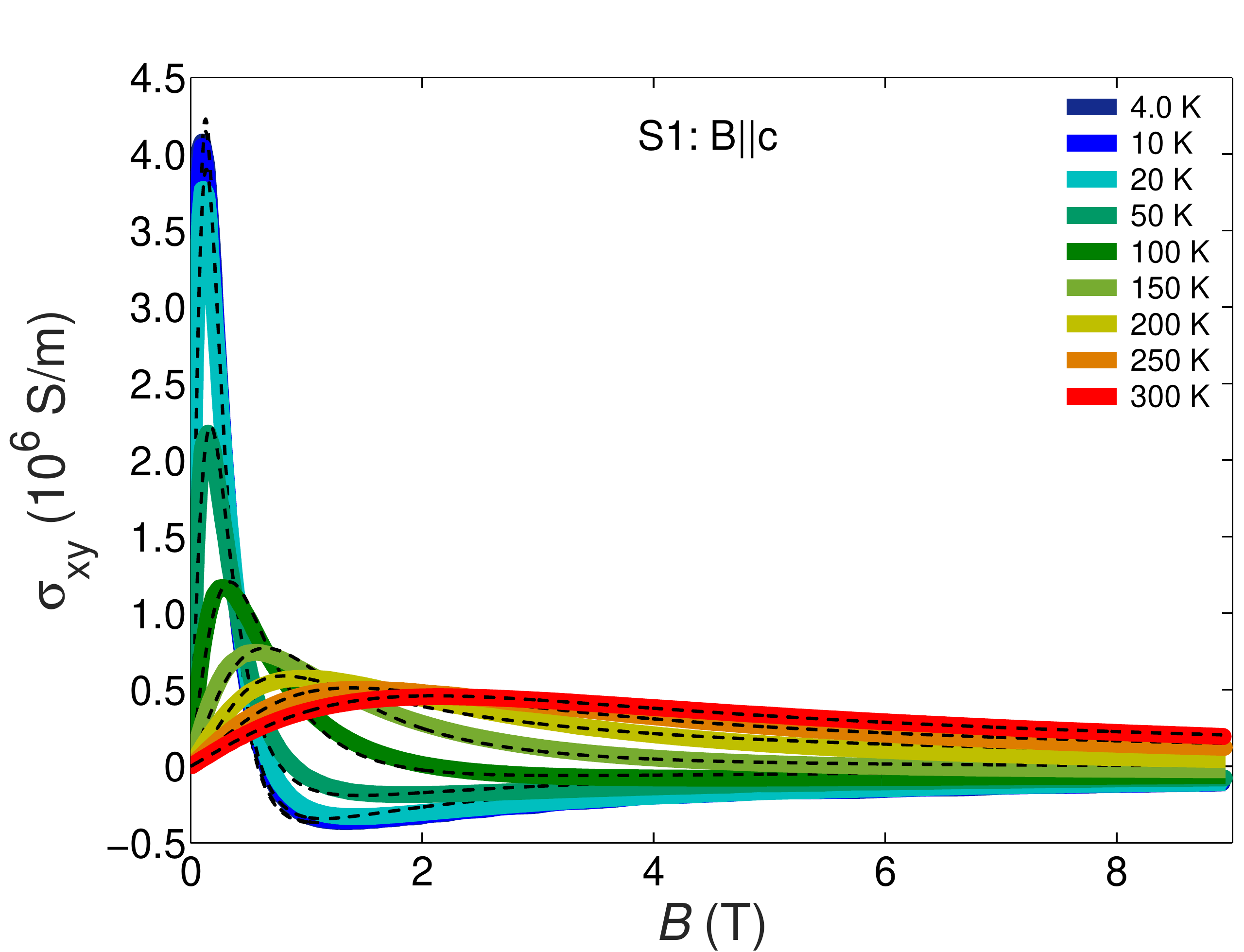}
\includegraphics[angle=-90,width=0.5\textwidth]{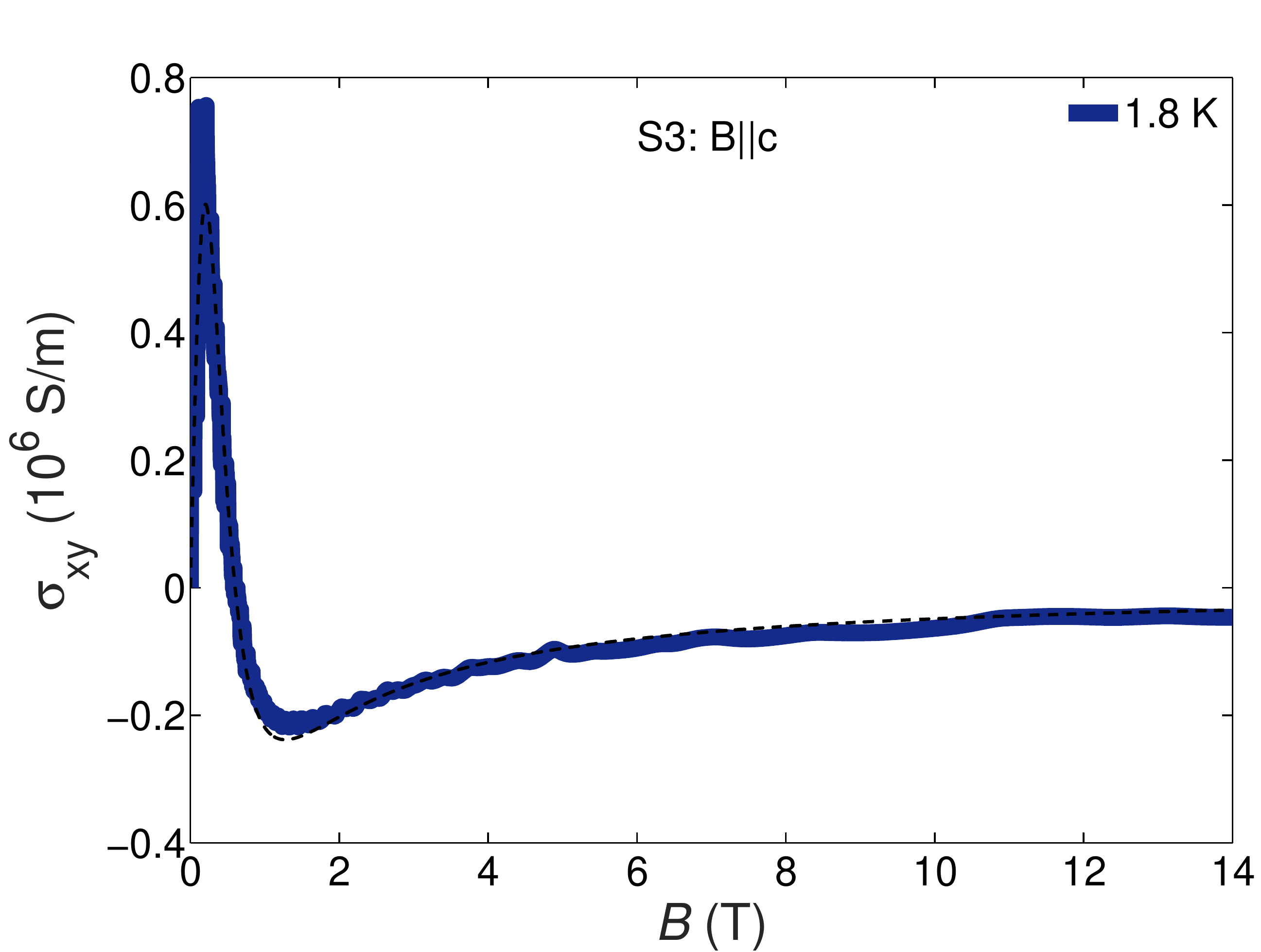}
}
\caption{{\bf Hall conductivity and two-band model fits}. The left and right-hand side graph show the Hall conductivity, $\sigma_\mathrm{xy}$ of sample S1 and S3 respectively for $B||c$. Solid lines correspond to experimental data whereas dashed lines denote fits to the two-band model.}
\label{fig:TBM-Fit}
\end{figure}

Figure \ref{fig:PositiveMRandHall} shows the tranverse and Hall resistivity, $\rho_\mathrm{xy}$, of the samples S1. The Hall coefficients of both TaP samples are initially positive and change sign above 1~T. The Hall sign reversal field is strongly temperature dependent. This nonlinear magnetic field dependence shows the presence of two competing charge carrier types in TaP and can be used to obtain the respective charge carrier densities.
In the case where $B || c$, the transverse and Hall conductivities are related to the resistivities via \cite{Singleton}: 
\begin{eqnarray}
\sigma_\mathrm{xx} = \frac{\rho_\mathrm{xx}}{\rho^2_\mathrm{xy}+\rho^2_\mathrm{xx}} \hspace{2em}\textrm{and}\hspace{2em} \sigma_\mathrm{xy} = - \frac{\rho_\mathrm{xy}}{\rho^2_\mathrm{xy}+\rho^2_\mathrm{xx}}
\label{eqn:ConductivityTensor}
\end{eqnarray}
The Hall conductivities are fitted by a two-band model using: 
\begin{eqnarray} 
\sigma_\mathrm{xy} = \left[ n_\mathrm{h} \mu_\mathrm{h}^2 \frac{1}{1+(\mu_\mathrm{h} B)^2}  - n_\mathrm{e} \mu_\mathrm{e}^2 \frac{1}{1+(\mu_\mathrm{e} B)^2}\right]eB \\
\sigma_\mathrm{xx}(B=0) = [n_\mathrm{h}\mu_\mathrm{h} + n_\mathrm{e}\mu_\mathrm{e}]e,
\end{eqnarray}
where $n_\mathrm{h}$ and $n_\mathrm{e}$ are the hole and electron densities, $\mu_\mathrm{h}$ and $\mu_\mathrm{e}$ are the hole and electron mobilities respectively, and $e$ is the elementary charge. 
Figure \ref{fig:TBM-Fit} (a, b) show the two-band model fits of the Hall conductivities of samples S1 and S3. Their Hall conductivity is well described by the two-band model over the entire temperature and magnetic field range. Slight deviations of the fits from the data originate from the uncertainty of the contact geometry influencing the absolute values of $\rho_\mathrm{xx}$ and $\rho_\mathrm{xy}$. The resulting charge carrier densities and mobilities are shown in Fig. 2 of the main text. Here, the carrier densities remain almost constant within the error bars of our measurements. However, we observe signs of a cross over from electron to hole dominated transport at 150~K. This is also reflected in the shift of the Hall conductivity zero towards higher fields and beyond the field range studied in this article.

%%%%%%%%%%%%%%%%%%%%%%%%%%%%%%%%%%%%%%%%%%%%%%%%%%%%%%%%%%%%%%%%%%%%%%%%%%%%%%%%%%%%%%%%%%%%%%%%%%%%%%%%%%%%%%%%%%%%%%%%%%
\clearpage

\section{Quantum Oscillations}

The electronic structure of TaP has been characterized by means of quantum oscillations. Here, the Shubnikov-de Haas (SdH) and de Haas-van Alphen (dHvA) effect were measured by electrical resistivity, SQUID-VSM and torque magnetometry measurements\cite{Shoenberg}. For experimental details please see the Methods section of the main text.

\begin{figure}[b!]
\begin{center}
\includegraphics[angle=-90,width=0.65\textwidth]{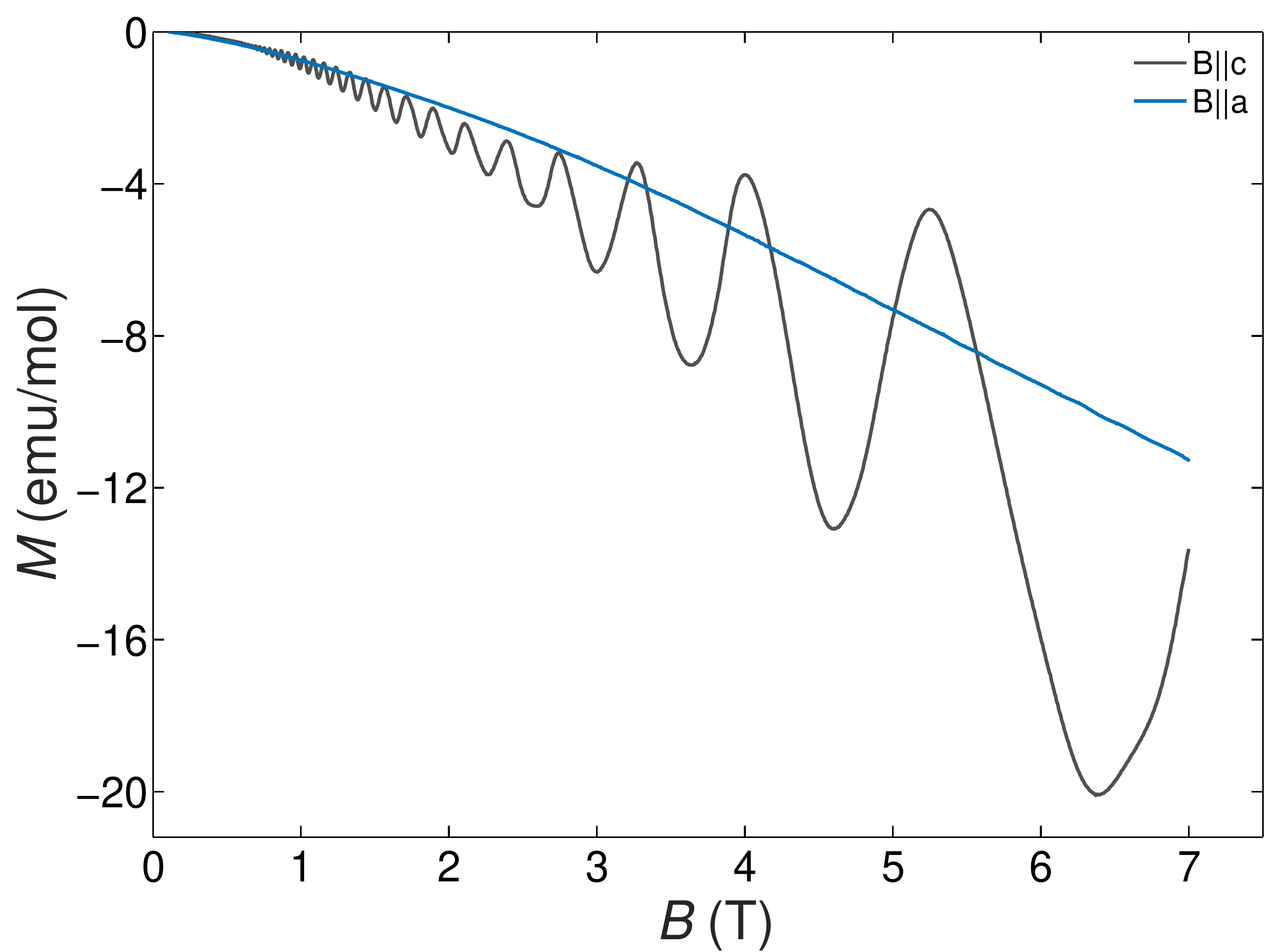}
\end{center}
\caption{{\bf Magnetization for the magnetic field applied along the main crystal axes.} Magnetization of TaP at $2\mathrm{K}$ for magnetic fields up to $7~\mathrm{T}$ applied along the crystallographic $a$ and $c$-direction. The diamagnetic magnetization is superimposed by strong quantum oscillation for field applied along the $c$-axis and very weak oscillations for $B||a$.}
\label{fig:VSM-Magnetisation}
\end{figure}

\noindent Figure \ref{fig:VSM-Magnetisation} shows the SQUID-VSM magnetizations of TaP for magnetic fields applied along the high symmetry axes at $2~\mathrm{K}$. Along the crystallographic $c$-direction the diamagnetic \clearpage\noindent magnetization of TaP is superimposed by prominent quantum oscillations starting at around $0.6~\mathrm{T}$. However, along the $a$-axis only weak quantum oscillations are visible. The frequency of these oscillations $F=A_\mathrm{ext}\times\hbar/2\pi\mathrm{e}$ is proportional to the extremal Fermi surface cross section $A_\mathrm{ext}$ perpendicular to the respective magnetic field direction\cite{Onsager52,Lifshitz56,Shoenberg}. Figures \ref{fig:ExtremalOrbits}-\ref{fig:SimulatedSignal} show the $k_\mathrm{z}$ dependence of the Fermi surface cross-section and the corresponding extremal orbits. Quantum oscillation spectra are obtained from magnetization and torque data by discrete Fourier transformation of the background subtracted oscillatory part of the respective signal. 

\begin{figure}[tb!] 
\centering
\includegraphics[angle=0,width=1.0\textwidth]{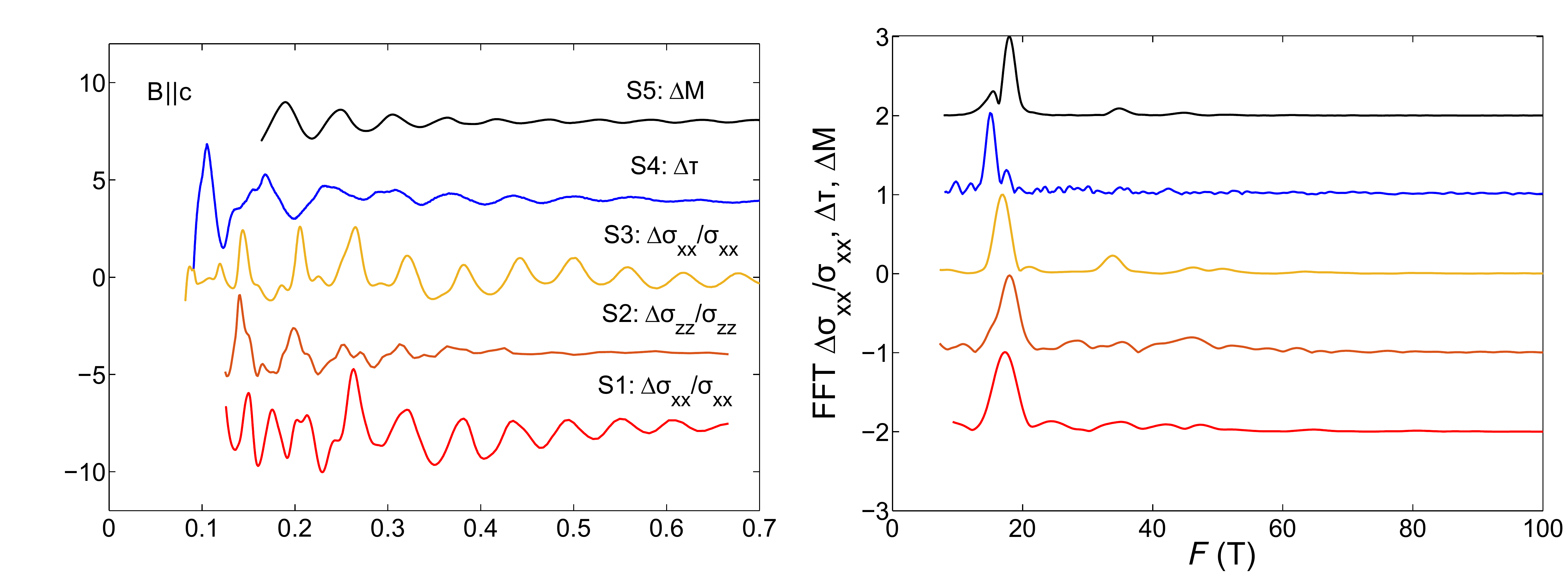}
\caption{{\bf Comparison of different quantum oscillation techniques and samples.} The left panel shows the background subtracted raw signals of the magnetization and resistivity measurements for $B||c$. The torque signal is given for an angle of 2.5 degrees off the c-axis as the magnetic torque is strictly zero for $B||c$. The right-hand side graph shows the Fourier transformation of the respective signals in the magnetic field window of 0.6 to $7~\mathrm{T}$.}
\label{fig:QOComparison}
\end{figure}

\noindent Figure \ref{fig:QOComparison} shows a comparison of the dHvA and SdH-oscillations of all five TaP samples for magnetic fields applied along the $c$-axis \cite{Singleton,Shoenberg,Lifshitz58}. As can be seen the magnetization and torque signal are dominated by two very close dHvA-frequencies, $F_\alpha=15~T$ and $F_\beta=18~T$ whilst the transport measurements can not distinguish between these two frequencies. As a consequence, the lowest frequency observed in the SdH-oscillations of S1, S2 and S3, F=17.5~T, is a superposition of these two frequencies. The relative amplitudes of the FFT peaks depend on the field range and the measurement method. Since resistivity and magnetization measurements were performed on samples from two different batches, finding the same dHvA-frequencies in all measurements is indicative of a similar chemical potential in all five samples.  

\noindent Throughout our measurements we observed strong second and third harmonics of $F_\alpha$ and $F_\beta$. Thus great care was taken when analysing the dHvA-spectra to exclude higher harmonics from the spectra of the fundamental dHvA-frequencies. Fourier transforms from various magnetic field ranges were correlated with each other to produce the angular dependence of Figure 1C (main text). The main results of the measured SdH and dHvA frequencies and their theoretical values along the main crystallographic axes are given in Table \ref{tab:dHvAConclusion}. Here the frequencies are named in accordance with the angular dependencies of Figures \ref{fig:dHvASignals} and \ref{fig:SdHSignals} and Figure 1C of the main text. 

%%%%%%%%%%%%%%%%%%%%%%%%%%%%%%%%%%%%%%%%%%%%%%%%%%%%%%%%%%%%%%%%%%%%%%%%%%%%%%%%%%%%%%%%%%%%%%%%%%%%%%%%%%%%%%%%%%%%%%%%%%

\section{Effective Mass}

\begin{figure}[tb!]
\centerline{
\includegraphics[angle=-90,width=0.5\textwidth]{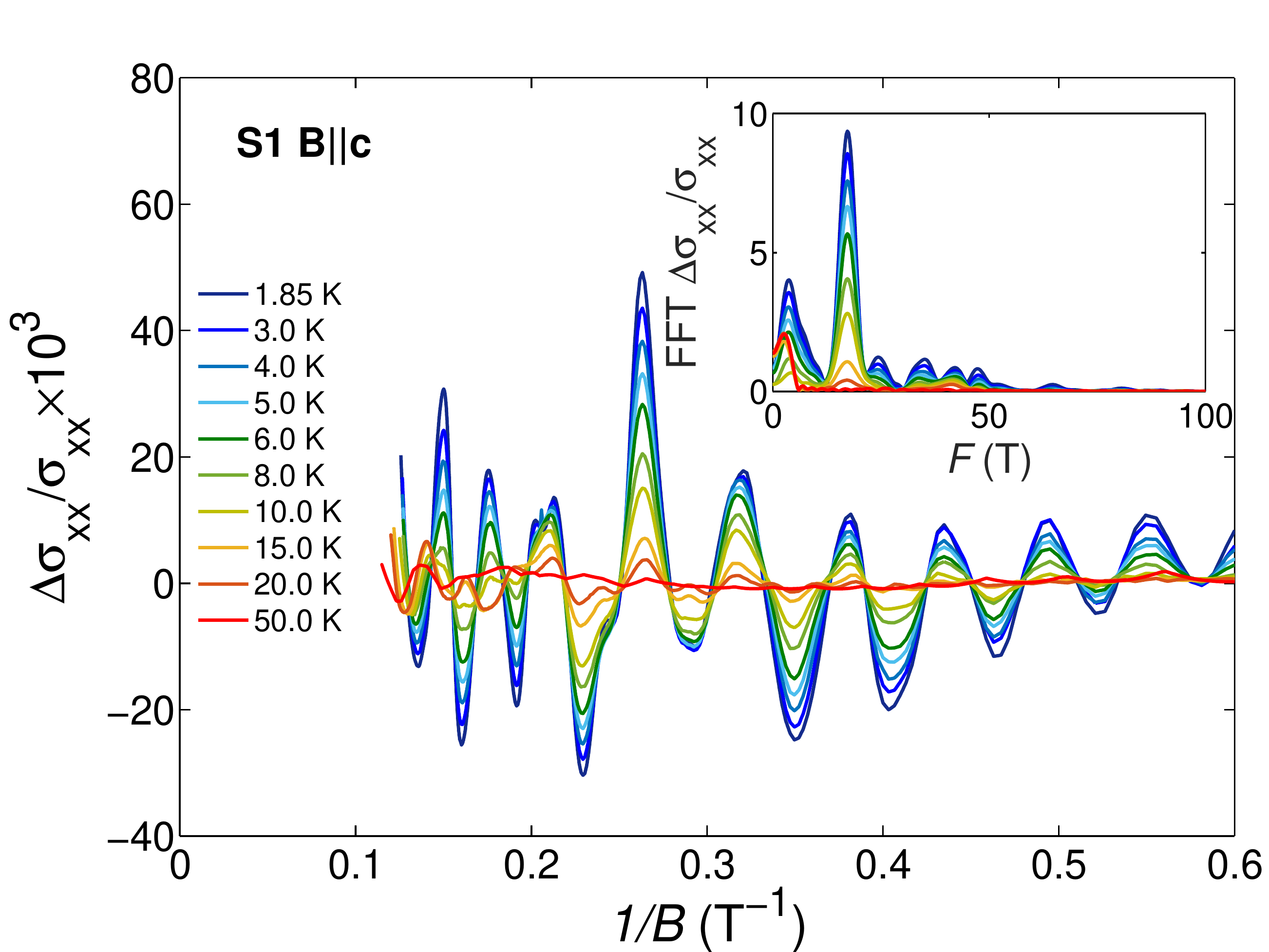}
\includegraphics[angle=-90,width=0.5\textwidth]{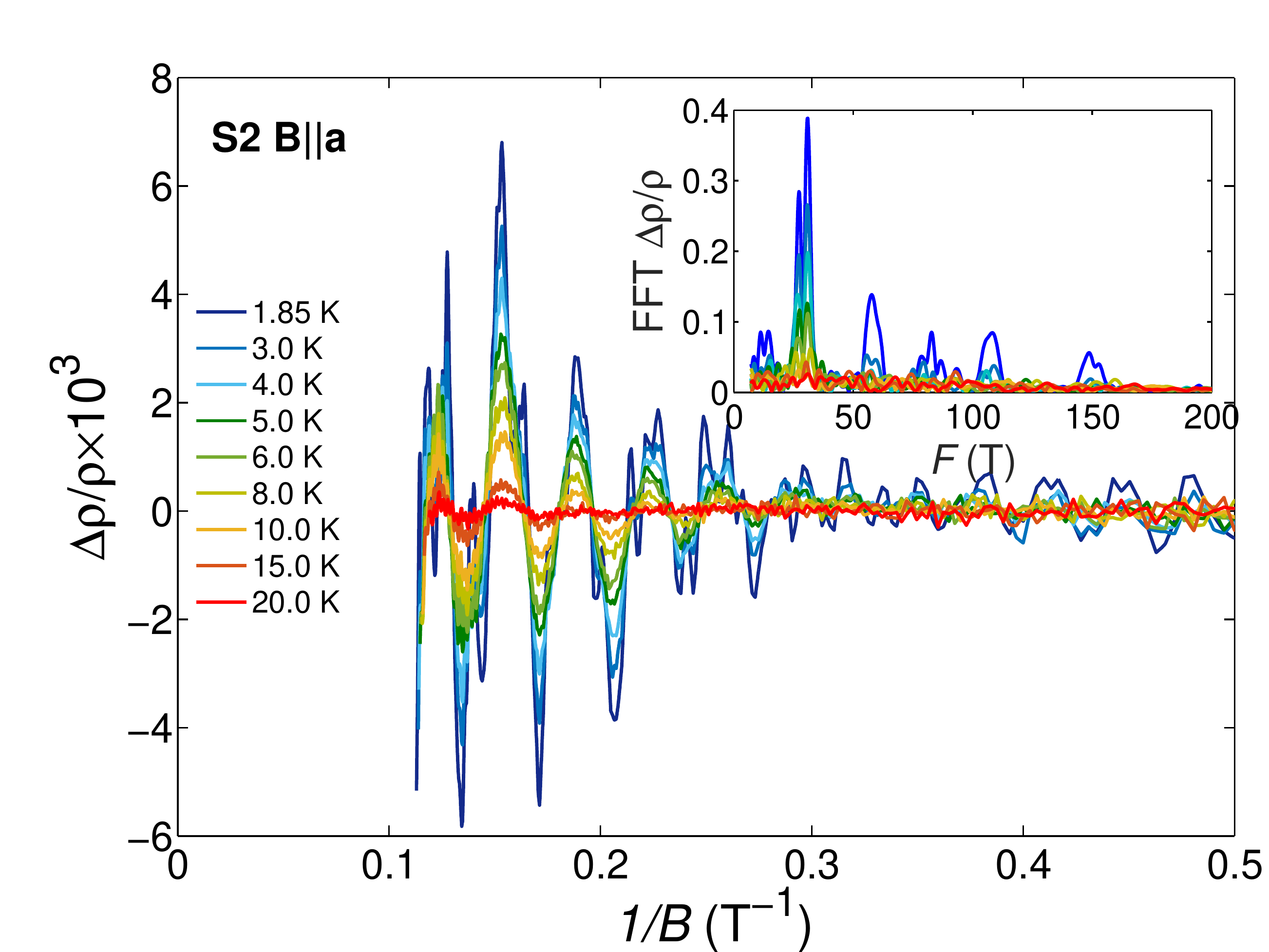}
}
\caption{{\bf Shubnikov de-Haas Oscillations (SdH).} SdH oscillations after background subtraction at different temperatures, left-hand side graph for $B||c$, right-hand side graph for $B||a$. The insets show the temperature dependent SdH Fourier transforms. }
\label{fig:SdHS1S2}
\end{figure}

\begin{figure}[tb!]
\centerline{
\includegraphics[angle=-90,width=0.5\textwidth]{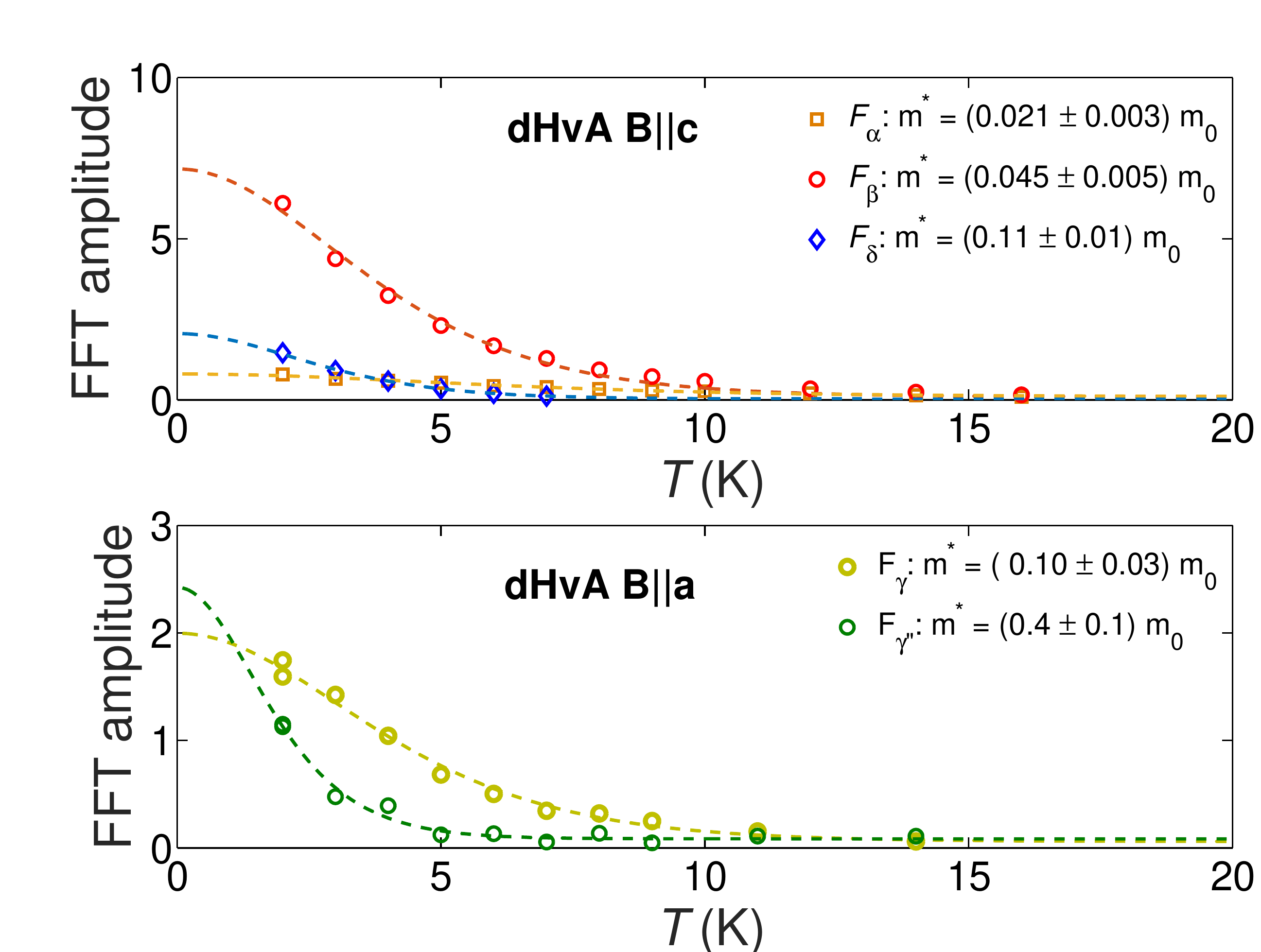}
\includegraphics[angle=-90,width=0.5\textwidth]{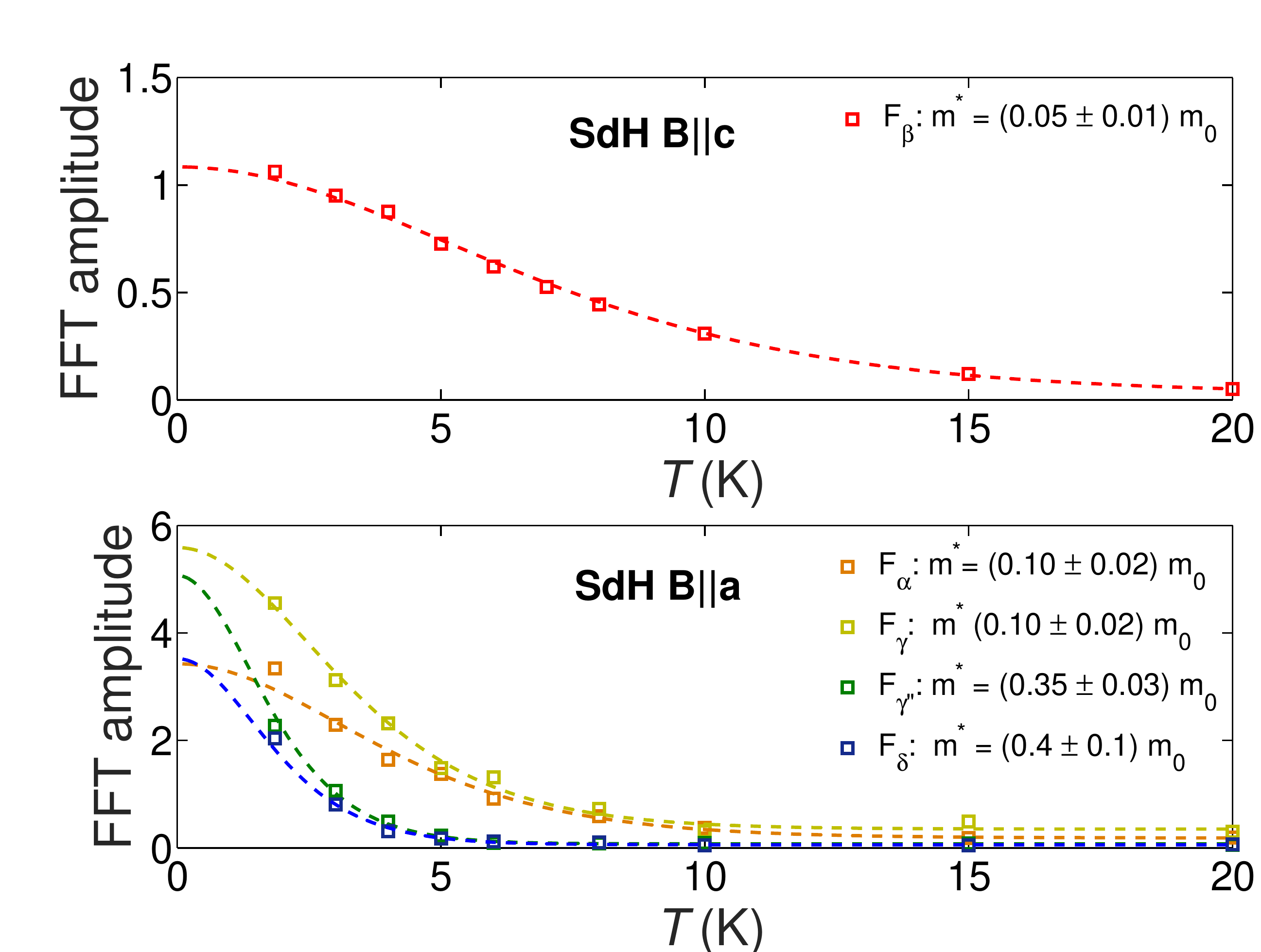}
}
\caption{{\bf Effective masses from de Haas-van Alphen and Shubnikov-de Haas amplitudes.} The left-hand side graph shows the temperature dependence of the de Haas-van Alphen FFT amplitudes in $B||c$ and $B||a$ direction (upper and lower graph respectively) as determined from SQUID-VSM measurements and the corresponding Lifshitz-Kosevich temperature reduction term fits\cite{Lifshitz56,Shoenberg}. The right-hand side shows the Shubnikov-de Haas amplitudes and temperature reduction term fits of samples S2 for $B||c$ and $B||a$ direction (upper and lower graph respectively). In order to increase the resolution of our data, different magnetic field ranges have been chosen to calculate the FFT amplitudes.}
\label{fig:EffectiveMass}
\end{figure}

By fitting the temperature dependence of the SdH and dHvA-amplitudes to the Lifshitz-Kosevich temperature reduction term\cite{Lifshitz56,Shoenberg}(see Fig. \ref{fig:EffectiveMass}):
\begin{eqnarray}
\frac{\Delta\sigma_\mathrm{xx}}{\sigma_\mathrm{xx}}(T) \propto \Delta \chi(T)\propto \frac{14.69\times m^* T/B}{\sinh(14.69\times m^* T/B)},
\end{eqnarray}
\noindent the effective masses of the strongest dHvA-signals were calculated for magnetic fields applied along both crystallographic axes. The resulting masses are shown in Table \ref{tab:dHvAConclusion} together with the values predicted by band structure calculations. Temperature dependencies of the raw SdH and dHvA signals can be found in Figure \ref{fig:SdHS1S2} - \ref{fig:EffectiveMass} and Figure 2B of the main text.

\noindent For magnetic fields applied along the $c$-axis, we find electron effective masses of $m_\delta^*= (0.11\pm0.01)~m_0$ and hole effective masses of $m_\alpha^*\approx (0.021\pm 0.003)~m_0$ and $m_\beta^*\approx (0.05\pm0.01)~m_0$ for the $\alpha$ and $\beta$-orbit respectively, where $m_0$ is the free electron mass. Both electron and the $\beta$-orbit effective masses are connected to extremal belly orbits around the center of their respective Fermi surface pockets in the $k_\mathrm{z}=0$-plane. On tilting the magnetic field away from the $c$-axis both effective masses increase with increasing extremal orbit size. We find that both hole effective masses double before we loose their signature $75^\circ$ out of the $c$-axis. The electron effective mass, on the other hand, is enhanced by a factor four along the $a$-axis.

\noindent The effective masses determined from band structure calculations (see Table \ref{tab:dHvAConclusion}) are of the same order as our experimentally obtained values. The theoretical band masses were directly determined from the slope of the extremal Fermi surface orbits with respect to the Fermi energy (see Fig. \ref{fig:EnergyDependence}), where $dF/d\epsilon_F=m^*/e\hbar$.

\begin{table}[tb!]
\centering
\caption{Quantum oscillation parameters as determined from resistivity (SdH) and magnetization (dHvA) measurements along the crystallographic $a$ and $c$-axis. The precise values in the second line are the theoretical frequencies and effective masses. The effective masses are given in units of the free electron mass  $m_0$.}
\begin{tabular}{|c c| c|c|c|c| c|c|}
\hline
\multicolumn{2}{|c|}{Direction} & \multicolumn{5}{c|}{Hole Pocket} & \multicolumn{1}{c|}{Electron Pocket}\\
  & & $\alpha$ & $\beta$& $\gamma$ & $\gamma'$ &$\gamma''$& $\delta$ \\
\hline
\multirow{5}{*}{$B||$c}& \multirow{3}{*}{$F$} & $(15\pm 1)~\mathrm{T}$  & $(18\pm 1) ~\mathrm{T}$  & $(25\pm 1) ~\mathrm{T}$ &  & & $(45\pm 2) ~\mathrm{T}$  \\
& & & &  &  & &$(51\pm 2) ~\mathrm{T}$\\
 & & 17.6~T & 17.8~T & 28.0~T &  &  & 52.7~T\\ \cline{2-8}
 &\multirow{2}{*}{$\frac{m^*}{m_0}$} & $0.021\pm 0.003$  & $0.05\pm 0.01$ & & &  &$0.11\pm 0.01$\\
 & & 0.07  & 0.06 & 0.16 & & & 0.18 \\
\hline
\multirow{4}{*}{$B||$a}&\multirow{2}{*}{$F$} & $(26\pm 1) ~\mathrm{T}$ &  & $(34\pm 2) ~\mathrm{T}$ &$(105\pm 4) ~\mathrm{T}$ & $(295\pm5)~\mathrm{T}$ & $(148\pm 5)~\mathrm{T}$\\ 
 & &  &  &  &  106 T& 340 T & 174 T \\ \cline{2-8}
 &\multirow{2}{*}{$\frac{m^*}{m_0}$} &  $0.10\pm 0.03$ &  & $0.13\pm 0.03$ & $0.35 \pm 0.03$& & $0.4\pm 0.1$\\
 & &  &  &  & 0.33 & 1.24 & 0.41 \\
\hline
\end{tabular}
\label{tab:dHvAConclusion}

\end{table}

\clearpage

%%%%%%%%%%%%%%%%%%%%%%%%%%%%%%%%%%%%%%%%%%%%%%%%%%%%%%%%%%%%%%%%%%%%%%%%%%%%%%%%%%%%%%%%%%%%%%%%%%%%%%%%%%%%%%%%%%%%
\section{Topology}

\begin{figure}[htb!]
\centerline{
\includegraphics[angle=-90,width=0.5\textwidth]{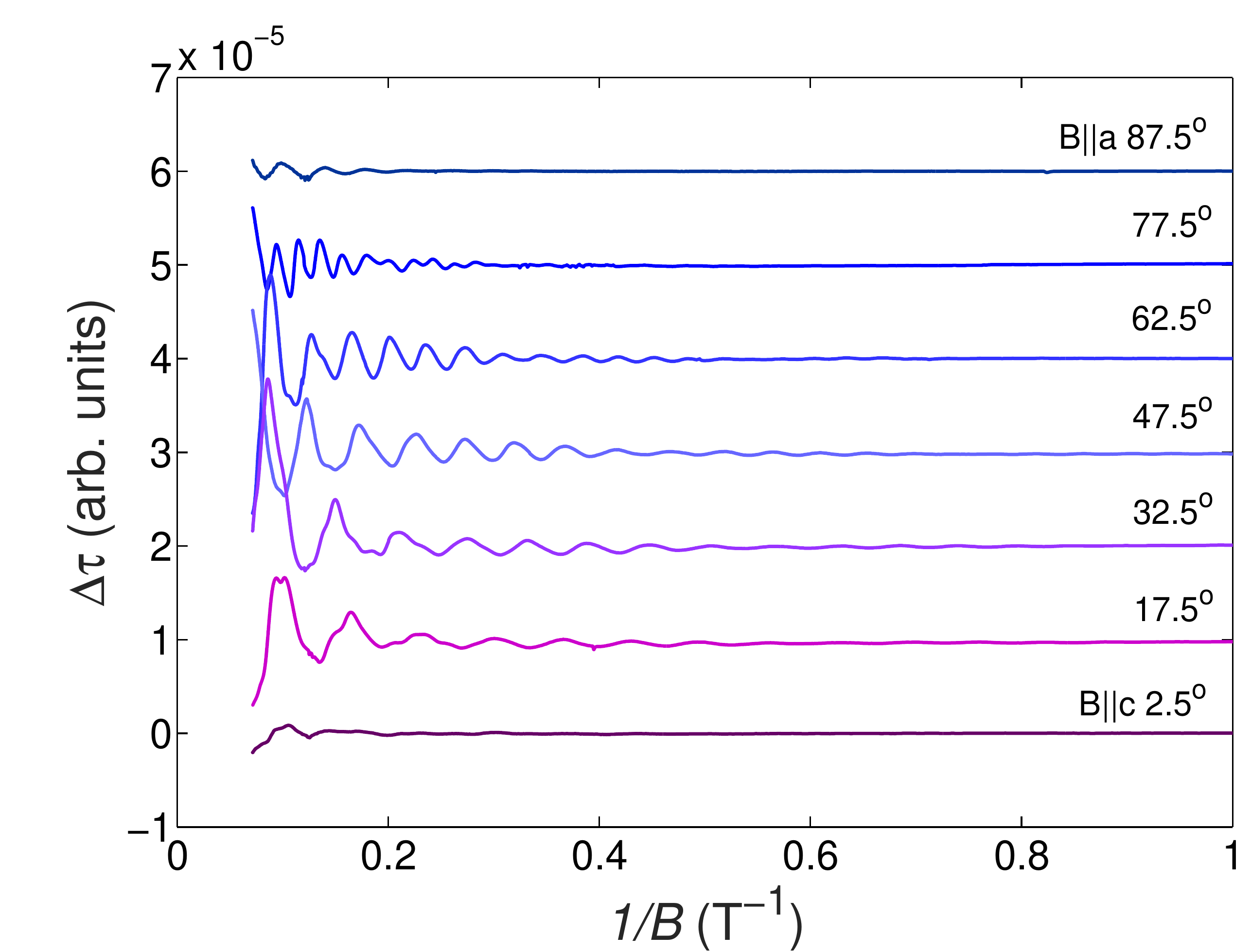}
\includegraphics[angle=-90,width=0.5\textwidth]{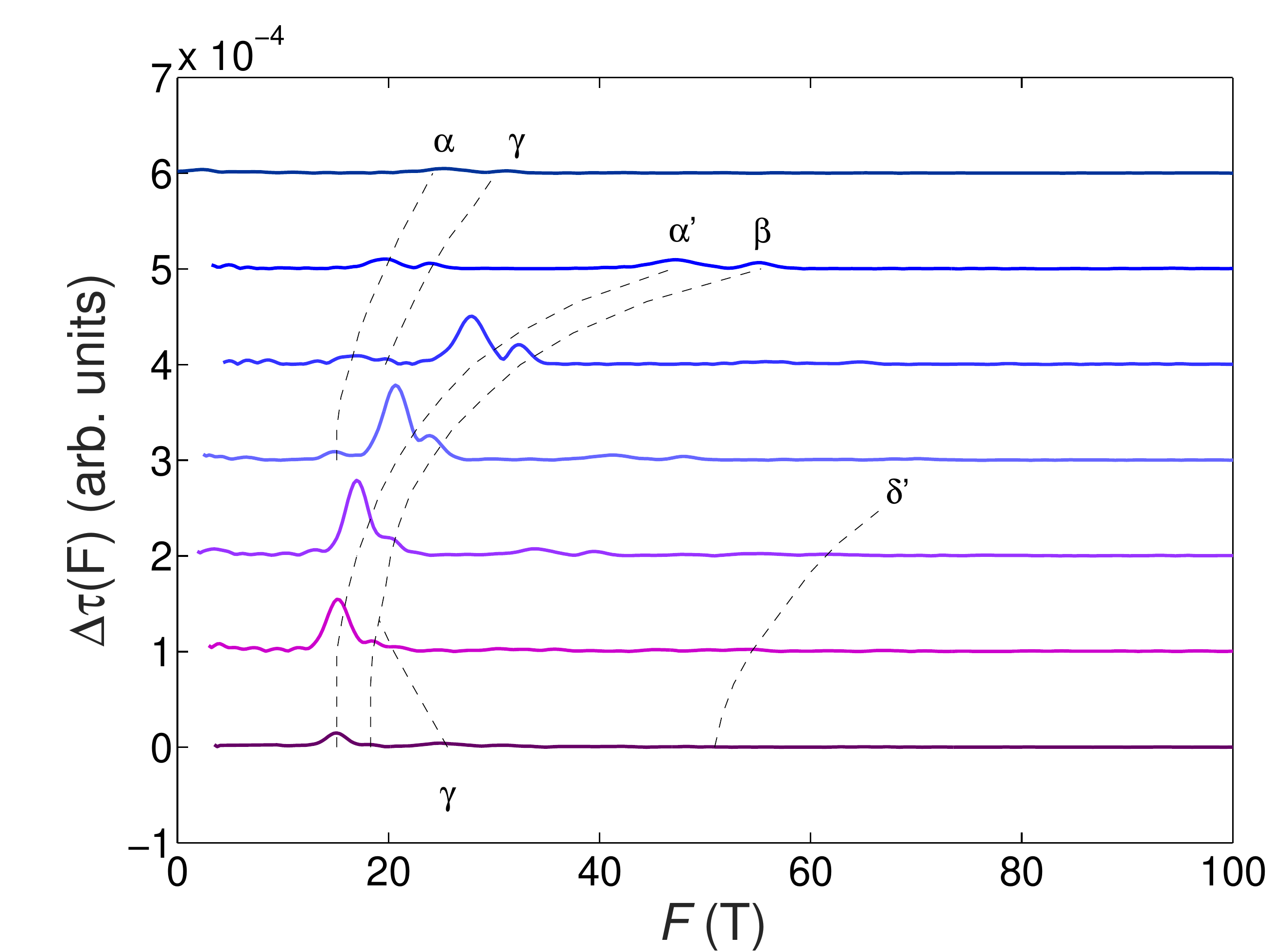}
}
\centerline{
\includegraphics[angle=-90,width=0.5\textwidth]{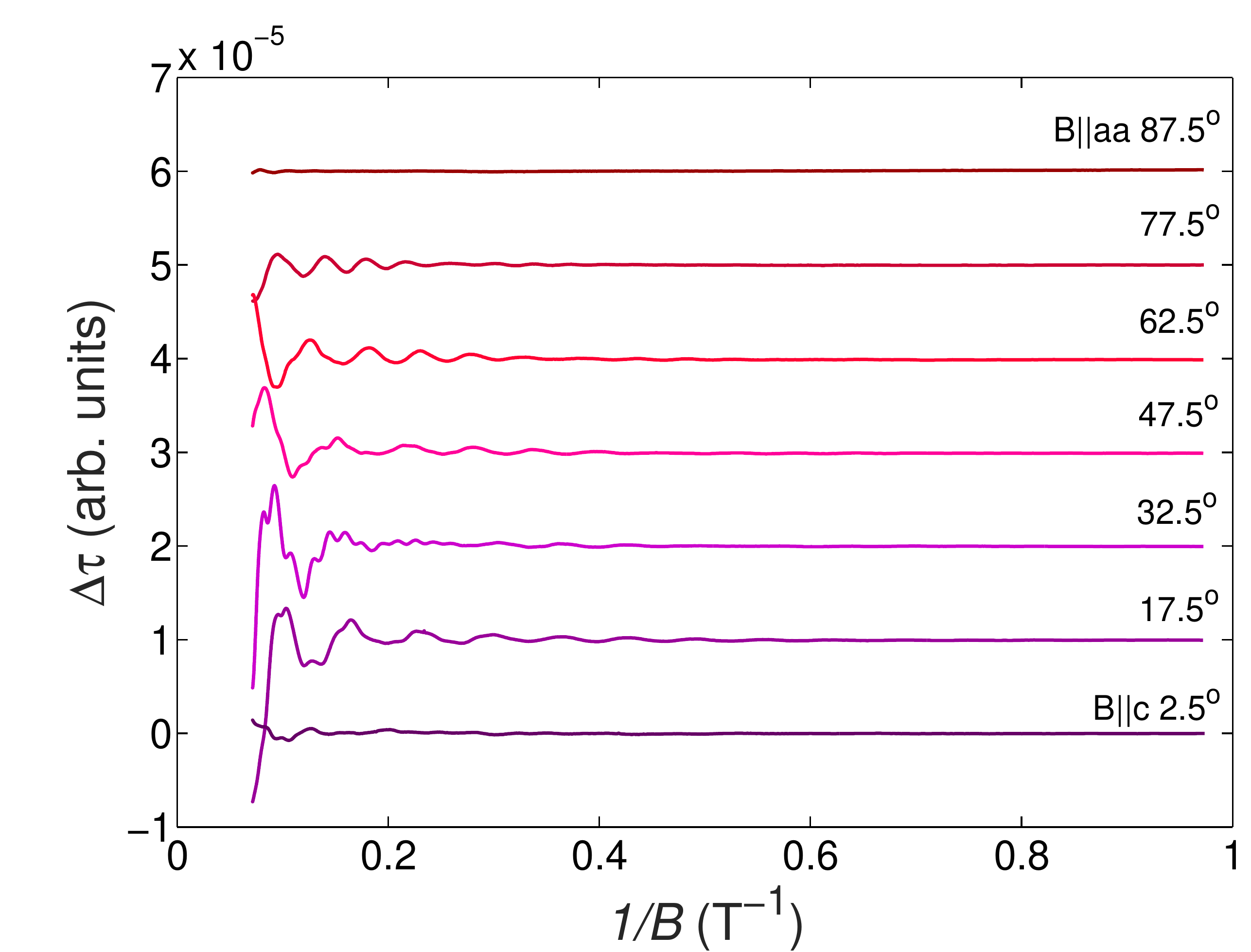}
\includegraphics[angle=-90,width=0.5\textwidth]{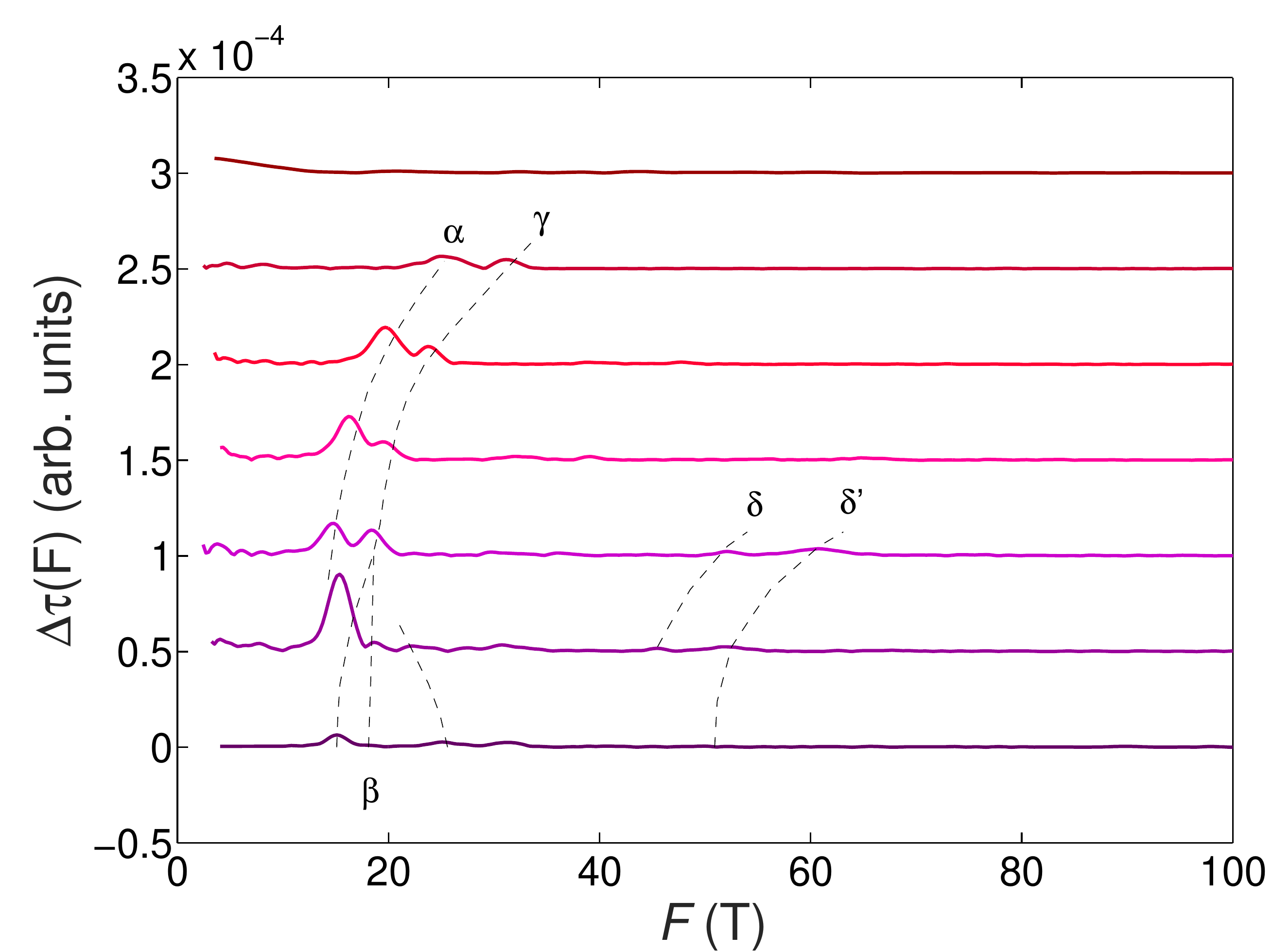}
}
\caption{{\bf Angular Dependence of dHvA Torque Signals.} The left graphs show the oscillatory component of the magnetic torque at $2~\mathrm{K}$ plotted versus the inverse magnetic field. The right hand side graphs show the respective Fourier transformations and characteristic spectra of the de Haas-van Alphen oscillations on the left. Dashed lines are guides to the eye, showing the angular dependence of the individual extremal orbits connected to the E1 and H1 Fermi surface. The upper and lower panels show the de Haas-van Alphen signals for magnetic fields applied in the (100) and (110)-plane respectively. The angles quoted are with respect to the crystallographic $c$-axis. A summary of the measured and theoretical quantum oscillation frequencies can be found in Fig. 1C of the main text.}
\label{fig:dHvASignals}
\end{figure}

\begin{figure}[htb!]
\centering
\includegraphics[angle=0,width=0.7\textwidth]{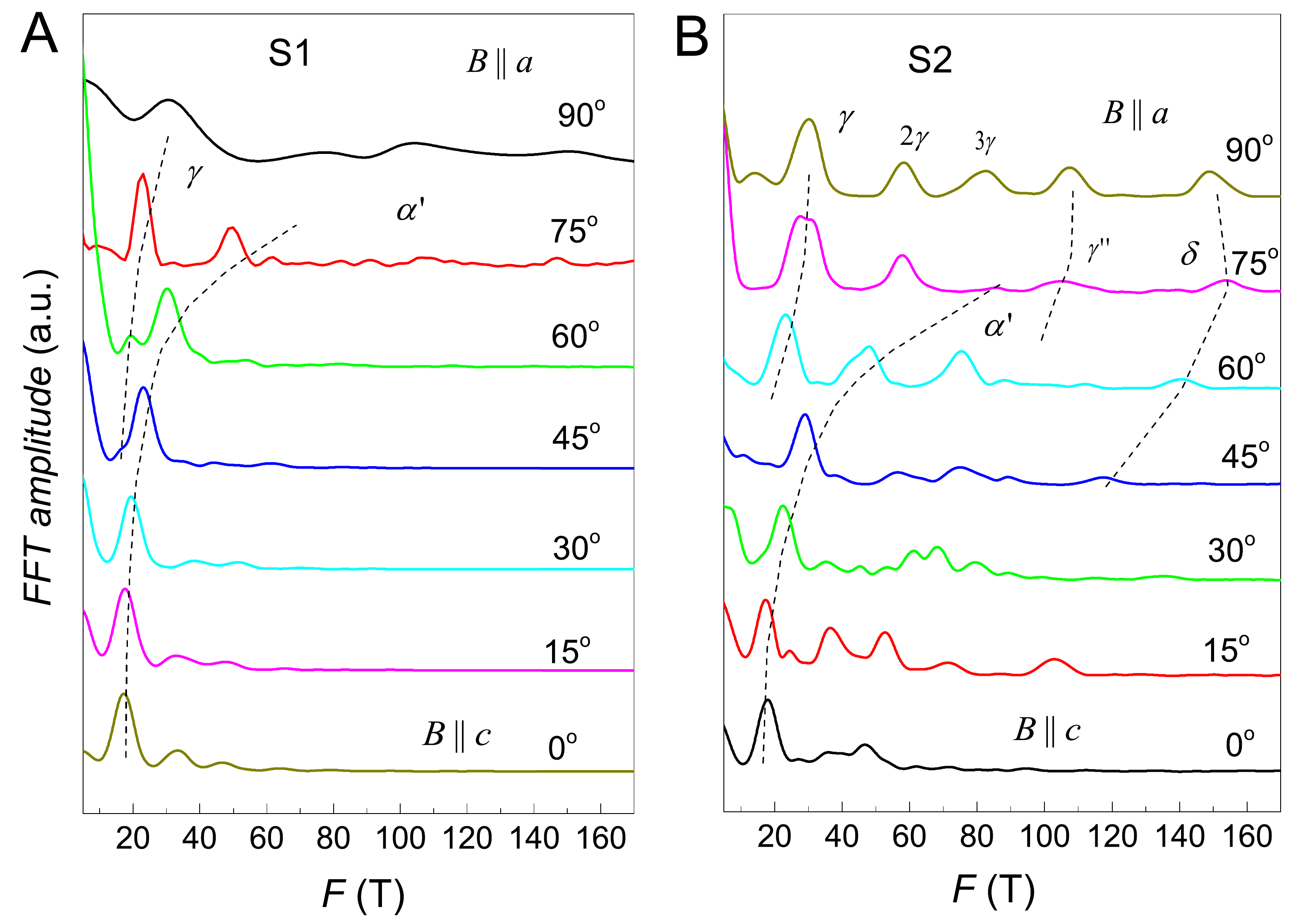}
\caption{{\bf Angular Dependence of SdH Signals for S1 and S2.} The left graph shows the Fourier transforms of S1 at different magnetic field angles within the (100)-plane. The right-hand side graph shows the same data for S2.}
\label{fig:SdHSignals}
\end{figure}

In order to reconstruct the full Fermi surface topology of TaP, the angular dependence of the quantum oscillation frequencies was determined from electrical resistance and magnetic torque measurements at $1.8~\mathrm{K}$ in magnetic fields up to $14~\mathrm{T}$. A PPMS rotator option was used to sample the extremal orbits for magnetic fields applied in the (100) and (110)-plane.

\noindent dHvA-signals from torque were obtained by background subtracting third order polynomials in the magnetic field range of 0.6 to $14.0~\mathrm{T}$ (left -hand side graphs of Fig. \ref{fig:dHvASignals}). Plotted against reciprocal magnetic field the remaining quantum oscillations were Fourier transformed to produce the quantum oscillation spectra (right-hand side graphs of Fig. \ref{fig:dHvASignals}). Figure 1C of the main text shows the frequencies determined from the angular dependent SdH and dHvA measurements. 

\noindent Using MBJ band structure calculations, we were able to identify the Fermi surface topology and Fermi energy matching our measured quantum oscillation frequencies. Figure \ref{fig:EnergyDependence} shows the MBJ band structure and theoretical energy dependence of the extremal orbits of TaP for B$||$c. As can be seen,  the energy of best fit lies $5~\mathrm{meV}$ above the charge neutrality point. Thus TaP is slightly electron doped as evidenced by the Hall measurements (see Fig. 2C and 2D of the main text).

\clearpage

\begin{figure}[htb!]
\centering
\includegraphics[width=1.0\textwidth]{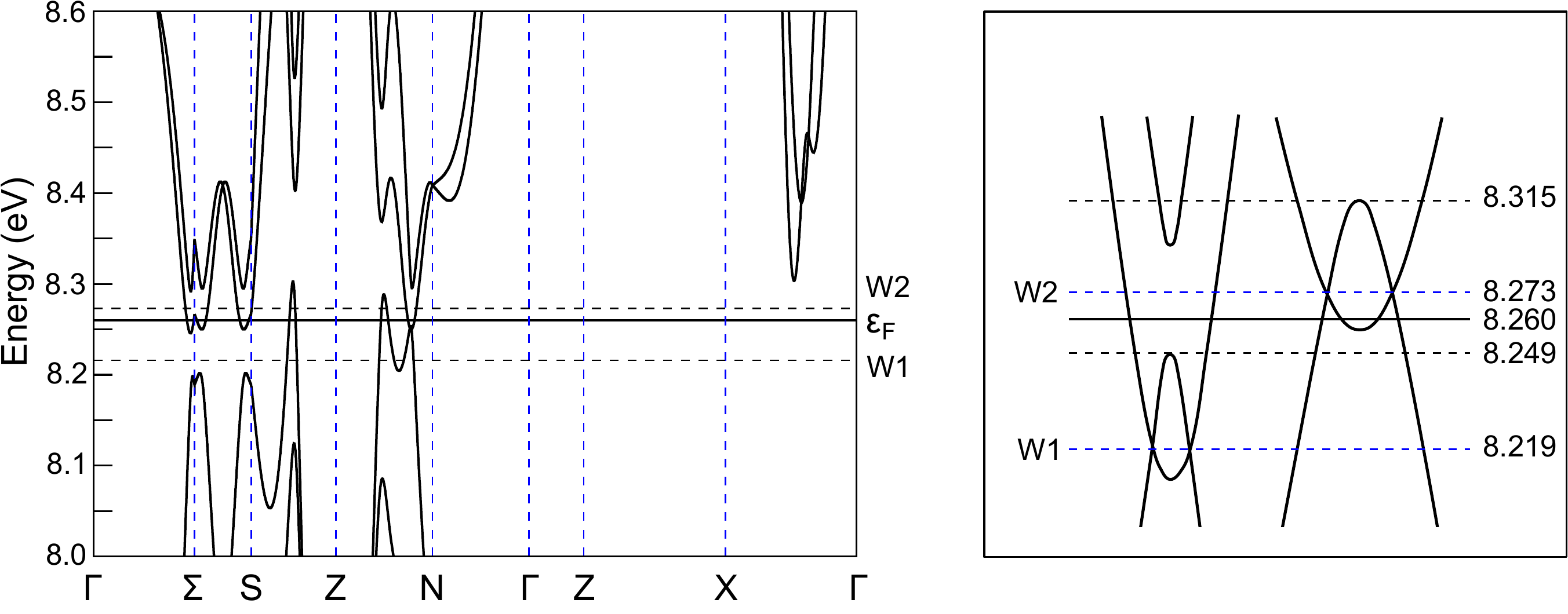}
\caption{{\bf MBJ Bandstructure of TaP.} The left graph shows the MBJ bandstructure along the major directions. Here the horizontal lines correspond to the energies of the W1 and W2 Weyl point as well as the Fermi energy of best fit. The right-hand side schematic shows a simplified version of the bandstructure, highlighting the position of the Fermi energy relative to the Weyl points.}
\label{fig:Bandstructure}
\end{figure}

\begin{figure}[htb!]
\centering
\includegraphics[angle=-90,width=0.7\textwidth]{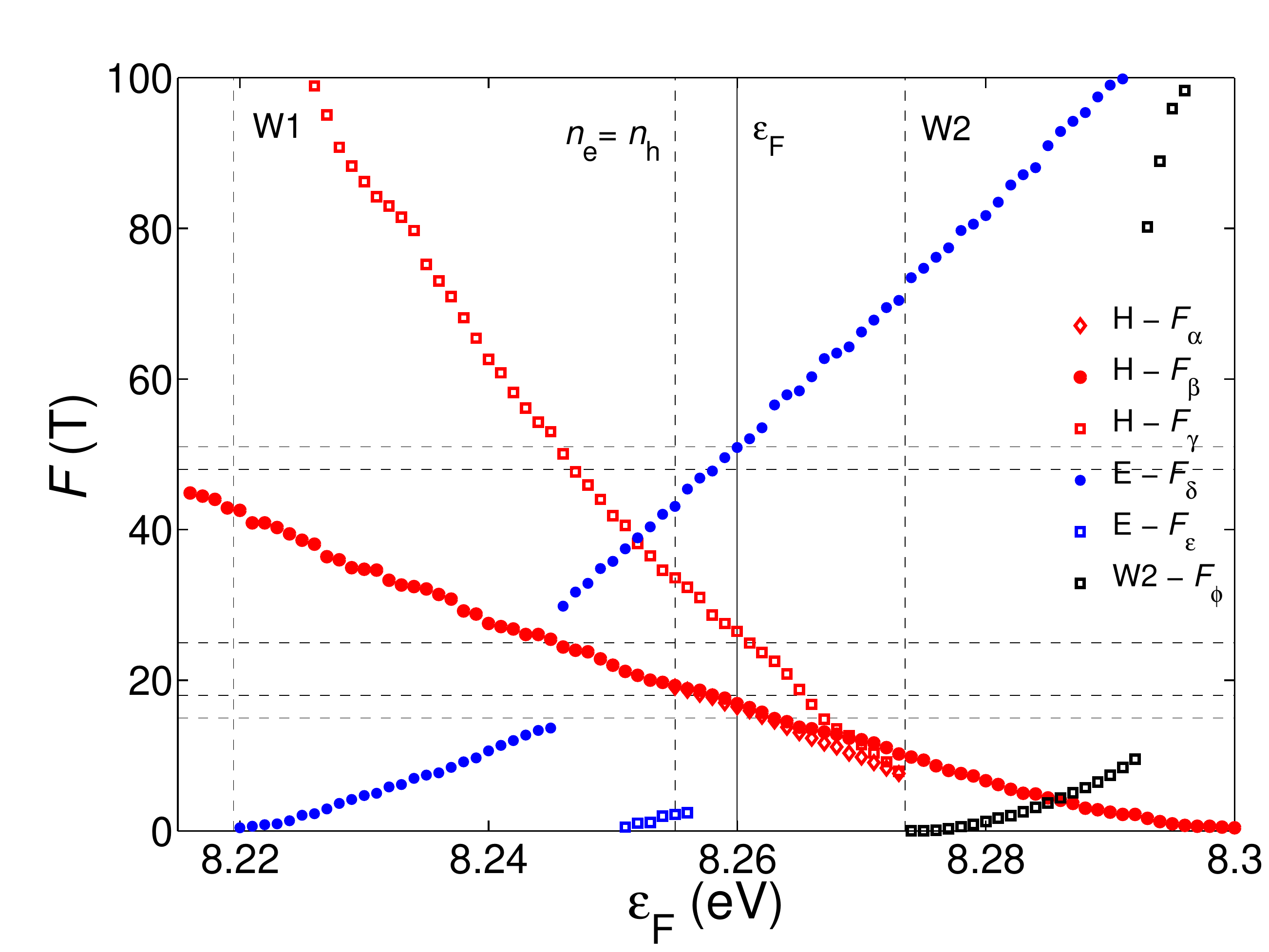}
\caption{{\bf Energy dependence of the extremal Fermi surface orbits for $B||c$.}  The graph shows the energy dependence of the extremal orbit size. Here the red, blue and green symbols are the respective extremal orbits of the banana shaped hole and electron pocket. The horizontal dashed lines mark the measured dHvA-frequencies for B$||$c, whilst the vertical dashed and solid lines represent the Weyl point energy of charge neutral point and energy of best fitting dHvA frequencies, as shown in Figure 1C of the main text.}
\label{fig:EnergyDependence}
\end{figure}

\begin{figure}[htb!]
	\centering
		\includegraphics[width=0.8\textwidth]{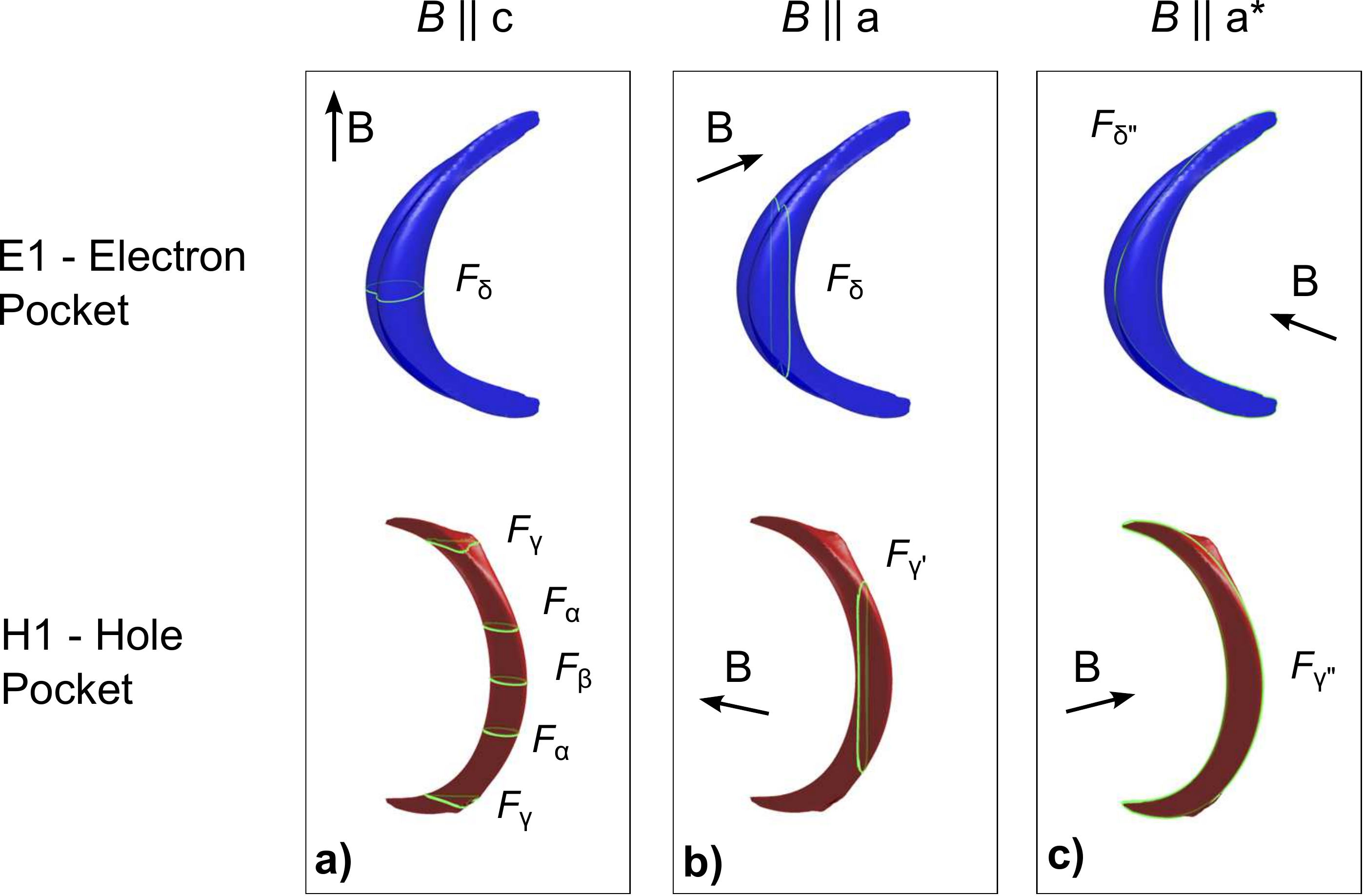}
		\caption{{\bf Extremal orbits of the E1 and H1 pocket for the magnetic field applied along the main crystallographic axes.} a) shows the extremal orbits for the magnetic field applied along the c-axis. Due to the four-fold rotational symmetry of the tetragonal Brillouin zone, Fermi surface pockets lying in the magnet field tilting plane and perpendicular to it yield the extremal orbits shown in b) and c) when the magnetic field is applied along the crystallographic a-axis.}
	\label{fig:ExtremalOrbits}
\end{figure}

\begin{figure}[htb!]
\begin{minipage}{0.50\textwidth}
\includegraphics[width=1.1\textwidth]{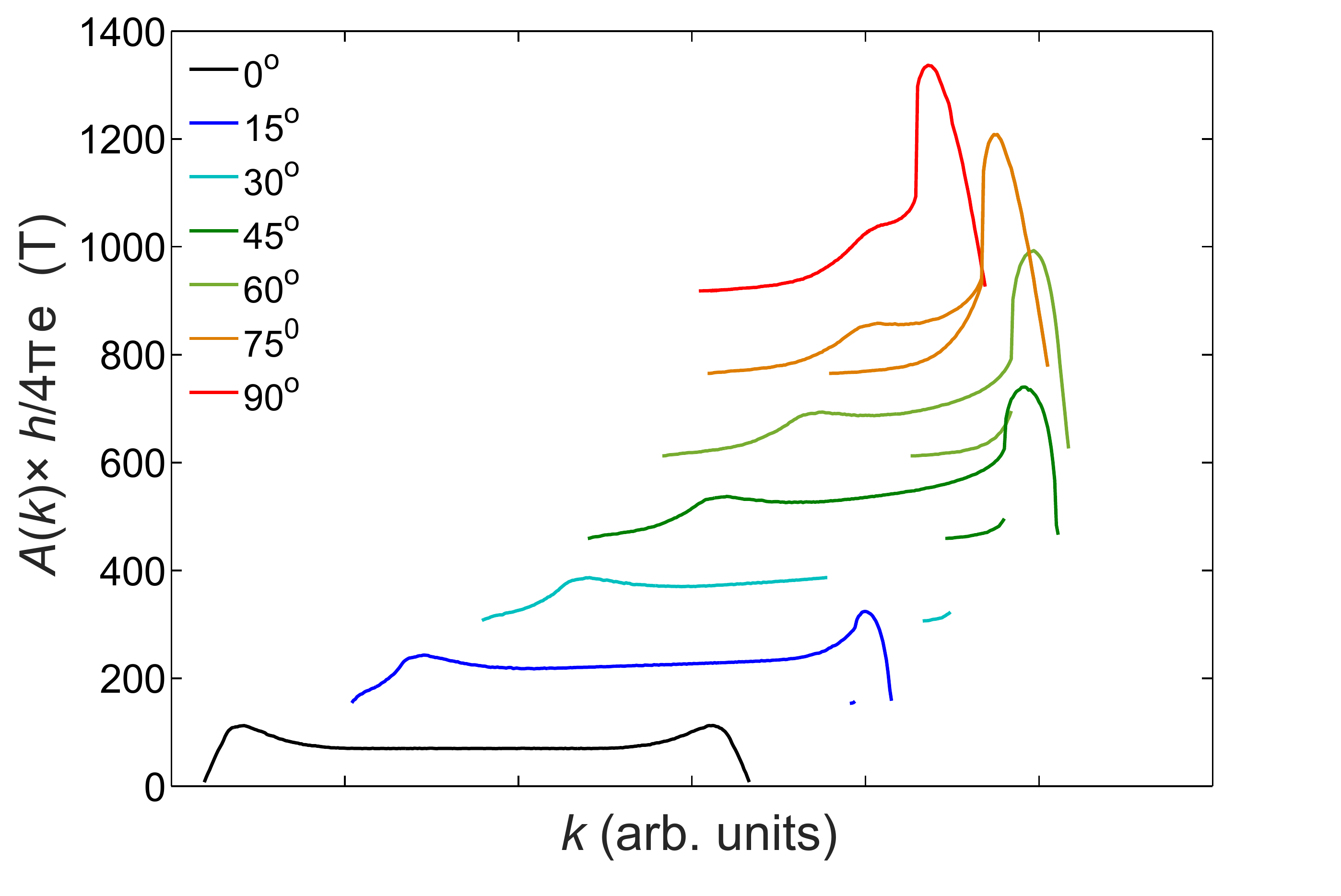}
\end{minipage}
\begin{minipage}{0.50\textwidth}
\includegraphics[width=1.1\textwidth]{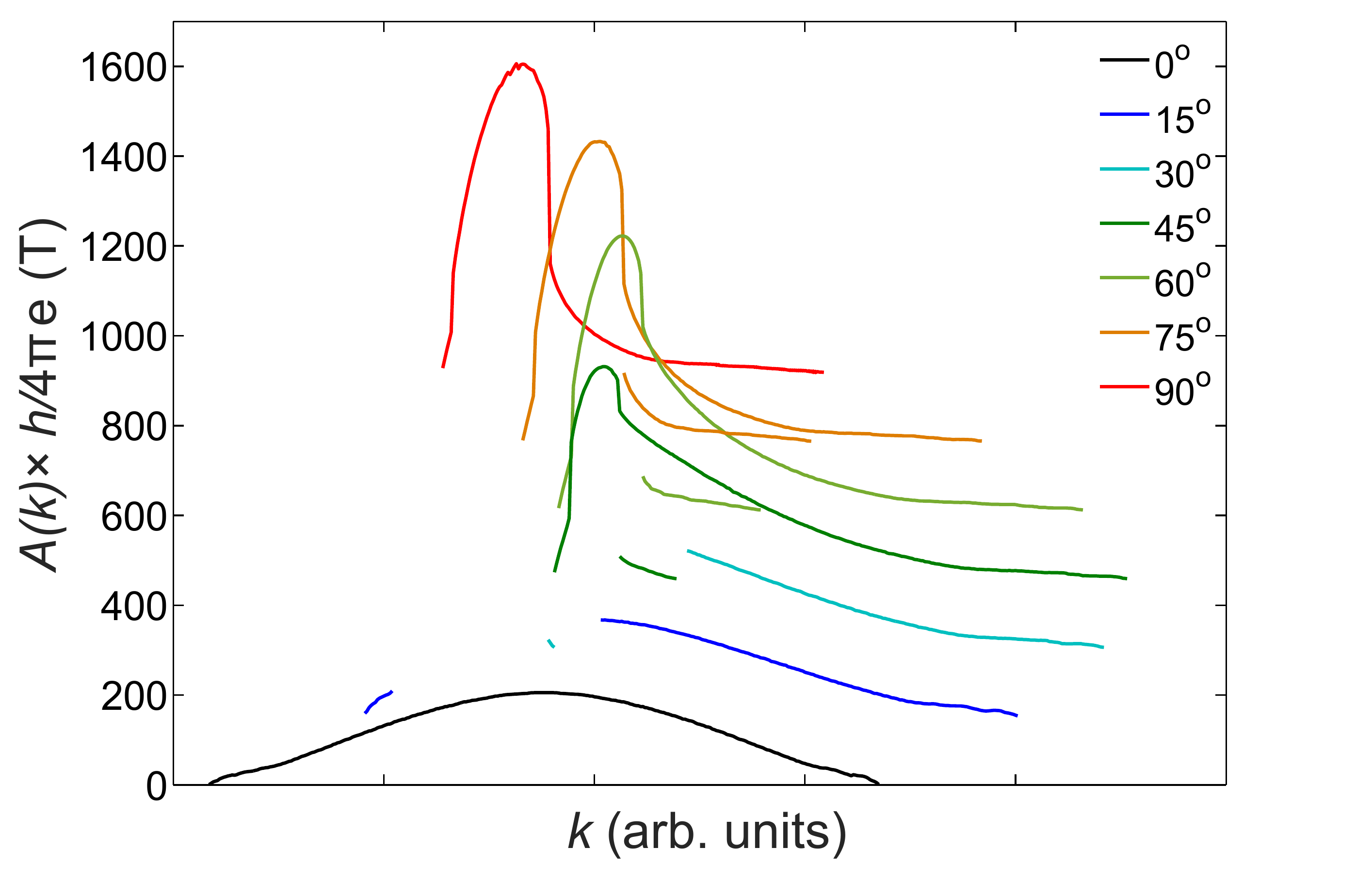}
\end{minipage}
\caption{{\bf Angular dependence of the Fermi surface cross sections for $B\in(100)$.} The left and right-hand side graphs show the $k||B$ dependence of the Fermi surface cross section of the hole (H) and electron (E) pocket for different magnetic field directions in the (100)-plane. Magnetic field angles are measured relative to the c-axis. Extremal Fermi surface orbits are marked with black dots and their respective names. Curves are shifted for clarity.}
\label{fig:AngularDependence}
\end{figure}

\noindent At a Fermi energy of $\epsilon_\mathrm{F}=8.260~\mathrm{eV}$ the Fermi surface of TaP consists of four banana shaped electron and hole pockets located along the [100]-directions. The $\beta$ and $\delta$ orbit correspond to the quasi-circular belly orbits around the centres of these pockets (see Fig. \ref{fig:ExtremalOrbits} and Fig. 3B of the main text). For $B||c$ these orbits lie in the $k_\mathrm{z}=0$-plane. On tilting the magnetic field away from the $c$-axis, these orbits tilt along the Fermi surface and become elliptical. The observed $1/cos(\theta)$-like behaviour for magnetic fields applied in the (010) and (110)-plane, is an indicator of the almost cylindrical shape near the belly and strong anisotropy of these pockets. The $\alpha$ and $\gamma$ frequencies, on the other hand, originate from extremal orbits of the H1-pocket closer to the W2 Weyl-points (see Fig. \ref{fig:ExtremalOrbits} - \ref{fig:SimulatedSignal} and Fig. 3B of the main text). Here the bumps of the H1 pocket, extending around the W2 Weyl-points, generate a maximal extremal orbit $\gamma$ running on top of the bumps and a neck orbit $\alpha$ closer to the centre of the Fermi surface. 

\noindent When the magnetic field is applied along the c-axis the two external $\alpha$-neck and $\gamma$-belly orbits, at either end of H1, are degenerate. By tilting the magnetic field in the plane of the Fermi surface pocket, away from the c-axis, this degeneracy is lifted and the orbits split into larger (prime orbits) and smaller orbits (see Fig. \ref{fig:AngularDependence}). $\beta$ being a belly orbit does not split. At a small critical angle of about $3^o$ the larger $\alpha'$-neck and central $\beta$-orbit merge and vanish. This is the analog of the first Yamaji angle in a quasi two-dimensional electron system \cite{Yamaji89}. On tilting the magnetic field even further, the splitting of the external $\gamma$ and $\gamma'$-orbits increases before the smaller $\gamma$-orbit merges with the remaining $\alpha$-neck orbit. Beyond $80^o$, the Fermi surface shows only one extremal orbit $\gamma'$.

\noindent Due to the fourfold symmetry of the tetragonal Brillouin zone, the hole pockets which are located at $90^o$ to the plane in which the magnetic field is tilted give rise to three more frequency branches (double primed, dashed lines in Fig. 1C of the main article). All of these orbits follow a quasi-two dimensional $1/cos(\theta)$ behaviour up to large angles. On increasing the polar field angle, the $\beta"$-orbit vanishes as the two $\alpha"$-orbits are pushed towards the centre of the Fermi surface where they merge. In close proximity to the $a$-axis the remaining $\alpha"$-orbit vanishes, as the two $\gamma"$-orbits merge. The new $\gamma"$-orbit encloses the entire H1 Fermi surface (see Fig. \ref{fig:ExtremalOrbits}). The E1 electron pockets on the other hand,  give rise to an ellipsoid like angular dependence of the extremal orbit size. Minor differences are only visible at large angles where the banana-shape leeds to a flatter angular dependence around the $a$-axis.

\noindent Compared to the band structure calculation we observe a more pronounced $F_\alpha$ neck-orbit. This leads to a larger splitting of $F_\alpha$ and $F_\beta$ and increases the critical angle for the $\alpha$ and $\gamma$-orbit. Both frequencies can observed up to B$||$a. The splitting of $F_\alpha$ and $F_\beta$, i.e. their frequency difference varies slightly between different band structure models and is beyond the resolution of our band structure calculations. Similarly the splitting observed in $\delta$ is not reproduced by the calculation. Here fine structures like a central waist in the electron pocket might give rise to the observed splitting.

%%%%%%%%%%%%%%%%%%%%%%%%%%%%%%%%%%%%%%%%%%%%%%%%%%%%%%%%%%%%%%%%%%%%%%%%%%%%%%%%%%%%%%%%%%%%%%%%%%%%%%%%%%%%%%%%%%%%%%%%%%%%%%%%
\section{Quantum Limit}

Due to the small bandwidth and low charge carrier effective mass in semi-metals like TaP, it is comparably easy to tune these system into their magnetic quantum limit by applying moderate magnetic fields\cite{Yang10,Arnold14Grp}. In the quantum limit the cyclotron energy $\hbar\omega_c$, i.e. the Landau level spacing, exceeds the bandwidth of the Fermi surface pocket. The bandwidth of the electron and hole pocket have been determined from our band structure calculation as $\epsilon\approx40~\mathrm{meV}$. Using the theoretical effective masses (Tab. \ref{tab:dHvAConclusion}), we obtain quantum limit fields of 62 and $23~\mathrm{T}$ for the electron and hole pocket respectively. Using the experimental values, however, we obtain slightly lower fields of 51 and $17~\mathrm{T}$. Given both results, we can conclude, that none of the Fermi surface pockets goes to its magnetic quantum limit in the magnetic field range studied in this letter.

%%%%%%%%%%%%%%%%%%%%%%%%%%%%%%%%%%%%%%%%%%%%%%%%%%%%%%%%%%%%%%%%%%%%%%%%%%%%%%%%%%%%%%%%%%%%%%%%%%%%%%%%%%%%%%%%%%%%%%%%%%%%%%%%

\section{Berry Phase}

A central interest in Weyl semi-metals with broken inversion symmetry is the determination of their Berry curvature or Berry phase $\Phi_\mathrm{B}$ by means of quantum oscillations\cite{He14,Weng15}. 

\noindent In general the total quantum oscillation phase $\phi$ at $1/B=0$ can be extracted from the back extrapolation of the Landau level index plotted against $1/B$ to zero. The total phase, which can be extracted this way is the sum of three phase factors\cite{Hubbard11,Shoenberg}:
\begin{eqnarray}
\phi = 2\pi(\gamma + \delta + \Phi_\mathrm{B}),
\label{eqn:BerryPhase}
\end{eqnarray}

\begin{figure}[htb!]
\centering
\includegraphics[angle=-90,width=0.65\textwidth]{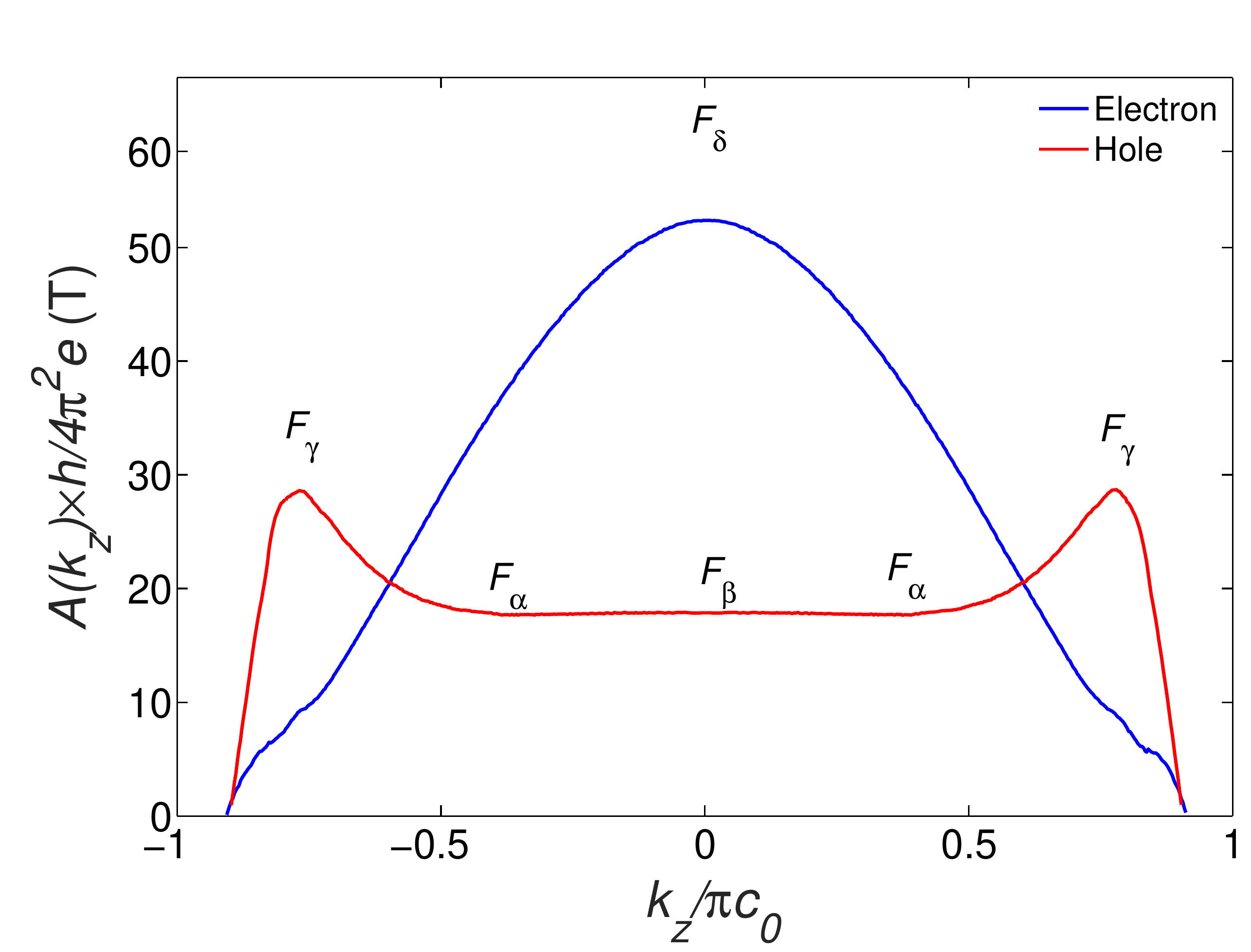}
\centering
\includegraphics[angle=-90,width=0.65\textwidth]{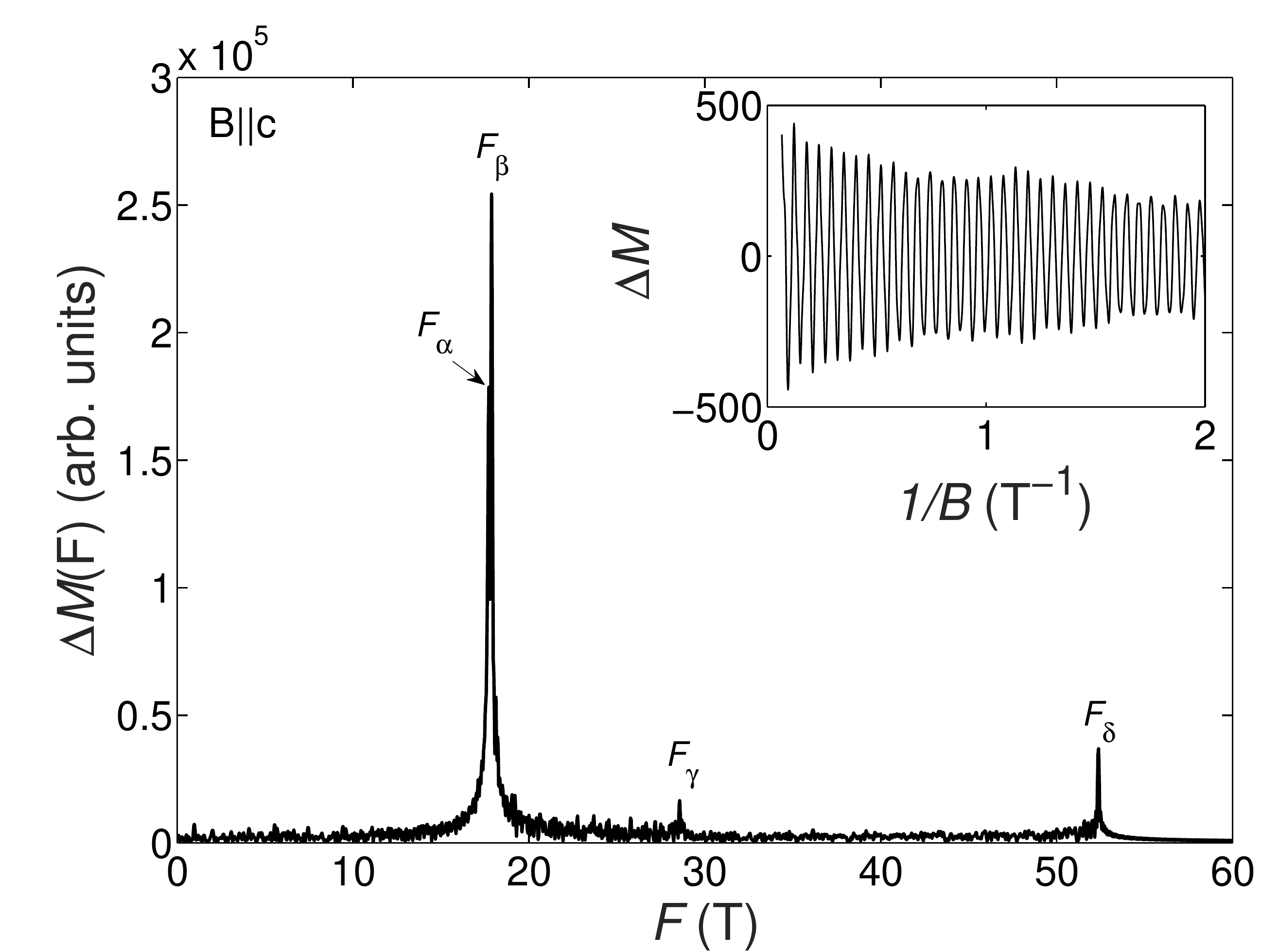}
\caption{{\bf Simulated de Haas-van Alphen signal.} The upper graph shows the $k_\mathrm{z}$ dependence of the Fermi surface cross-section of the electron and hole pocket for $B||c$. Here the extremal cross-sections are marked with the name of their respective quantum oscillation frequencies. The lower graph shows Fourier transform of the de Haas-van Alphen signal of the electron and hole pocket as determined by Eqn. \ref{eqn:QOInterference}. The inset shows the raw dHvA signal.}
\label{fig:SimulatedSignal}
\end{figure}

\clearpage

where $\gamma$ arises from the Onsager-Lifshitz quantisation and is $\pm^1/_2$ for massive and 0 for massless Fermions\cite{Onsager52}. The phase factor $\delta$ depends on the Fermi surface curvature, i.e. $d^2A_\mathrm{ext}/dk_\mathrm{B}^2$ and can take values between $+^1/_8$ and $-^1/_8$. It is zero for two-dimensional Fermi surfaces where $d^2A_\mathrm{ext}/dk_\mathrm{B}^2=0$. Thus, it is generally not possible to determine the Berry phase $\Phi_B$ without a precise knowledge of the Fermi surface topology\cite{Wang15,Luo15,Yang15}.

\noindent In order to exclude the influence of $\delta$ in the determination of $\Phi_B$, we have simulated the dHvA-signal for the electron and hole pocket (see Fig. \ref{fig:SimulatedSignal}). Here the oscillatory dHvA-magnetization for magnetic field applied along the $c$-axis is defined as:
\begin{eqnarray}
\tilde{M} \propto \int_{-\pi/c_0}^{\pi/c_0}{\sin\left(\frac{\hbar}{e}\frac{A(k_\mathrm{z})}{B}\right)dk_\mathrm{z}},
\label{eqn:QOInterference}
\end{eqnarray}
where the $A(k_\mathrm{z})$ (Fig. \ref{fig:SimulatedSignal}) have been calculated from our band structure using the Fermi energy of $\epsilon_F=8.260$~eV.

\begin{figure}[htb!]
\centering
\includegraphics[angle=-90,width=0.65\textwidth]{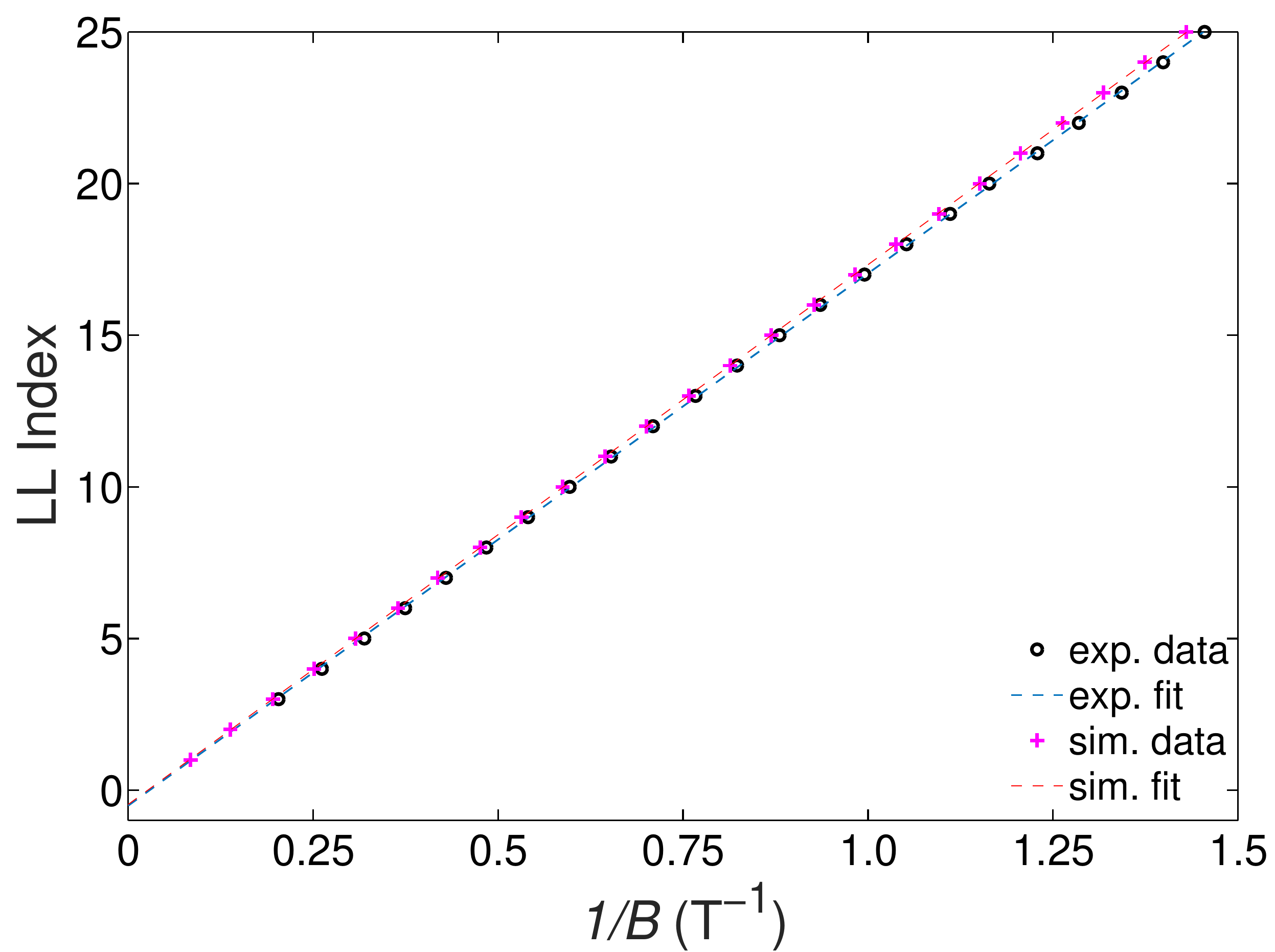}
\caption{{\bf Berry Phase plot for $B||c$.} The graph shows the Landau level indices of experimental and simulated de Haas-van Alphen oscillations as a function of the inverse magnetic field. Both curves extrapolate to $\phi/2\pi=-0.50\pm0.03$ at infinite field. The slope of both curves represents the dominant $F_\beta$ orbit and is 17.55 and $17.80~\mathrm{T}$ for the experimental and theoretical data respectively.}
\label{fig:ZeroPhase}
\end{figure}

\noindent In Figure \ref{fig:ZeroPhase} the Landau level indices of the measured and simulated dHvA-oscillations for $B||c$ are plotted as function of the inverse magnetic field. Here the Landau level was assigned to the falling zero-crossing of the oscillatory magnetization when plotted as a function of $1/B$. Both data are in excellent agreement and extrapolate to a $\phi/2\pi=0.50\pm0.03$. We conclude that $\gamma+\delta+\phi_\mathrm{B}=-0.5$, which is indicative of a quasi-two dimensional orbit of massive charge carriers with $\gamma=-^1/_2$ and $\delta=0$. This is also evidenced by the $1/\cos(\theta)$-like angular dependence of the corresponding $F_\beta$ orbit. Furthermore, using Eqn. \ref{eqn:BerryPhase}, we can exclude additional phases, such as the Berry phase $\phi_\mathrm{B}$, as the measured total phase is in quantitative agreement with our semi-classical model. Due to the negligible dHvA amplitudes of the other Fermi surface orbits it was not possible to determine their phase.

%%%%%%%%%%%%%%%%%%%%%%%%%%%%%%%%%%%%%%%%%%%%%%%%%%%%%%%%%%%%%%%%%%%%%%%%%%%%%%%%%%%%%%%%%%%%%%%%%%%%%%%%%%%%%%%%%%%%%%%%%%%%%%%%
\section{Scattering Times and Mobilities}

The scattering time $\tau$, mean free path $l$ and charge carrier mobility $\mu$ can be determined from the magnetic field dependence of the quantum oscillation amplitude i.e. the Dingle reduction term\cite{Dingle52,Shoenberg}:
\begin{eqnarray}
\tilde{M}(B)\propto B^{-1/2}\exp\left(-\frac{14.69\times m^*T_\mathrm{D}}{B}\right).
\end{eqnarray}
Here $T_D=\hbar/2\pi k_\mathrm{B}\tau$ is the Dingle temperature, which is inversely proportional to the scattering times. In reciprocal field and corrected for the $\sqrt{B}$-prefactor, the Dingle term represents an exponentially decaying envelope function, which results in a Lorentzian line shape observed in the quantum oscillation spectra. As the full-width half-maximum of these Lorentzians lines $\Delta F_{1/2}$ is directly related to the Dingle temperature by $T_D=\Delta F_{1/2}/(14.69 4\pi m^*)$, the line width can be used to determine the Dingle temperature and scattering times in systems with complex dHvA-spectra.

\noindent The Dingle temperature of the hole pocket has been extracted from the Lorentzian line width. By varying the magnetic field window, over which the data is Fourier transformed, at low fields, we confirmed that the observed linewidths show no significant sinc-function contribution originating from a finite window size. Here, the lowest magnetic field was chosen such that the quantum oscillation amplitude has decayed to beneath the noise level. A double Lorentzian fit was applied to the $F_\alpha$ -$F_\beta$ double peak to extract the linewidth of both signals. Simulations showed that for the given window size and decay constants the sinc-contribution accounts for 1 and $21 \%$ of the total linewidths of $F_\alpha$ and $F_\beta$ respectively. Based on the intrinsic linewidths, Dingle temperatures of $1.5\pm0.3 \mathrm{K}$ and scattering times of $(8\pm2)\times10^{-13} \mathrm{s}$ were calculated for both orbits. Using the fitted effective masses we can calculate the charge carrier mean free path and mobility $\mu=\tau e/m^*$ of the hole charge carriers. We find that the hole mobility is $\mu_h\approx(3.2\pm0.5)\times10^{4}~\mathrm{cm}^2/\mathrm{Vs}$. Taking into account the measured dHvA-frequencies, effective masses and assuming a parabolic band, the mean free path of the holes can be calculated by applying: $l=v_F\times\tau$ with $v_F=\sqrt{\frac{2e\hbar}{m^{*2}}F}$. Here, we find a mean free path of $l_h=0.4~\mu\mathrm{m}$. Performing the same analysis on the electron pocket was not possible due to its comparably small signal size.


\begin{thebibliography}{10}
\expandafter\ifx\csname url\endcsname\relax
  \def\url#1{\texttt{#1}}\fi
\expandafter\ifx\csname urlprefix\endcsname\relax\def\urlprefix{URL }\fi
\providecommand{\bibinfo}[2]{#2}
\providecommand{\eprint}[2][]{\url{#2}}

\bibitem{Wan2011}
\bibinfo{author}{Wan, X.~G.}, \bibinfo{author}{Turner, A.~M.},
  \bibinfo{author}{Vishwanath, A.} \& \bibinfo{author}{Savrasov, S.~Y.}
\newblock \bibinfo{title}{{Topological semimetal and Fermi-arc surface states
  in the electronic structure of pyrochlore iridates}}.
\newblock \emph{\bibinfo{journal}{Phys. Rev. B}} \textbf{\bibinfo{volume}{83}},
  \bibinfo{pages}{205101} (\bibinfo{year}{2011}).

\bibitem{Liu:2014bf}
\bibinfo{author}{Liu, Z.~K.} \emph{et~al.}
\newblock \bibinfo{title}{{Discovery of a Three-Dimensional Topological Dirac
  Semimetal, Na3Bi}}.
\newblock \emph{\bibinfo{journal}{Science}} \textbf{\bibinfo{volume}{343}},
  \bibinfo{pages}{864--867} (\bibinfo{year}{2014}).

\bibitem{Weng:2014ue}
\bibinfo{author}{Weng, H.}, \bibinfo{author}{Fang, C.}, \bibinfo{author}{Fang,
  Z.}, \bibinfo{author}{Bernevig, B.~A.} \& \bibinfo{author}{Dai, X.}
\newblock \bibinfo{title}{{Weyl Semimetal Phase in Noncentrosymmetric
  Transition-Metal Monophosphides}}.
\newblock \emph{\bibinfo{journal}{Phys. Rev. X}} \textbf{\bibinfo{volume}{5}},
  \bibinfo{pages}{011029} (\bibinfo{year}{2015}).

\bibitem{Huang:2015uu}
\bibinfo{author}{Huang, S.-M.} \emph{et~al.}
\newblock \bibinfo{title}{{A Weyl Fermion semimetal with surface Fermi arcs in
  the transition metal monopnictide TaAs class}}.
\newblock \emph{\bibinfo{journal}{Nat. Comms.}} \textbf{\bibinfo{volume}{6}},
  \bibinfo{pages}{7373} (\bibinfo{year}{2015}).

\bibitem{Lv2015TaAs}
\bibinfo{author}{Lv, B.~Q.} \emph{et~al.}
\newblock \bibinfo{title}{{Experimental Discovery of Weyl Semimetal TaAs}}.
\newblock \emph{\bibinfo{journal}{Phys. Rev. X}} \textbf{\bibinfo{volume}{5}},
  \bibinfo{pages}{031013} (\bibinfo{year}{2015}).

\bibitem{Xu2015TaAs}
\bibinfo{author}{Xu, S.-Y.} \emph{et~al.}
\newblock \bibinfo{title}{{Discovery of a Weyl fermion semimetal and
  topological Fermi arcs}}.
\newblock \emph{\bibinfo{journal}{Science}} \textbf{\bibinfo{volume}{349}},
  \bibinfo{pages}{613} (\bibinfo{year}{2015}).

\bibitem{Yang2015TaAs}
\bibinfo{author}{Yang, L.~X.} \emph{et~al.}
\newblock \bibinfo{title}{{Weyl Semimetal Phase in non-Centrosymmetric Compound
  TaAs}}.
\newblock \emph{\bibinfo{journal}{Nat. Phys.}} \textbf{\bibinfo{volume}{11}},
  \bibinfo{pages}{728?--732} (\bibinfo{year}{2015}).

\bibitem{Adler1969}
\bibinfo{author}{Adler, S.~L.}
\newblock \bibinfo{title}{{Axial-Vector Vertex in Spinor Electrodynamics}}.
\newblock \emph{\bibinfo{journal}{Phys. Rev.}} \textbf{\bibinfo{volume}{177}},
  \bibinfo{pages}{2426--2438} (\bibinfo{year}{1969}).

\bibitem{Bell1969}
\bibinfo{author}{Bell, J.~S.} \& \bibinfo{author}{Jackiw, R.}
\newblock \bibinfo{title}{{A PCAC puzzle: $\pi^0 \rightarrow \gamma\gamma$ in
  the $\sigma$-model}}.
\newblock \emph{\bibinfo{journal}{Nuov Cim A}} \textbf{\bibinfo{volume}{60}},
  \bibinfo{pages}{47--61} (\bibinfo{year}{1969}).

\bibitem{Nielsen1983}
\bibinfo{author}{Nielsen, H.~B.} \& \bibinfo{author}{Ninomiya, M.}
\newblock \bibinfo{title}{{The Adler-Bell-Jackiw anomaly and Weyl fermions in a
  crystal}}.
\newblock \emph{\bibinfo{journal}{Phys. Lett. B}}
  \textbf{\bibinfo{volume}{130}}, \bibinfo{pages}{389--396}
  (\bibinfo{year}{1983}).

\bibitem{Chang2015}
\bibinfo{author}{Chang, M.-C.} \& \bibinfo{author}{Yang, M.-F.}
\newblock \bibinfo{title}{Chiral magnetic effect in the absence of weyl node}.
\newblock \emph{\bibinfo{journal}{Phys. Rev. B}} \textbf{\bibinfo{volume}{92}},
  \bibinfo{pages}{205201} (\bibinfo{year}{2015}).

\bibitem{Ma2015}
\bibinfo{author}{Ma, J.} \& \bibinfo{author}{Pesin, D.}
\newblock \bibinfo{title}{Chiral magnetic effect and natural optical activity
  in metals with or without weyl points}.
\newblock \emph{\bibinfo{journal}{Phys. Rev. B}} \textbf{\bibinfo{volume}{92}},
  \bibinfo{pages}{235205} (\bibinfo{year}{2015}).

\bibitem{Zhong2015}
\bibinfo{author}{Zhong, S.}, \bibinfo{author}{Moore, J.} \&
  \bibinfo{author}{Souza, I.}
\newblock \bibinfo{title}{Gyrotropic magnetic effect and the orbital moment on
  the fermi surface}.
\newblock \emph{\bibinfo{journal}{arXiv:1510.02167}}  (\bibinfo{year}{2015}).

\bibitem{Liu2015NbP}
\bibinfo{author}{Liu, Z.~K.} \emph{et~al.}
\newblock \bibinfo{title}{{Evolution of the Fermi surface of Weyl semimetals in
  the transition metal pnictide family}}.
\newblock \emph{\bibinfo{journal}{Nature Materials}}
  \textbf{\bibinfo{volume}{15}}, \bibinfo{pages}{27--31}
  (\bibinfo{year}{2016}).

\bibitem{Huang2015anomaly}
\bibinfo{author}{Huang, X.} \emph{et~al.}
\newblock \bibinfo{title}{{Observation of the chiral anomaly induced negative
  magneto-resistance in 3D Weyl semi-metal TaAs}}.
\newblock \emph{\bibinfo{journal}{arXiv}}  (\bibinfo{year}{2015}).
\newblock \eprint{1503.01304}.

\bibitem{Zhang2015ABJ}
\bibinfo{author}{Zhang, C.} \emph{et~al.}
\newblock \bibinfo{title}{{Observation of the Adler-Bell-Jackiw chiral anomaly
  in a Weyl semimetal}}.
\newblock \emph{\bibinfo{journal}{arXiv}}  (\bibinfo{year}{2015}).
\newblock \eprint{1503.02630}.

\bibitem{Yoshida1976}
\bibinfo{author}{Yoshida, K.}
\newblock \bibinfo{title}{A geometrical transport model for inhomogeneous
  current distribution in semimetals under high magnetic fields}.
\newblock \emph{\bibinfo{journal}{J. Phys. Soc. Jap.}}
  \textbf{\bibinfo{volume}{40}}, \bibinfo{pages}{1027} (\bibinfo{year}{1976}).

\bibitem{volovik2007quantum}
\bibinfo{author}{Volovik, G.}
\newblock \bibinfo{title}{Quantum analogues: From phase transitions to black
  holes and cosmology}.
\newblock \emph{\bibinfo{journal}{eds. William G. Unruh and Ralf Schutzhold,
  Springer Lecture Notes in Physics}} \textbf{\bibinfo{volume}{718}},
  \bibinfo{pages}{31--73} (\bibinfo{year}{2007}).

\bibitem{Burkov:2011de}
\bibinfo{author}{Burkov, A.~A.} \& \bibinfo{author}{Balents, L.}
\newblock \bibinfo{title}{{Weyl Semimetal in a Topological Insulator
  Multilayer}}.
\newblock \emph{\bibinfo{journal}{Phys. Rev. Lett.}}
  \textbf{\bibinfo{volume}{107}}, \bibinfo{pages}{127205}
  (\bibinfo{year}{2011}).

\bibitem{murakami2007}
\bibinfo{author}{Murakami, S.}
\newblock \bibinfo{title}{{Phase transition between the quantum spin Hall and
  insulator phases in 3D: emergence of a topological gapless phase}}.
\newblock \emph{\bibinfo{journal}{New J. Phys.}} \textbf{\bibinfo{volume}{10}},
  \bibinfo{pages}{029802} (\bibinfo{year}{2008}).

\bibitem{Young:2012kz}
\bibinfo{author}{Young, S.~M.} \emph{et~al.}
\newblock \bibinfo{title}{{Dirac Semimetal in Three Dimensions}}.
\newblock \emph{\bibinfo{journal}{Phys. Rev. Lett.}}
  \textbf{\bibinfo{volume}{108}} (\bibinfo{year}{2012}).

\bibitem{Fang:2012ga}
\bibinfo{author}{Fang, C.}, \bibinfo{author}{Gilbert, M.~J.},
  \bibinfo{author}{Dai, X.} \& \bibinfo{author}{Bernevig, B.~A.}
\newblock \bibinfo{title}{{Multi-Weyl Topological Semimetals Stabilized by
  Point Group Symmetry}}.
\newblock \emph{\bibinfo{journal}{Phys. Rev. Lett.}}
  \textbf{\bibinfo{volume}{108}} (\bibinfo{year}{2012}).

\bibitem{bertlmann2000anomalies}
\bibinfo{author}{Bertlmann, R.~A.}
\newblock \emph{\bibinfo{title}{Anomalies in quantum field theory}},
  vol.~\bibinfo{volume}{91} (\bibinfo{publisher}{Oxford University Press},
  \bibinfo{year}{2000}).

\bibitem{Nielsen1981}
\bibinfo{author}{Nielsen, H.~B.} \& \bibinfo{author}{Ninomiya, M.}
\newblock \bibinfo{title}{{Absence of neutrinos on a lattice}}.
\newblock \emph{\bibinfo{journal}{Nuclear Physics B}}
  \textbf{\bibinfo{volume}{185}}, \bibinfo{pages}{20--40}
  (\bibinfo{year}{1981}).

\bibitem{Son:2013jz}
\bibinfo{author}{Son, D.~T.} \& \bibinfo{author}{Spivak, B.~Z.}
\newblock \bibinfo{title}{{Chiral anomaly and classical negative
  magnetoresistance of Weyl metals}}.
\newblock \emph{\bibinfo{journal}{Phys. Rev. B}} \textbf{\bibinfo{volume}{88}},
  \bibinfo{pages}{104412} (\bibinfo{year}{2013}).

\bibitem{Xu2011}
\bibinfo{author}{Xu, G.}, \bibinfo{author}{Weng, H.}, \bibinfo{author}{Wang,
  Z.}, \bibinfo{author}{Dai, X.} \& \bibinfo{author}{Fang, Z.}
\newblock \bibinfo{title}{{Chern Semimetal and the Quantized Anomalous Hall
  Effect in HgCr$_{2}$Se$_{4}$}}.
\newblock \emph{\bibinfo{journal}{{Phys. Rev. Lett.}}}
  \textbf{\bibinfo{volume}{107}}, \bibinfo{pages}{186806}
  (\bibinfo{year}{2011}).

\bibitem{Yang2011QHE}
\bibinfo{author}{Yang, K.-Y.}, \bibinfo{author}{Lu, Y.-M.} \&
  \bibinfo{author}{Ran, Y.}
\newblock \bibinfo{title}{{Quantum Hall effects in a Weyl semimetal: Possible
  application in pyrochlore iridates}}.
\newblock \emph{\bibinfo{journal}{Phys. Rev. B}} \textbf{\bibinfo{volume}{84}},
  \bibinfo{pages}{075129} (\bibinfo{year}{2011}).

\bibitem{Grushin2012}
\bibinfo{author}{Grushin, A.~G.}
\newblock \bibinfo{title}{{Consequences of a condensed matter realization of
  Lorentz-violating QED in Weyl semi-metals}}.
\newblock \emph{\bibinfo{journal}{Phys. Rev. D}} \textbf{\bibinfo{volume}{86}},
  \bibinfo{pages}{045001} (\bibinfo{year}{2012}).

\bibitem{Parameswaran2014}
\bibinfo{author}{Parameswaran, S.~A.}, \bibinfo{author}{Grover, T.},
  \bibinfo{author}{Abanin, D.~A.}, \bibinfo{author}{Pesin, D.~A.} \&
  \bibinfo{author}{Vishwanath, A.}
\newblock \bibinfo{title}{{Probing the chiral anomaly with nonlocal transport
  in three-dimensional topological semimetals}}.
\newblock \emph{\bibinfo{journal}{Phys. Rev. X}} \textbf{\bibinfo{volume}{4}},
  \bibinfo{pages}{031035} (\bibinfo{year}{2014}).

\bibitem{Zhang:2015ub}
\bibinfo{author}{Zhang, C.} \emph{et~al.}
\newblock \bibinfo{title}{{Detection of chiral anomaly and valley transport in
  Dirac semimetals}}.
\newblock \emph{\bibinfo{journal}{arXiv}}  (\bibinfo{year}{2015}).
\newblock \eprint{1504.07698}.

\bibitem{Behrends:2015ux}
\bibinfo{author}{Behrends, J.}, \bibinfo{author}{Grushin, A.~G.},
  \bibinfo{author}{Ojanen, T.} \& \bibinfo{author}{Bardarson, J.~H.}
\newblock \bibinfo{title}{{Visualizing the chiral anomaly in Dirac and Weyl
  semimetals with photoemission spectroscopy}}.
\newblock \emph{\bibinfo{journal}{arXiv}}  (\bibinfo{year}{2015}).
\newblock \eprint{1503.04329}.

\bibitem{Kim2013}
\bibinfo{author}{Kim, H.-J.} \emph{et~al.}
\newblock \bibinfo{title}{{Dirac versus weyl fermions in topological
  insulators: Adler-bell-jackiw anomaly in transport phenomena}}.
\newblock \emph{\bibinfo{journal}{{Phys. Rev. Lett.}}}
  \textbf{\bibinfo{volume}{111}}, \bibinfo{pages}{246603}
  (\bibinfo{year}{2013}).

\bibitem{Li2014ZrTe5}
\bibinfo{author}{Li, Q.} \emph{et~al.}
\newblock \bibinfo{title}{{Observation of the chiral magnetic effect in
  ZrTe5}}.
\newblock \emph{\bibinfo{journal}{arXiv}}  (\bibinfo{year}{2014}).
\newblock \eprint{1412.6543}.

\bibitem{Xiong2015}
\bibinfo{author}{Xiong, J.} \emph{et~al.}
\newblock \bibinfo{title}{{Signature of the chiral anomaly in a Dirac
  semimetal: a current plume steered by a magnetic field}}.
\newblock \emph{\bibinfo{journal}{arXiv}}  (\bibinfo{year}{2015}).
\newblock \eprint{1503.08179}.

\bibitem{Wang:2015wm}
\bibinfo{author}{Wang, Z.} \emph{et~al.}
\newblock \bibinfo{title}{{Helicity protected ultrahigh mobility Weyl fermions
  in NbP}}.
\newblock \emph{\bibinfo{journal}{arXiv}}  (\bibinfo{year}{2015}).
\newblock \eprint{1506.00924}.

\bibitem{Yang:2015vz}
\bibinfo{author}{Yang, X.}, \bibinfo{author}{Liu, Y.}, \bibinfo{author}{Wang,
  Z.}, \bibinfo{author}{Zheng, Y.} \& \bibinfo{author}{Xu, Z.-A.}
\newblock \bibinfo{title}{{Chiral anomaly induced negative magnetoresistance in
  topological Weyl semimetal NbAs}}.
\newblock \emph{\bibinfo{journal}{arXiv}}  (\bibinfo{year}{2015}).
\newblock \eprint{1506.03190}.

\bibitem{Xu2015NbAs}
\bibinfo{author}{Xu, S.-Y.} \emph{et~al.}
\newblock \bibinfo{title}{{Discovery of Weyl semimetal NbAs}}.
\newblock \emph{\bibinfo{journal}{arXiv}}  (\bibinfo{year}{2015}).
\newblock \eprint{1504.01350}.

\bibitem{Yoshida1975}
\bibinfo{author}{Yoshida, K.}
\newblock \bibinfo{title}{Anomalous electric fields in semimetals under high
  magnetic fields}.
\newblock \emph{\bibinfo{journal}{J. Phys. Soc. Jap.}}
  \textbf{\bibinfo{volume}{39}}, \bibinfo{pages}{1473} (\bibinfo{year}{1975}).

\bibitem{Edel1976}
\bibinfo{author}{Edel'Man, V.}
\newblock \bibinfo{title}{Electrons in bismuth}.
\newblock \emph{\bibinfo{journal}{Advances in Physics}}
  \textbf{\bibinfo{volume}{25}}, \bibinfo{pages}{555--613}
  (\bibinfo{year}{1976}).

\bibitem{Shekhar2015}
\bibinfo{author}{Shekhar, C.} \emph{et~al.}
\newblock \bibinfo{title}{{Extremely large magnetoresistance and ultrahigh
  mobility in the topological Weyl semimetal NbP}}.
\newblock \emph{\bibinfo{journal}{Nat. Phys.}} \textbf{\bibinfo{volume}{11}},
  \bibinfo{pages}{645} (\bibinfo{year}{2015}).

\bibitem{Kapitza1928}
\bibinfo{author}{Kapitza, P.}
\newblock \bibinfo{title}{The study of the specific resistance of bismuth
  crystals and its change in strong magnetic fields and some allied problems}.
\newblock \emph{\bibinfo{journal}{Proc. Roy. Soc. Lond. A}}
  \textbf{\bibinfo{volume}{119}}, \bibinfo{pages}{358--386}
  (\bibinfo{year}{1928}).

\bibitem{Kopelevich:2003aa}
\bibinfo{author}{Kopelevich, Y.} \emph{et~al.}
\newblock \bibinfo{title}{Reentrant metallic behavior of graphite in the
  quantum limit}.
\newblock \emph{\bibinfo{journal}{Phys. Rev. Lett.}}
  \textbf{\bibinfo{volume}{90}}, \bibinfo{pages}{156402}
  (\bibinfo{year}{2003}).

\bibitem{shoenberg1984magnetic}
\bibinfo{author}{Shoenberg, D.}
\newblock \emph{\bibinfo{title}{Magnetic oscillations in metals}}
  (\bibinfo{publisher}{Cambridge University Press}, \bibinfo{year}{1984}).

\bibitem{Rosenman1969}
\bibinfo{author}{Rosenman, I.}
\newblock \bibinfo{title}{Effet shubnikov de haas dans cd 3 as 2: Forme de la
  surface de fermi et modele non parabolique de la bande de conduction}.
\newblock \emph{\bibinfo{journal}{J. Phys. Chem. Solid}}
  \textbf{\bibinfo{volume}{30}}, \bibinfo{pages}{1385--1402}
  (\bibinfo{year}{1969}).

\bibitem{novoselov2005two}
\bibinfo{author}{Novoselov, K.} \emph{et~al.}
\newblock \bibinfo{title}{Two-dimensional gas of massless dirac fermions in
  graphene}.
\newblock \emph{\bibinfo{journal}{nature}} \textbf{\bibinfo{volume}{438}},
  \bibinfo{pages}{197--200} (\bibinfo{year}{2005}).

\bibitem{Singleton2001}
\bibinfo{author}{Singleton, J.}
\newblock \emph{\bibinfo{title}{Band theory and electronic properties of
  solids}}, vol.~\bibinfo{volume}{2} (\bibinfo{publisher}{Oxford University
  Press}, \bibinfo{year}{2001}).

\bibitem{Besara2015}
\bibinfo{author}{Besara, T.} \emph{et~al.}
\newblock \bibinfo{title}{Non-stoichiometry and defects in the weyl semimetals
  taas, tap, nbp, and nbas}.
\newblock \emph{\bibinfo{journal}{arXiv:1511.03221v2}}  (\bibinfo{year}{2015}).

\bibitem{Pippardbook}
\bibinfo{author}{Pippard, A.~B.}
\newblock \emph{\bibinfo{title}{Magnetoresistance in Metals}}.
\newblock No.~\bibinfo{number}{2} in \bibinfo{series}{Cambridge Studies in Low
  Temperature Physics} (\bibinfo{publisher}{Cambridge University Press},
  \bibinfo{year}{1989}).

\bibitem{Reed1971}
\bibinfo{author}{Reed, W.~A.}, \bibinfo{author}{Blount, E.~I.},
  \bibinfo{author}{Marcus, J.~A.} \& \bibinfo{author}{Arko, A.~J.}
\newblock \bibinfo{title}{Anomalous longitudinal magnetoresistance in metals}.
\newblock \emph{\bibinfo{journal}{J. of Appl. Phys.}}
  \textbf{\bibinfo{volume}{42}}, \bibinfo{pages}{5453} (\bibinfo{year}{1971}).

\bibitem{Yoshida1976b}
\bibinfo{author}{Yoshida, K.}
\newblock \bibinfo{title}{An anomalous behavior of the longitudinal
  magnetoresistance in semimetals}.
\newblock \emph{\bibinfo{journal}{J. Phys. Soc. Jap.}}
  \textbf{\bibinfo{volume}{41}}, \bibinfo{pages}{574} (\bibinfo{year}{1976}).

\bibitem{Yoshida1980}
\bibinfo{author}{Yoshida, K.}
\newblock \bibinfo{title}{Transport of spatially inhomogeneous current in a
  compensated metal under magnetic fields. iii. a case of bismuth in
  longitudinal and transverse magnetic fields}.
\newblock \emph{\bibinfo{journal}{J. Appl. Phys.}}
  \textbf{\bibinfo{volume}{51}}, \bibinfo{pages}{4226} (\bibinfo{year}{1980}).

\bibitem{martin1988chemischen}
\bibinfo{author}{Martin, J.} \& \bibinfo{author}{Gruehn, R.}
\newblock \bibinfo{title}{{Zum Chemischen Transport von Monophosphiden MP (M=
  Zr, Hf, Nb, Ta, Mo, W) und Diposphiden MP2 (M= Ti, Zr, Hf)}}.
\newblock \emph{\bibinfo{journal}{Z. Kristallogr}}
  \textbf{\bibinfo{volume}{182}}, \bibinfo{pages}{180} (\bibinfo{year}{1988}).

\bibitem{Rossel1996}
\bibinfo{author}{Rossel, C.} \emph{et~al.}
\newblock \bibinfo{title}{Active microlevers as miniature torque
  magnetometers}.
\newblock \emph{\bibinfo{journal}{J. Appl. Phys.}}
  \textbf{\bibinfo{volume}{79}}, \bibinfo{pages}{8166--8173}
  (\bibinfo{year}{1996}).

\bibitem{kresse1996}
\bibinfo{author}{Kresse, G.} \& \bibinfo{author}{Furthm{\"u}ller, J.}
\newblock \bibinfo{title}{Efficient iterative schemes for ab initio
  total-energy calculations using a plane-wave basis set}.
\newblock \emph{\bibinfo{journal}{Phys. Rev. B}} \textbf{\bibinfo{volume}{54}},
  \bibinfo{pages}{11169} (\bibinfo{year}{1996}).

\bibitem{Becke2006}
\bibinfo{author}{Becke, A.~D.} \& \bibinfo{author}{Johnson, E.~R.}
\newblock \bibinfo{title}{A simple effective potential for exchange}.
\newblock \emph{\bibinfo{journal}{J. Chem. Phys.}}
  \textbf{\bibinfo{volume}{124}}, \bibinfo{pages}{221101}
  (\bibinfo{year}{2006}).

\bibitem{Tran2009}
\bibinfo{author}{Tran, F.} \& \bibinfo{author}{Blaha, P.}
\newblock \bibinfo{title}{Accurate band gaps of semiconductors and insulators
  with a semilocal exchange-correlation potential}.
\newblock \emph{\bibinfo{journal}{Phys. Rev. Lett.}}
  \textbf{\bibinfo{volume}{102}}, \bibinfo{pages}{226401}
  (\bibinfo{year}{2009}).

\bibitem{Mostofi2008}
\bibinfo{author}{Mostofi, A.~A.} \emph{et~al.}
\newblock \bibinfo{title}{{wannier90: A tool for obtaining maximally-localised
  Wannier functions}}.
\newblock \emph{\bibinfo{journal}{Compu. Phys. Commun}}
  \textbf{\bibinfo{volume}{178}}, \bibinfo{pages}{685--699}
  (\bibinfo{year}{2008}).

\end{thebibliography}

\begin{thebibliography}{10}

\bibitem{Uher83}
Uher, C. and Sander, L.
\newblock {\em Phys. Rev. B}{ \bf 27}(2), 1326--1332 (1983).

\bibitem{Reed71}
Reed, W., Blount, E., Marcus, J., and Arko, A.
\newblock {\em J. Appl. Phys.}{ \bf 42}(5453) (1971).

\bibitem{Yoshida76}
Yoshida, K.
\newblock {\em J. Phys. Soc. Jpn.}{ \bf 40}(4) (1976).

\bibitem{Ueda80}
Ueda, Y. and Kino, T.
\newblock {\em J. Phys. Soc. Jpn.}{ \bf 48}(5) (1980).

\bibitem{Pippard}
Pippard, A.
\newblock {\em Magnetoresistance in Metals}.
\newblock Cambridge University Press,  (1989).

\bibitem{Yoshida80}
Yoshida, K.
\newblock {\em J. Appl. Phys.}{ \bf 51}(4226) (1980).

\bibitem{Singleton}
Singleton, J.
\newblock {\em Band Theory and Electronic Properties of Solids}.
\newblock Oxford Master Series in Physics,  (2001).

\bibitem{Shoenberg}
Shoenberg, D.
\newblock {\em Magnetic Oscillations in Metals}.
\newblock Cambridge University Press,  (2009).

\bibitem{Onsager52}
Onsager, L.
\newblock {\em Phil. Mag.}{ \bf 43}(344), 1006--1008 (1952).

\bibitem{Lifshitz56}
Lifshitz, I. and Kosevich, A.
\newblock {\em Sov. Phys. JETP}{ \bf 2}, 636 (1956).

\bibitem{Lifshitz58}
Lifshitz, I. and Kosevich, A.
\newblock {\em J. Phys. Chem. Solids.}{ \bf 4}, 1--10 (1958).

\bibitem{Yamaji89}
Yamaji, K.
\newblock {\em J. Phys. Soc. Jpn.}{ \bf 58}, 1520 (1989).

\bibitem{Yang10}
Yang, H., Fauque, B., Malone, L., Antunes, A., Zhu, Z., Uher, C., and Behnia,
  K.
\newblock {\em Nat. Comms.}{ \bf 47}(1) (2010).

\bibitem{Arnold14Grp}
Arnold, F., Isidori, A., Kampert, E., Yager, B., Eschrig, M., and Saunders, J.
\newblock {\em ArXiv}{ \bf cond-mat:str-el}, 1411.3323v1 (2014).

\bibitem{He14}
He, L., Hong, X., Dong, J., Pan, J., Zhang, Z., Zhang, J., and Li, S.
\newblock {\em Rhys. Rev. Lett.}{ \bf 113}(246402) (2014).

\bibitem{Weng15}
Weng, H., Fang, C., Fang, Z., Bernevig, B., and Dai, X.
\newblock {\em Phys. Rev. X}{ \bf 5}(011029) (2015).

\bibitem{Hubbard11}
Hubbard, S., Kershaw, T., Usher, A., Savchenko, A., and Shytov, A.
\newblock {\em Phys. Rev. B}{ \bf 83}(035122) (2011).

\bibitem{Wang15}
Wang, Z., Zheng, Y., Shen, Z., Zhou, Y., Yang, X., Li, Y., Feng, C., and Xu,
  Z.-A.
\newblock {\em ArXiv}{ \bf cond-mat:str-el}, 1506.00924 (2015).

\bibitem{Luo15}
Luo, Y., Ghimire, N., Wartenbe, M., Neupane, M., McDonald, R., Thompson, E.
  B.~J., and Ronning, F.
\newblock {\em ArXiv}{ \bf cond-mat:str-el}, 1506.01751v1 (2015).

\bibitem{Yang15}
Yang, X., Li, Y., Wang, Z., Zhen, Y., and Xu, Z.
\newblock {\em ArXiv}{ \bf cond-mat:str-el}, 1506.02283v1 (2015).

\bibitem{Dingle52}
Dingle, R.
\newblock {\em Proc. Roy. Soc. A}{ \bf 211}(1107), 517--525 (1952).

\end{thebibliography}
\end{document}